\documentclass[aps,prb,reprint,twocolumn,superscriptaddress,floatfix,nofootinbib]{revtex4-1}
\usepackage{epsfig,amsmath,amssymb,color,comment,physics}
\usepackage[makeroom]{cancel}
\usepackage[caption=false]{subfig}
\usepackage{mathrsfs}
\usepackage[countmax]{subfloat}
\usepackage[normalem]{ulem}
\usepackage[english]{babel}
\usepackage{dsfont}
\usepackage[bookmarks=true,colorlinks,linkcolor=OrangeRed,urlcolor=NavyBlue,citecolor=RoyalBlue]{hyperref}
\usepackage[dvipsnames]{xcolor}
\usepackage{pgfplots}
\pgfplotsset{compat=1.15}
\usetikzlibrary{
        pgfplots.colorbrewer,
    }

\usepackage{multirow} 
\usepackage{graphicx}  

\usepackage{bbold} 
\usepackage{dcolumn}
\usepackage{bm}

\usepackage{siunitx}   
\usepackage{tikz}
\usepackage{calc} 
\usepackage{wasysym} 
\usepackage{mathdesign}

\DeclareMathSymbol{\mhexagon}{\mathord}{wasy}{57}

\definecolor{mygold}{rgb}{0.93,0.59,0.13}

\renewcommand{\vec}[1]{\mathbf{#1}} 
\newcommand{\rots}{{\overline{C}_{6}}}
\newcommand{\epsI}{\epsilon_I} 
\newcommand{\epsS}{\epsilon_S} 
\newcommand{\epsSig}{\epsilon_\Sigma}
\newcommand{\inv}{^{-1}}

\newcommand{\phase}[3][\mathbf{r}_\mu]{\phi_{#2}[\ifthenelse{\equal{#3}{}}{#1}{#3(#1)}]}
 \newcommand{\Ztwo}{\mathbb{Z}_2}
\newcommand{\kvec}{\mathbf{k}}

\newcommand{\hc}{\text{h.c.}}
\newcommand{\UO}{U_{\mathcal{O}}}
\newcommand{\Arg}{\text{Arg}}
\begin{document}

\title{Projective symmetry group classification of chiral $\mathbb{Z}_2$ spin liquids on the pyrochlore lattice: application to the spin-$1/2$ XXZ Heisenberg model}

\author{Benedikt Schneider}
\affiliation{Physics Department, Arnold Sommerfeld Center for Theoretical Physics, and Center for NanoScience, Ludwig Maximilian University Munich, Germany}
\author{Jad C.~Halimeh}
\affiliation{ INO-CNR BEC Center and Department of Physics, University of Trento, Via Sommarive 14, I-38123 Trento, Italy}
\author{Matthias Punk}
\affiliation{Physics Department, Arnold Sommerfeld Center for Theoretical Physics, and Center for NanoScience, Ludwig Maximilian University Munich, Germany}

\begin{abstract}
We give a complete classification of fully symmetric as well as chiral $\Ztwo$ quantum spin liquids on the pyrochlore lattice using a projective symmetry group analysis of Schwinger boson mean-field states. We find 50 independent ansätze, including the 12 fully symmetric nearest-neighbor $\Ztwo$ spin liquids that have been classified by Liu \emph{et al.} [\href{https://journals.aps.org/prb/abstract/10.1103/PhysRevB.100.075125}{Phys.~Rev.~B \textbf{100}, 075125 (2019)}]. For each class we specify the most general symmetry-allowed mean-field Hamiltonian. Additionally, we test the properties of a subset of the spin liquid ansätze by solving the mean-field equations for the spin-$1/2$ XXZ model near the antiferromagnetic Heisenberg point. We find four chiral spin liquids that break the screw symmetry of the lattice modulo time reversal symmetry. These states have a different symmetry than the previously studied monopole flux state and their unique characteristic is a $\frac{\pi}{3}$ flux enclosed by every rhombus of the lattice. 
\end{abstract}

\date{\today}
\maketitle
\tableofcontents

\section{Introduction}

Quantum spin liquids are phases of frustrated magnets which do not exhibit long-range magnetic order down to zero temperature and cannot be classified based on Landau's theory of spontaneous symmetry breaking. In contrast to trivial paramagnetic phases, they exhibit topological order \cite{wen_1990} with long range entanglement and excitations that carry fractional quantum numbers and can have anyonic exchange statistics \cite{Savary_Balents_review_2017}.

A promising platform to study such exotic forms of quantum magnetism are materials where the interplay between electronic correlations and strong spin-orbit coupling gives rise to spin-orbital moments interacting via frustrated exchange interactions \cite{Takagi2019, rau_frustrated_2019}.
The $3d$ rare-earth pyrochlore magnets are an interesting family of frustrated quantum magnets in this class. 
They have the structure $R_2M_2O_7$, with $R$ a trivalent rare-earth ion and $M$ a non-magnetic tetravalent transition metal ion. The former are arranged on a pyrochlore lattice, which consists of corner-sharing tetrahedra. For a subclass of these materials the strong spin-orbit coupling together with the crystal field splitting of the $4f$ orbitals leads to a $j=1/2$ doublet \cite{rau_frustrated_2019}. The small effective spin and the geometrically frustrated pyrochlore lattice enhance spin fluctuations and suppress magnetic ordering in these systems. Prominent examples include $Yb_2Ti_2O_7$ and $Tb_2Ti_2O_7$, which show interesting paramagnetic behavior down to very low temperatures and potentially realize an exotic quantum spin-ice phase \cite{Gardner1999, Molavian2007, ross_quantum_2011, Thompson2011, fennell_power-law_2012}, where the spin dynamics is strongly constrained, following the “two in, two out” ice rule on each tetrahedron. Their low energy properties are described by compact $\mathrm{U}(1)$ gauge theories, which feature magnetic monopole excitations \cite{gingras_quantum_2014}.
 While the microscopic details of these materials are rather complex, their low-energy physics is governed by effective spin-$1/2$ moments, coupled by various symmetry-allowed exchange interactions. Minimal models exhibit dominant Heisenberg interactions, often with an easy axis exchange anisotropy \cite{gingras_quantum_2014}. 
 
 In this work we study the spin-$1/2$ nearest-neighbor XXZ Hamiltonian on the pyrochlore lattice as a minimal model for the description of the above mentioned quantum spin-ice phases. Since quantitatively reliable numerical methods to study frustrated quantum magnets in three dimensions for large system sizes are not available, several properties of its phase diagram are still under debate.

So far, most attention has been focused at the quantum spin-ice phase in the vicinity of the classical Ising limit, where antiferromagnetic easy-axis interactions dominate and transverse exchange interactions are small. 
Recent studies also found a nematic spin liquid for strong antiferromagnetic transverse exchange interactions, which breaks the $\mathrm{U}(1)$ spin rotation symmetry of the XXZ Hamiltonian in the easy-plane, as well as the $C_3$ rotation symmetry of the pyrochlore lattice \cite{benton_quantum_2018, taillefumier_competing_2017}. The nature of the ground-state in the vicinity of the $SU(2)$ symmetric Heisenberg point is still unclear, however. Various possible ground states have been suggested, including dimer-ordered \cite{harris_ordering_1991, berg_singlet_2003, tsunetsugu_antiferromagnetic_2001, tsunetsugu_spin-singlet_2001, canals_quantum_2000, Hagymasi_Possible_2021} and symmetric \cite{canals_pyrochlore_1998, Igbal_Quantum_2019} and symmetry broken \cite{astrakhantsev2021brokensymmetry} spin liquid states, as well as chiral spin liquid states \cite{burnell_monopole_2009, kim_chiral_2008}, which break time reversal symmetry.

In this work we use a projective symmetry group (PSG) approach together with a Schwinger boson representation of the spin operators to provide a complete classification of symmetric as well chiral $\Ztwo$ spin liquid states on the pyrochlore lattice. Here, chiral $\Ztwo$ spin liquids are gapped spin liquids which break time-reversal symmetry. Moreover, lattice symmetries can be broken up to a time reversal transformation. For the PSG construction of chiral spin liquid states we follow the work of Messio \emph{et al.} \cite{messio_time_2013}. As a byproduct we recover the fully symmetric $\Ztwo$ spin liquids previously classified by Liu \emph{et al.} \cite{liu_competing_2019}. In order to characterize the newly constructed chiral ansätze we use Schwinger boson mean-field theory (SBMFT) and solve the mean-field equations to compare their ground-state energies. Furthermore, we calculate static spin structure factors to characterize spin correlations in these states.

The outline of the paper is as follows: In Sec.~\ref{sec:Model} we introduce the local XXZ model and develop a general mean-field decoupling in terms of bond operators within SBMFT. 
In Sec.~\ref{sec:MeanFieldAnsatz} we use PSG to systematically classify  symmetric and chiral mean-field ansätze. The detailed calculations can be found in Appendix \ref{Appendix:SolChiralPSG} and \ref{Appendix:FluxTransform}. After choosing reasonable ansätze (Sec.~\ref{Sec:ChoosingAnsätze}) we diagonalize the Hamiltonian and calculate free energies in Sec.~\ref{Sec:Diagonalization}. In Sec.~\ref{sec:SpinStructureFactors} we calculate static spin structure factors. Finally, we discuss our results in Sec.~\ref{sec:Discussion}.

\section{Model and methods}
\label{sec:Model}

\subsection{Pyrochlore lattice}

\begin{figure}
\subfloat[\label{fig:PyrochloreLatticeStdUnitCell}]{%
  \includegraphics[width=.49\linewidth]{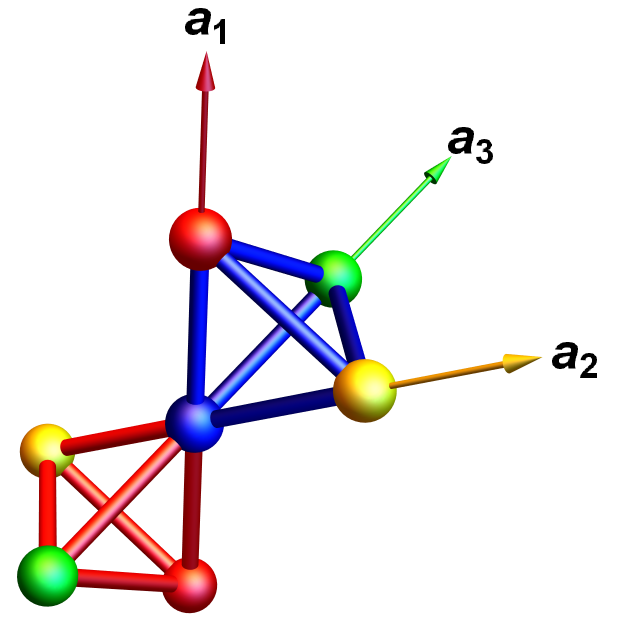}%
} 
\subfloat[ ]{%
  \includegraphics[width=.49\linewidth]{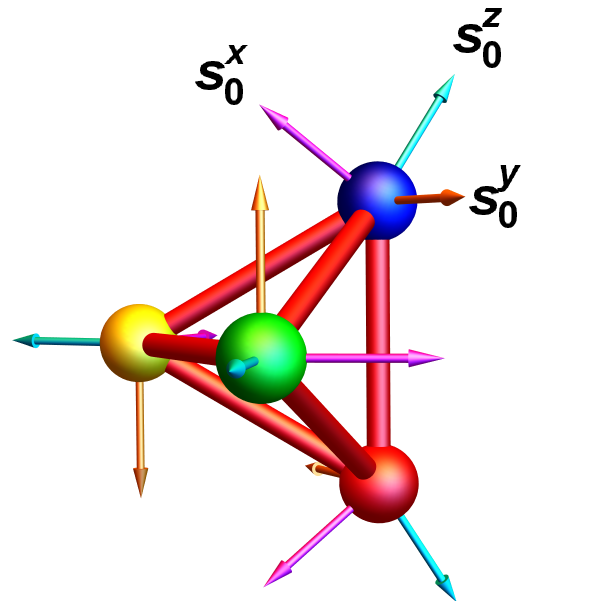}%
}\hfill
\subfloat[ ]{%
  \includegraphics[width=.49\linewidth]{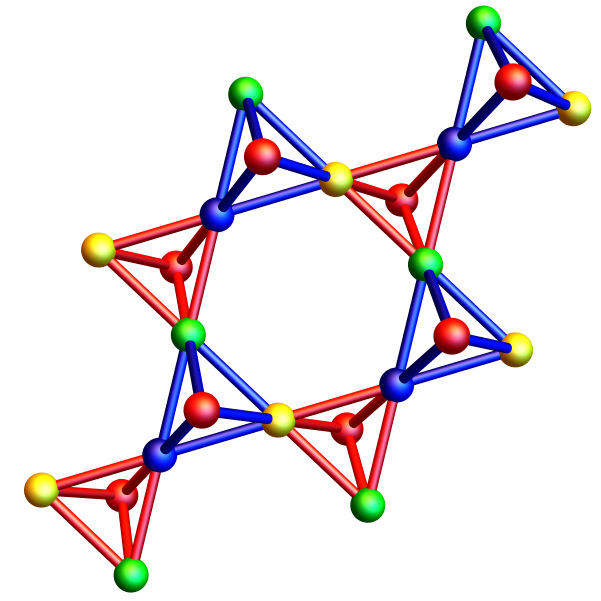}%
}
\caption{(Color online). (a): A unit cell of the pyrochlore lattice. nearest-neighbor sites are connected by bonds. Blue bonds (main tetrahedron) are within a unit cell and red bonds (inverse tetrahedron) are between neighboring unit cells.  (b): The local spin basis of the inverse tetrahedron \eqref{equation:localbasis}. (c): The enlarged unit cell consisting of eight tetrahedra.}
\label{fig:PyrochloreLattice}
\end{figure}

The lattice is spanned by the FCC-basis vectors  $\mathbf{a}_1 = \frac{1}{2}(0,1,1)$,  $\mathbf{a}_2 = \frac{1}{2}(1,0,1)$,  $\mathbf{a}_3 = \frac{1}{2}(1,1,0)$ and has four sublattices. We will include the zero vector as a fourth lattice vector for later convenience: $\mathbf{a}_0 = (0,0,0)$.  Also, for easier use of symmetries, we introduce the sublattice coordinates:
\begin{align}
    \mathbf{r}_\mu &= (r_1,r_2,r_3)_\mu := r_1\mathbf{a}_1 + r_2\mathbf{a}_2 + r_3\mathbf{a}_3 +\frac{1}{2}\mathbf{a_\mu} \\
    &= \frac{1}{2}(r_2+r_3,r_1+r_2,r_1+r_3)+ \frac{1}{2}\mathbf{a}_\mu. \nonumber
\end{align}
We will refer to the tetrahedra spanned by  $\{\mathbf{r}_\mu:\mu = 0,1,2,3\}$  and $\{\mathbf{r}_\mu-\vec{a_\mu}:\mu = 0,1,2,3\}$ as the main and inverse tetrahedron, respectively, and label them with $\mathbf{r}$. The spin operators in the local basis are defined as:
\begin{equation}
\label{eq:GeneralSpinBasis}
    \mathbf{\hat{S}}_{\vec{r_\mu}} = (\hat{S}^x,\hat{S}^y,\hat{S}^z)_\mu = \hat{S}_{\vec{r_\mu}}^x \mathbf{s}^x_\mu +  \hat{S}_{\vec{r_\mu}}^y \mathbf{s}^y_\mu + \hat{S}^z_{\vec{r_\mu}} \mathbf{s}^z_\mu.
\end{equation}
The local basis vectors $\mathbf{s}^\gamma_\mu$ are defined in Appendix \ref{appendix:SpinBasis}. 
The space group of the lattice is  $Fd\Bar{3}m$ (No.227) which we will later refer to as $\chi$. It is generated by the translations $T_1$,$T_2$,$T_3$ along the lattice vectors, a sixfold rotoreflection $\rots $ around the $\mathbf{s}^z_0$ axis and a screw operation $S$ around the $\vec{a}_3$ axis.\cite{liu_competing_2019}  The rotoreflection can be constructed by inversion $I$ and $C_3$ rotation around the $\mathbf{s}^z_0$ axis:  $\rots = C_3 I$. We denote time reversal symmetry as $\mathcal{T}$.
In Appendix \ref{appendix:SpinBasis} we list how the sublattice coordinates and local spin basis transform under symmetries of the space group as well as the algebraic group relations.

\subsection{Model and Schwinger-boson mean-field theory}
\label{subsec:SBMFT}
The XXZ model is given by the following Hamiltonian
\begin{align}
    H =& \sum_{<ij>}J_{\perp}(\hat{S}^x_i\hat{S}^x_j+\hat{S}^y_i\hat{S}^y_j)+ J_{zz} \hat{S}^z_i\hat{S}^z_j, \label{eq:ModelHamiltonian}
\end{align}
where $\hat{S}^\gamma_i$ is the $\gamma \in \{x,y,z\}$ component of the spin-$1/2$ operator on lattice site $i$ in the local basis and the sums run over nearest-neighbor bonds on the pyrochlore lattice. For $J_{zz} = J_\perp$ the model reduces to the Heisenberg model in the local spin basis, which is the Klein dual of the Heisenberg model in the global spin basis \cite{rau_frustration_2018}. This will later enable us to compare results from both models. \\
We use the parametrization 
\begin{align}
    &J_{zz} = J\cos{(\theta)}, &J_\perp = J\sin{(\theta)}, \label{eq:Parametrization}
\end{align}
and set  $J = \sqrt{J_{zz}^2+J_{\perp}^2} = 1$.
The spin operators can be represented in terms of Schwinger bosons
\begin{align}
    \hat{S}^\gamma_i =  \frac{1}{2}\hat{b}^\dagger_{i}\sigma^\gamma\hat{b}_{i}, \label{eq:PartonRepresentation}
\end{align}
where $\hat{b}^\dagger = (\hat{b}_{i,\uparrow}^\dagger , \hat{b}_{i,\downarrow}^\dagger)$ are bosonic creation operators satisfying $\comm{b_{i,\alpha}}{b\dagger_{j,\beta}}=\delta_{ij}\delta_{\alpha,\beta}$ and $\sigma^\gamma$ are the Pauli matrices. The Schwinger boson representation is invariant under $\mathrm{U}(1)$ gauge transformations
\begin{equation}
\label{eq:GaugeTransformation}
    G: \hat{b}_j \xrightarrow{} e^{i\phi_G[j]}\hat{b}_j,
\end{equation}
with $\phi_G[j]$ a lattice site dependent phase. To ensure that the operators in Eq.~\eqref{eq:PartonRepresentation} obey the spin algebra
we have to constrain the boson density per site to $2\mathcal{S}$
\begin{equation}
    \hat{n_i}=\hat{b}_{i}^\dagger\hat{b}_{i}=2\mathcal{S}. \label{eq: Boson Density Constraint}
\end{equation}
This projection can be achieved by adding a site-dependent Lagrange multiplier $\sum_i\lambda_i(\hat{n}_i-2\mathcal{S})$\cite{wang_spin_2006} to the Hamiltonian.\\
At this point it is possible to decouple the Hamiltonian in terms of the hopping singlet $\hat{B}_{ij}$ and triplet $\hat{t}_{ij}^{h,\gamma}$ as well as pairing singlet $\hat{A}_{ij}$ and triplet $\hat{t}_{ij}^{p,\gamma}$ operators
\begin{align}
    \hat{B}_{ij} &= \frac{1}{2}b_{i }^\dagger b_{j }, && \hat{A}_{ij}  = \frac{1}{2}b_{i}(i\sigma^2)b_{j},\\
    \hat{t}_{ij}^{h,\gamma} & =\frac{i}{2}  b_{i }^\dagger\sigma^\gamma b_{j }, &&
    \hat{t}_{ij}^{p,\gamma}   =-\frac{i}{2}b_{i }(\sigma^\gamma\cdot i\sigma^2) b_{j }, \nonumber
\end{align}
with $\gamma \in \{x,y,z\}$ by using the identities
\begin{subequations}
\begin{align}
    \mathbf{\hat{S}}_i\mathbf{\hat{S}}_j    =& :\hat{B}_{ij}^\dagger \hat{B}_{ij}: -\hat{A}_{ij}^\dagger \hat{A}_{ij},\\
        \hat{S}^\gamma_i \hat{S}^\gamma_j  =& :\hat{B}_{ij}^\dagger \hat{B}_{ij}: -\hat{t}_{ij}^{p, \gamma\dagger} \hat{t}_{ij}^{p,\gamma}\\
                    =& :\hat{t}_{ij}^{h, \gamma\dagger}\hat{t}_{ij}^{h, \gamma}: - \hat{A}_{ij}^\dagger \hat{A}_{ij},
\end{align}
\end{subequations}
where $:\;\;:$ denotes normal ordering.
We arrive at a Hamiltonian of the form
\begin{align}
    H = \sum_{<i,j>}:\mathbf{\hat{h}}_{ij}^\dagger J^h_{ij} \mathbf{\hat{h}}_{ij}: + \mathbf{\hat{p}}_{ij}^\dagger J^p_{ij} \mathbf{\hat{p}}_{ij} + C_{ij} + \sum_i\lambda_i(\hat{n}_i-2\mathcal{S}) \label{eq:HoppingPairingHamiltonian},
\end{align}
where $\mathbf{\hat{h}}_{ij}^\dagger = (\hat{B}^\dagger_{ij}, \hat{t}^{h,x\dagger}_{ij},\hat{t}^{h,y \dagger}_{ij}, \hat{t}^{h,z \dagger}_{ij})$ and $\mathbf{\hat{p}}_{ij}^\dagger = (\hat{A}^\dagger_{ij}, \hat{t}^{p,x\dagger}_{ij},\hat{t}^{p,y \dagger}_{ij}, \hat{t}^{p,z \dagger}_{ij})$ are vectors of the hopping and pairing operators. $J^h_{ij} $ and $J^p_{ij} $ are the hopping and pairing coupling matrices that depend on $J_{ij}$ and $C_{ij}$ is a constant.\\
On an empty lattice $\ket{0}$ that satisfies $\hat{b}_{i,\uparrow}\ket{0} = \hat{b}_{i,\downarrow}\ket{0}=0$ for all $i$, $\hat{A}_{ij}^\dagger$ creates a spin singlet between sites $i$ and $j$. $\hat{t}_{ij}^{p, \gamma\dagger}$ creates a spin triplet with direction $\gamma$ between sites $i$ and $j$. The hopping operator $\hat{B}_{kj}$ moves these singlet or triplet from the sites $i$ and $j$ to site $i$ and $k$: $\hat{B}_{kj}\hat{A}_{ij}^\dagger\ket{0} \sim \hat{A}_{ik}^\dagger\ket{0}$. Finally, the triplet hopping operators also move the spin singlets and triplets around while simultaneously changing their flavor.
For example: $\hat{t}_{kj}^{h, z}\hat{A}_{ij}^\dagger\ket{0} \sim \hat{t}_{ik}^{p, z\dagger}\ket{0}$ or $\hat{t}_{kj}^{h, z}  \hat{t}_{ij}^{p, x\dagger}\ket{0} \sim \hat{t}_{ik}^{p, y \dagger}\ket{0}$.\\
Note, that the decoupling is not unique since for $i \neq j$ :
\begin{align}
    &:\hat{B}_{ij}^\dagger \hat{B}_{ij}: + \hat{A}_{ij}^\dagger \hat{A}_{ij} \nonumber \\
    =\; & :\hat{t}^{h,\gamma \dagger}_{ij} \hat{t}^{h,\gamma}_{ij}: +  \hat{t}^{p,\gamma \dagger}_{ij} \hat{t}^{p,\gamma}_{ij} = \frac{1}{4}\hat{n}_i\hat{n}_j = \mathcal{S}^2 \label{eq:BondRelation},
\end{align}
where in the last equality we explicitly used the boson density constraint from Eq ~\eqref{eq: Boson Density Constraint}. It is therefore possible to set either  $J^h_{ij} $ or $J^p_{ij} $ to zero and only describe the system in terms of hopping or pairing terms. However, once we apply the mean-field approximation our choice of parametrization greatly effects our results. A theory with $J^p_{ij}=0$ can only describe magnetically ordered states while a theory with $J^h_{ij}=0$ has been shown to lead to quantitatively worse results for ground-state energy and dynamical spin structure factor \cite{mezio_test_2011, Flint_Symplectic_2009}. 

We therefore choose to keep both terms. Another operator identity is:
\begin{equation}
\hat{A}_{ij}^\dagger \hat{A}_{ij} = -\hat{t}^{p,x \dagger}_{ij} \hat{t}^{p,x}_{ij} - \hat{t}^{p,y \dagger}_{ij} \hat{t}^{p,y}_{ij} - \hat{t}^{p,z \dagger}_{ij} \hat{t}^{p,z}_{ij} + 2\mathcal{S}^2. \label{eq:OperatorIdentityAtoT} 
\end{equation}
We deal with this ambiguity by choosing parametrizations that preserve the SU(2) symmetry at the Heisenberg point explicitly (see Sec.~\ref{Sec:ChoosingAnsätze}).\\
To treat the parametrized Hamiltonian \eqref{eq:HoppingPairingHamiltonian} we make two standard approximations. Firstly, we consider only a site independent Lagrange multiplier $\lambda_i = \lambda$. This results in the boson density constraint \eqref{eq: Boson Density Constraint} being fulfilled only on average. Secondly, we apply a mean-field approximation: 
\begin{subequations}
\label{eq:MeanFieldApproximationBondOperators}
\begin{align}
    \mathbf{\hat{h}}_{ij}^\dagger J^h_{ij}\mathbf{\hat{h}}_{ij} & \approx  \mathbf{\hat{h}}_{ij}^\dagger J^h_{ij}\mathbf{h}_{ij} +
     \mathbf{h}_{ij}^\dagger J^h_{ij}\mathbf{\hat{h}}_{ij} -
      \mathbf{h}_{ij}^\dagger J^h_{ij}\mathbf{h}_{ij}, \\
          \mathbf{\hat{p}}_{ij}^\dagger J^p_{ij}\mathbf{\hat{p}}_{ij} & \approx  \mathbf{\hat{p}}_{ij}^\dagger J^p_{ij}\mathbf{p}_{ij} +
     \mathbf{p}_{ij}^\dagger J^p_{ij}\mathbf{\hat{p}}_{ij} -
      \mathbf{p}_{ij}^\dagger J^p_{ij}\mathbf{p}_{ij},   
\end{align}
\label{eq:MeanFieldApprox}
\end{subequations}
where
\begin{subequations}
\begin{align}
    \mathbf{h}^\dagger_{ij} &= \expval{\mathbf{\hat{h}^\dagger} _{ij}}=(\mathcal{B}^*_{ij},t^{h,x*}_{ij},t^{h,y*}_{ij},t^{h,z*}_{ij}),\\
    \mathbf{p}^\dagger_{ij} &= \expval{\mathbf{\hat{p}^\dagger} _{ij}}=(\mathcal{A}^*_{ij},t^{p,x*}_{ij},t^{p,y*}_{ij},t^{p,z*}_{ij}).
\end{align} 
\end{subequations}
This leaves us with a Hamiltonian that is quadratic in boson operators
\begin{align}
    H =& \sum_{<i,j>}\hat{b}_i^\dagger u^h_{ij} \hat{b_j} +\hat{b}^\dagger_i u^p_{ij} \hat{b}_j^\dagger + \hc +  f(\mathbf{h}_{ij},\mathbf{p}_{ij}) \label{eq:MFH}\\
    &+ \lambda \sum_i(\hat{b}_i^\dagger \hat{b}_i-2\mathcal{S}). \nonumber 
\end{align} 
Here, $f$ is given by 
\begin{equation}
    f(\mathbf{h}_{ij},\mathbf{p}_{ij}) = - \mathbf{h}_{ij}^\dagger J^h_{ij} \mathbf{h}_{ij} - \mathbf{p}_{ij}^\dagger J^p_{ij} \mathbf{p}_{ij} + C_{ij},
\end{equation} where $C_{ij}$ is constant
while $u^h_{ij}$ and $u^p_{ij}$ are complex $2\times2$ matrices defined by 
\begin{subequations}
\label{eq:uMatrixShortNotation}
\begin{align}
    u^h_{ij} &= \frac{1}{2}\sum_{m = 0}^3 i^{1-\delta_{m,0}}(\mathbf{h}^\dagger_{ij}J^h_{ij})^m\sigma^m &  \nonumber\\
    &\equiv a^h_{ij}\sigma^0 + i (b^h_{ij}\sigma^1 + c^h_{ij}\sigma^2 + d^h_{ij}\sigma^3) \nonumber \\
    &\equiv(a^h_{ij} ,b^h_{ij} ,c^h_{ij} ,d^h_{ij} ),\\
    u^p_{ij} &= \frac{1}{2}\sum_{m = 0}^3 i^{1-\delta_{m,0}}(J^p_{ij}\mathbf{p}_{ij})^m\sigma^m(i\sigma^2) \nonumber\\
        &\equiv a^p_{ij}i\sigma^2 + i (b^p_{ij}\sigma^1 + c^p_{ij}\sigma^2 + d^p_{ij}\sigma^3)(i\sigma^2) \nonumber \\ &\equiv(a^p_{ij},b^p_{ij},c^p_{ij},d^p_{ij}).
\end{align}
\end{subequations}
 This notation is adapted from Liu \emph{et al.}\cite{liu_competing_2019} and is particularly helpful, since $a^h_{ij}$ and $a^p_{ij}$ transform as scalars while $(b^h_{ij} ,c^h_{ij} ,d^h_{ij})$ and $(b^p_{ij} ,c^p_{ij} ,d^p_{ij})$ transform as $SO(3)$ vectors. The parameters $a^h_{ij},\ldots,d^h_{ij}$ are functions of the mean-fields $\mathbf{h}_{ij}$ and the coupling matrix $J^h_{ij}$ and appear as prefactors to the operators $\hat{B}_{ij},\ldots,\hat{t}^{h,z}_{ij}$ in the Hamiltonian. Similarly, the parameters $a^p_{ij},\ldots,d^p_{ij}$ are functions of the mean-fields $\mathbf{p}_{ij}$ and the coupling matrix $J^p_{ij}$ and appear as prefactors to the operators $\hat{A}_{ij},\ldots,\hat{t}^{p,z}_{ij}$ in the Hamiltonian. When exchanging $i \leftrightarrow{} j$ the matrices $u^h_{ji},u^p_{ji}$ transform like  $u^h_{ji}=(u^h_{ij})^\dagger$ and $u^p_{ji}=(u^p_{ij})^T$ and the parameters transform like $(a^h_{ji} ,b^h_{ji} ,c^h_{ji} ,d^h_{ji}) = (a^{h*}_{ij} ,- b^{h*}_{ij} ,-c^{h*}_{ij} ,- d^{h*}_{ij})$ and $(a^p_{ji},b^p_{ji},c^p_{ji},d^p_{ji})= (-a^p_{ij},b^p_{ij},c^p_{ij},d^p_{ij})$.
The set of matrices $u^h_{ij}$ and $u^p_{ij}$ or rather the set of expectation values $\mathbf{h}_{ij}$ and $\mathbf{p}_{ij}$ are known as the mean-field ansatz.\\
The Hamiltonian \eqref{eq:MFH} is the most general nearest-neighbor mean-field Hamiltonian. To investigate spin liquid states in the XXZ model we have to choose a mean-field decoupling of the model Hamiltonian \eqref{eq:ModelHamiltonian} which fixes $J^h_{ij}$, $J^p_{ij}$ and $C_{ij}$ and an ansatz which fixes $\mathbf{h}_{ij}$ and $\mathbf{p}_{ij}$. \\
Once an ansatz is chosen the mean-field Hamiltonian \eqref{eq:MFH} can be diagonalized by a Bogoliubov transform, a ground state can be constructed and the values of  $\mathbf{h}_{ij}$ and $\mathbf{p}_{ij}$ have to be solved self consistently:
\begin{align}
    \mathbf{h}_{ij} = \expval{\mathbf{\hat{h}}_{ij}}, &&  \mathbf{p}_{ij} = \expval{\mathbf{\hat{p}}_{ij}}, && 2\mathcal{S} = \frac{1}{N}\sum_i \expval{\hat{n}_i}. \label{eq:selfconsistency}
\end{align}

\section{Mean-field ansätze}
\label{sec:MeanFieldAnsatz}
Motivated by the
chiral spin liquid states found by Burnell \emph{et al.}~ \cite{burnell_monopole_2009} and Kim \emph{et al.}~\cite{kim_chiral_2008} using a fermionic parton construction, we consider general chiral ansätze that fulfill all lattice symmetries modulo time reversal. We classify all possible ansätze with the PSG method introduced by Wen \cite{wen_1990} for fermionic partons and later generalized by Wang and Vishvanath to Schwinger bosons \cite{wang_spin_2006} to symmetric  spin liquids. In particular, we follow the strategy from Messio \emph{et al.} \cite{messio_time_2013}, who generalized the bosonic PSG to chiral ansätze where time reversal symmetry and lattice symmetries modulo time reversal are broken. Due to the $\mathrm{U}(1)$ gauge symmetry of the Schwinger boson representation \eqref{eq:PartonRepresentation} the mean-field ansatz does not have to be strictly symmetric under all lattice symmetries $\mathcal{O}$ but can in general be symmetric under the gauge enriched lattice symmetries $ \widetilde{\mathcal{O}}$:
\begin{equation}
    \widetilde{\mathcal{O}} = G_\mathcal{O}\mathcal{O}: \hat{b}_i \xrightarrow{} e^{i\phi_\mathcal{O}[\mathcal{O}(i)]}\UO^\dagger \hat{b}_{\mathcal{O}(i)} .
\end{equation}
The set of gauge inequivalent phases $\phi_\mathcal{O}[i]$ are defined by the algebraic PSG. The gauge transformations $G_\mathbb{1}$ are elements of the so called Invariant Gauge Group (IGG).
Since the pyrochlore lattice is not bipartite and we are interested in ansätze with both hopping and pairing terms we have to consider an IGG of $\Ztwo$ \cite{wang_spin_2006}. Before classifying the chiral ansätze it is useful to first revisit the fully symmetric ansätze classified by Liu \emph{et al.} \cite{liu_competing_2019}.

\subsection{Symmetric ansätze}
Fully symmetric ansätze can be constructed by fixing $u^h_{\Vec{0}_0\Vec{0}_1}=(a^h,b^h,c^h,d^h)$ and $u^p_{\Vec{0}_0\Vec{0}_1}=(a^p,b^p,c^p,d^p)$ on the bond $\Vec{0}_0\xrightarrow{} \Vec{0}_1$ and then mapping them onto all other bonds by symmetry operations:
\begin{subequations}
\label{eq:TransformationMeanFieldAnsatz}
\begin{align}
    u^h_{\mathcal{O}(ij)} &= \UO u^h_{ij} \UO^\dagger e^{-i(\phi_\mathcal{O}[\mathcal{O}(i)] - \phi_\mathcal{O}[\mathcal{O}(j)])}, \\
    u^p_{\mathcal{O}(ij)} &= \UO u^h_{ij} \UO^T e^{-i(\phi_\mathcal{O}[\mathcal{O}(j)] + \phi_\mathcal{O}[\mathcal{O}(i)])} . 
\end{align}
\end{subequations}
$\UO$ are the $SU(2)$ matrices associated with the symmetry operations $\mathcal{O}$ (Appendix \ref{appendix:SU2Matrices}). The algebraic PSG has been solved by Liu \emph{et al.}~\cite{liu_competing_2019}.
They found 16 different $\Ztwo$ PSG equivalent classes defined by the phases:
\begin{subequations}
\label{eq:PhaseEquationsSymmetricSL}
\begin{align}
        \phase{T_1}{} =& 0,\\
        \phase{T_2}{} =& n_1\pi r_1,\\
        \phase{T_3}{}  =& n_1\pi(r_1 +r_2),\\
        \phase{\mathcal{T}}{} =& 0,\\
        \phase{\rots}{} =&  \delta_{\mu,1,2,3}(n_{ST_1}-n_1)\pi               
                     -  r_1\delta_{\mu,2,3}n_1\pi \nonumber\\
                    &-  r_2 n_{\rots T_1}\pi                         
                     -  r_3\delta_{\mu,2}n_1\pi \nonumber\\   
                    &-   n_1\pi(r_1r_2+r_1r_3),  \\
        \phase{S}{} =& ((-)^{\delta_{\mu,1,2,3}}\frac{(n_{ST_1}-n_1)}{2} + \delta_{\mu,2}n_{\rots S})\pi  \nonumber\\ 
                &  +r_1\pi(n_1\delta_{\mu,1,2} - n_{ST_1}) \nonumber\\
                &  + r_2\pi(n_1\delta_{\mu,2}   - n_{ST_1})  
                   + r_3\pi n_1\delta_{\mu,1,2} \nonumber\\           
                &  - \frac{n_1\pi}{2}(r_1 + r_2)(r_1 + r_2 + 1),
\end{align}
\end{subequations}
where $n_1,\; n_{ST_1} \; n_{\rots S}, \; n_\rots$ are all $\Ztwo$ parameters that are either $0$ or $1$. The ansätze will be labeled by $n_1\pi-(n_{\rots S}n_{ST_1}n_\rots)$. When $n_1= 1$, translation symmetry is realized projectively and the unit cell is enlarged. Depending on the PSG equivalence class the ansatz forces some of the mean-field parameters $a^h,b^h,c^h,d^h,a^p,b^p,c^p,d^p$ to be zero. Liu \emph{et al.}~\cite{liu_competing_2019} give a table of all independent non zero parameters in the global spin basis. We are, however, interested in the local basis. We transform their solution to the local spin basis by 
\begin{equation}
    (a^h_l,b^h_l,c^h_l,d^h_l)^h = U_0 (a^h_g,b^h_g,c^h_g,d^h_g)^h U_1^\dagger, \label{eq:FromGlobalToLocal}
\end{equation}
where the subscripts $l$ and $g$ are for "local" and "global" respectively.
The matrices $U_\mu$ are the $SU(2)$ matrices corresponding to the transformation from global to local spin basis on sublattice $\mu$. They are specified in Appendix \ref{appendix:SU2Matrices}. Eq.~\eqref{eq:FromGlobalToLocal} gives us explicitly
\begin{subequations}
\label{eq:ParameterGlobaltoLocal}
\begin{align}
    a_l &= -b_g,\\
    b_l &= \frac{1}{\sqrt{6}}(-2a_g+c_g-d_g), \\
    c_l &= \frac{1}{\sqrt{2}}(c_g+d_g), \\
    d_l &= \frac{1}{\sqrt{3}}(a_g+c_g-d_g),
\end{align}
\end{subequations}
Based on Eq.~\eqref{eq:ParameterGlobaltoLocal} we can translate their solution into Table \ref{Table:AllowedPSGParameters}. It lists all independent non zero nearest-neighbor parameters. We use the parameters $n_1$-$(n_{\rots S}n_{S\rots}n_\rots)$ to label the states while Liu \emph{et al.} use $n_1$-$(n_{\rots S}n_{ST_1}n_\rots)$, where $n_{S\rots} = n_1 + n_{ST_1} + n_\rots $. The four classes $n_1$-(00$n_\rots$)  have an accidental IGG of $U(1)$ at nearest-neighbor level, since they don't allow any non-zero nearest-neighbor pairing fields. Therefore, one can construct 12 different fully symmetric $\Ztwo$ spin liquid ansätze at nearest-neighbor level.

\begin{table}[h!]
    \centering
    \begin{tabular}{cl}
         $n_1$-$(n_{\rots S}n_{S\rots}n_\rots)$ & \; NN \\
         \hline\\
          $n_1$-(00$n_\rots$)& $b^h,d^h$        \\
         $n_1$-(01$n_\rots$)&  $b^h,d^h,a^p$    \\
         $n_1$-(10$n_\rots$)&  $a^h,c^p$        \\
         $n_1$-(11$n_\rots$)&  $a^h,b^p,d^p$    \\
    \end{tabular}
    \caption{All independent non-zero nearest-neighbor mean-field parameters for the 16 different $\Ztwo$ PSG equivalence classes in the local spin basis. Fields are fixed on bond $\mathbf{0}_0 \xrightarrow{} \mathbf{0}_1 $. All other nearest-neighbor parameters are constrained to be zero. The table is translated from Liu \emph{et al.}\cite[Table II]{liu_competing_2019}  by using Eq.~\eqref{eq:ParameterGlobaltoLocal}. }
    \label{Table:AllowedPSGParameters}
\end{table}

\subsection{Chiral ansätze}
\label{sec:Weakly Symmetric ansätze}
Chiral spin liquids break time reversal symmetry and some lattice symmetries modulo a global spin flip (action of time reversal symmetry) \cite{messio_time_2013}. In the classical limit $\mathcal{S}\xrightarrow{} \infty$ they correspond to non-coplanar spin states (i.e., they have non zero scalar spin chirality  $\expval{\hat{S}_i\cdot(\hat{S}_j\times\hat{S}_k)} \neq 0$). To construct a chiral ansatz we start by defining a parity $\epsilon_{\mathcal{O}}$ for each symmetry operator $\mathcal{O}\in \chi$ in the lattice space group $\chi = Fd\Bar{3}m$. $\epsilon_{\mathcal{O}}= 1$  when an ansatz respects the symmetry and it is  $\epsilon_{\mathcal{O}}= -1$ when it only respects the ansatz modulo a time reversal. Let us define the subgroup $\chi_e$ of all lattice symmetries that necessarily have even parity $\epsilon_{\mathcal{O}}= 1$ and and the set of operators  with undetermined parity as $\chi_o = (\chi-\chi_e)$. $\chi_e$ contains at least all squares of symmetry operators $T_1^2,T_2^2,T_3^2,S^2$ , $\rots^2 = I^2 C_3^2 = C_3\inv$ since their parities are $\epsilon_{\mathcal{O}} = (\pm1)^2=1$. We can translate the algebraic group relations (Eq.~\eqref{eq:algebraicRelations}) into equations for the parity to find more generators of $\chi_e$.
The nontrivial equations are: 
\begin{subequations}
\label{eq:parity equations}
\begin{align}
    \epsilon_{S^2}\epsilon_{T_3} &= 1, \label{eq:parityEQ1}\\
    \epsilon_{C_3} \epsilon_{T_i} &= \epsilon_{T_{i+1}}\epsilon_{C_3} .\label{eq:parityEQ2}
\end{align}
\end{subequations}
Eq.~\eqref{eq:parityEQ1} shows that $T_3$ has even parity. Therefore, Eq.~\eqref{eq:parityEQ2} implies that this is also true for $T_1$ and $T_2$. The parities of $\rots$ and $S$ stay undetermined. This concludes the analysis following Messio \emph{et al.} \cite{messio_time_2013}. We are, however, still missing one generator of $\chi_e$. In general, once generators of even and undetermined parity are found by inspecting the generators of the full symmetry group $\chi$, we also have to consider operators of the form  $\mathcal{O}_o^{-1}\mathcal{O}_e\mathcal{O}_o$ where $\mathcal{O}_o \in \chi_o$ and $\mathcal{O}_e \in \chi_e$. This can be repeated until no new generators of $\chi_e$ are found.  With this approach we can construct the symmetry operator $C_3':=I S C_3 I S = S^{-1} C_3 S$ which has even parity $\epsilon_{C_3'}=\epsS^2\epsI^2\epsilon_{C_3}=1$. $C_3'$ is a $\frac{2\pi}{3}$ rotation about the $\mathbf{s}_3^z$ axis on the inverse tetrahedron. Since $C_3'$ cannot be written as a product of the operators $\{T_1, T_2, T_3, C_3\}$ we have to add it to the set of generators. $I C_3 I = C_3$ gives no new generator and therefore $\chi_e$ is generated by $\{T_1, T_2, T_3, C_3, C_3'\}$  while $\rots,S\in \chi_o$.  
The algebraic relations of $\chi_e$ are
\begin{subequations}
\label{eq:AlgebraicRelationsChie}
\begin{align}
    T_iT_{i+1}T_i\inv T_{i+1}\inv   & = 1,\\
                            C_3^3     & = 1,\\
                            C_3'^3     & = 1, \\
            (C_3  C_3')^2           & = 1,\\
            C_3 T_i C_3\inv T_{i+1}\inv & = 1, \\
            C_3' T_1 (C_3') \inv T_{1} T_{2}\inv & = 1,\\
            C_3' T_2 (C_3')\inv T_{1} & = 1,\\
            C_3' T_3 (C_3') \inv T_{1} T_{3}\inv & = 1,
\end{align}
\end{subequations}
where $i= i+3$.
The chiral algebraic PSG is then defined as the algebraic PSG of $\chi_e$. We solve the chiral algebraic PSG in Appendix \ref{Appendix:SolChiralPSG}.  The phases are given by: 
\begin{subequations}
 \begin{align}
        \phase{T_1}{} &= 0,\\
        \phase{T_2}{} &= n_1\pi r_1,\\
        \phase{T_3}{} &= n_1\pi(r_1 +r_2),\\
    \phase{C_3}{} &= \frac{2\pi \xi}{3} \delta_{\mu, 0}   + n_1\pi(r_1r_2+r_1r_3),\\
    \phase{C_3'}{} &= -\frac{2\pi \xi}{3} \delta_{\mu, 3} \nonumber\\
                    & +(\frac{2\pi \xi}{3}+n_{C_3C_3'}+n_{C_3'T_2}) (-\delta_{\mu, 0}+\delta_{\mu, 2})\pi  \nonumber\\
                    &  + r_1\pi n_{C_3'T_2} +  r_3\pi\frac{r_3-1}{2}n_1 + n_1\pi r_1r_2 \nonumber\\
    &  +  r_2 \pi(\frac{r_2-1}{2}n_1+n_{C_3'T_2}),
\end{align}
\end{subequations}
where $\xi \in \{-1,0,1\}$, $n_1, n_{C_3C_3'}, n_{C_3'T_2} \in \{0,1\}$. $n_1$ once again determines the size of the unit cell.\\
The next step is to find all compatible  ansätze. Since elements of $\chi_e$ cannot map between main and inverse tetrahedra but from one bond on one main tetrahedron to every other bond on any main tetrahedron we have two independent bonds: One on a main and one on an inverse tetrahedron. We choose the bonds $01$ ($\mathbf{0}_0 \xrightarrow{} \mathbf{0}_1$) and $I01$ ($\mathbf{0}_0 \xrightarrow{} \mathbf{0}_1-\mathbf{a}_1 $). We label the mean-field parameters $ (a^t_1,b^t_1,c^t_1,d^t_1)$ on bond $01$ and $ (a^t_2,b^t_2,c^t_2,d^t_2)$ on bond $I01$ with $t\in \{h,p\}$. With Eqs.~\eqref{eq:TransformationMeanFieldAnsatz} the mean-field parameters of all other bonds can be calculated. The chiral ansätze can break $\mathcal{T}$, $I$ and $S$ while satisfying $\mathcal{T}I$ and $\mathcal{T}S$. Therefore, the mean-field parameters are complex numbers in general: $ a^h_i \xrightarrow{} a^h_i e^{i\phi_{a^h_i}}, \ldots, a^p_i \xrightarrow{} a^p_i e^{-i\phi_{a^p_i}},\ldots$, where $a^h_i, \ldots, a^p_i, \ldots \in  \mathbb{R}$.    The different sign convention of the phases comes from the fact that $a^h$ depends on $\mathbf{h}_{ij}^*$ while $a^p$ depends on $\mathbf{p}_{ij}$. First we find all possible ansätze that respect the PSG of $\chi_e$ by mapping the bonds $01$ and $I01$ onto themselves with $S^{-1}C_3SC_3$  (note, that this also flips the bond). For the $01$ bond this results in     
\begin{align}
    (a^h_1,b^h_1,c^h_1,d^h_1) &=  (-a^{h*}_1,b^{h*}_1,c^{h*}_1,d^{h*}_1)e^{-i\pi(\frac{4 \xi}{3}+n_{C_3C_3'}+n_{C_3'T_2})},\\
        (a^p_1,b^p_1,c^p_1,d^p_1) &=  (a^{p}_1,-b^{p}_1,-c^{p}_1,-d^{p}_1)e^{i\pi(n_{C_3C_3'}+n_{C_3'T_2})}.
\end{align}
For the $I01$ bond this results in     
\begin{align}
    (a^h_2,b^h_2,c^h_2,d^h_2) &=  (-a^{h*}_2,b^{h*}_2,c^{h*}_2,d^{h*}_2)e^{-i\pi(\frac{4\pi \xi}{3}+n_{C_3C_3'} )},\\
    (a^p_2,b^p_2,c^p_2,d^p_2) &=  (a^{p}_2,-b^{p}_2,-c^{p}_2,-d^{p}_2)e^{i\pi n_{ C_3C_3'}}.
\end{align}

\begin{table}[]
\caption{All independent non-zero nearest-neighbor mean-field parameters for the different PSG equivalence classes that respect the symmetries of $\chi_e$ in the local spin basis. Parameters with index 1 and 2 are fixed on bond $\mathbf{0}_0 \xrightarrow{} \mathbf{0}_1$ and $\mathbf{0}_0 \xrightarrow{} \mathbf{0}_1-\mathbf{a}_1 $, respectively. All parameters not mentioned in the table are forced to be 0. The ansätze are labeled by the parameters $\xi\in \{-1,0,1\}$, $n_1, n_{C_3C_3'}, n_{C_3'T_2} \in \{0,1\}$}
\label{tab:EquivalenceClasseChiE}
\begin{tabular}{cl}
$ n_1-(n_{C_3C_3'}n_{C_3'T_2}\xi)$  & NN \\ \hline \\[-3pt]
$ n_1-(00\xi)$  & $a^h_1,b^h_1,c^h_1,d^h_1,a^p_1 $ \\
                &  $a^h_2,b^h_2,c^h_2,d^h_2,a^p_2$ \\[10pt]
$ n_1-(10 \xi)$  & $a^h_1,b^h_1,c^h_1,d^h_1,b^p_1,c^p_1,d^p_1$ \\
                 & $a^h_2,b^h_2,c^h_2,d^h_2,b^p_2,c^p_2,d^p_2$ \\[10pt]
$ n_1-(01 \xi)$  & $a^h_1,b^h_1,c^h_1,d^h_1,b^p_1,c^p_1,d^p_1$ \\
                 & $a^h_2,b^h_2,c^h_2,d^h_2,a^p_2$ \\[10pt]
$n_1-(11 \xi)$  & $a^h_1,b^h_1,c^h_1,d^h_1,a^p_1 $ \\
                & $a^h_2,b^h_2,c^h_2,d^h_2,b^p_2,c^p_2,d^p_2$ \\
Constraints: &   \\\hline \\[-3pt]
$\Re [a^h_1 e^{i\frac{\pi}{2}(\frac{4 \xi}{3}+n_{C_3C_3'}+n_{C_3'T_2})}]=0$, & $\Re [a^h_2 e^{i\frac{\pi}{2}(\frac{4 \xi}{3}+n_{C_3C_3'} )}]=0$, \\
$\Im [b^h_1 e^{i\frac{\pi}{2}(\frac{4 \xi}{3}+n_{C_3C_3'}+n_{C_3'T_2})}]=0$, & $\Im [b^h_2 e^{i\frac{\pi}{2}(\frac{4 \xi}{3}+n_{C_3C_3'} )}]=0$, \\
$\Im [c^h_1 e^{i\frac{\pi}{2}(\frac{4 \xi}{3}+n_{C_3C_3'}+n_{C_3'T_2})}]=0$, & $\Im [c^h_2 e^{i\frac{\pi}{2}(\frac{4 \xi}{3}+n_{C_3C_3'} )}]=0$, \\
$\Im [d^h_1 e^{i\frac{\pi}{2}(\frac{4 \xi}{3}+n_{C_3C_3'}+n_{C_3'T_2})}]=0$, & $\Im [d^h_2 e^{i\frac{\pi}{2}(\frac{4 \xi}{3}+n_{C_3C_3'} )}]=0$.
\end{tabular}
\end{table}
Table \ref{tab:EquivalenceClasseChiE} lists all allowed nearest-neighbor mean-field parameters for ansätze respecting the symmetries of $\chi_e$.\\
To get to all chiral ansätze we have to impose rotoreflection and screw symmetry modulo time reversal. Therefore, we have to fix the moduli of the mean-field parameters on the bonds $01$ and $I01$ to be the same such that:
$a_1^h=a_2^h = a^h, \ldots, a_1^p = a_2^p = a^p, \ldots$. Notice, that for $ n_{C_3'T_2}=1$  this is not possible for the pairing fields. Therefore, such ansätze either break $\rots$ and $S$ as well as $\mathcal{T}\rots$ and $\mathcal{T}S$ or have no pairing field and therefore an accidental IGG of $U(1)$. Either way, they correspond to ansätze that we do not want to consider and we set $ n_{C_3'T_2}= 0$ in the rest of this work. This means that $a^p$ cannot appear in an ansatz together with  $b^p ,c^p ,d^p $.
\noindent
All further restrictions to the ansätze can be found by 
transformation of expectation values of gauge invariant loop operators \cite{messio_time_2013}. For example: $ \hat{B}_{ij}\hat{B}_{jk}\hat{B}_{ki} $ or $ \hat{A}_{ij}\hat{B}_{jk}\hat{A}^\dagger_{ki} $. These are analogous to the Wilson loop operators in gauge theory. The loop operators are directly related to products of spin operators and therefore have a straight forward physical interpretation. For example the triple product of the spins at sites i,j,k can be written using two of these loops:
\begin{align}
    \hat{S}_i \cdot (\hat{S}_j\times\hat{S}_k)  &= -2i:(\hat{B}_{ij}\hat{B}_{jk}\hat{B}_{ki} -  \hat{B}^\dagger_{ij}\hat{B}^\dagger_{jk}\hat{B}^\dagger_{ki} ): \label{eq:tripleproductLoopOperator}\\
    &= \phantom{-} 2i:(\hat{A}_{ij}\hat{A}_{jk}^\dagger\hat{B}_{ki}-\hat{A}_{ij}^\dagger\hat{A}_{jk}\hat{B}_{ki}^\dagger): .
\end{align}
In SBMFT the expectation values of loop operators can be written as products of the mean-fields: $\expval{\hat{B}_{ij}\hat{B}_{jk}\hat{B}_{ki}}\approx \mathcal{B}_{ij}\mathcal{B}_{ij}\mathcal{B}_{ki}$.
Using Eq.~\eqref{eq:tripleproductLoopOperator} we can directly see that ansätze that respect time reversal symmetry, i.e.,~where the mean-fields are real, do not give rise to non-coplanar spin configurations. \\
The complex argument of the loops, called fluxes, boil down to a sum of complex arguments of the mean-field parameters, e.g:
\begin{equation}
  \Arg\left(\expval{\hat{B}_{ij}\hat{B}_{jk}\hat{B}_{ki}}\right) = \Arg(\mathcal{B}_{ij}) + \Arg(\mathcal{B}_{jk}) + \Arg(\mathcal{B}_{ki}). \label{eq:MeanFieldFlux} 
\end{equation}
Under the action of an operator $\mathcal{O}_o\in \chi_o$, Eq.~\eqref{eq:MeanFieldFlux} transforms like
\begin{align}
 \mathcal{O}_o \Arg\left(\expval{\hat{B}_{ij}\hat{B}_{jk}\hat{B}_{ki}}\right) &= \epsilon_{\mathcal{O}_o}\big[\Arg(\mathcal{B}_{\mathcal{O}_o(ij)})\\ &+\Arg(\mathcal{B}_{\mathcal{O}_o(jk)})+\Arg(\mathcal{B}_{\mathcal{O}_o(ki)})\big]\nonumber. \label{eq:FluxEquation}
 \end{align}
 
\noindent
The flux is invariant under $\mathcal{O}_o $ if  $\epsilon_{\mathcal{O}_o} = 1$ and the flux changes its sign if $\epsilon_{\mathcal{O}_o} = -1$. We can write down equations like Eq.~\eqref{eq:FluxEquation} for all independent fluxes on the lattice and then solve for the phases $\Arg(B_{ij})=\phi_{ \mathcal{B}_{ij}}$ depending on the parities of all elements in $\chi_0$. We choose to study the flux transformations under action of inversion $I = \rots^3$ and mirror symmetry $\Sigma = I S$ since the resulting phase equations have a particularly nice form. The calculations are performed in Appendix \ref{Appendix:FluxTransform}. The solutions are presented in Tab.~\ref{tab:WeaklySymmetricAnsaetze}. \\

\begin{table*}[]
\centering
\caption{All independent non-zero nearest-neighbor mean-field parameters for all nearest-neighbor $\Ztwo$ chiral PSG equivalence classes in the local spin basis. Parameters with index 1 and 2 are fixed on bond $\mathbf{0}_0 \xrightarrow{} \mathbf{0}_1$ and $\mathbf{0}_0 \xrightarrow{} \mathbf{0}_1-\mathbf{a}_1 $, respectively. All nearest-neighbor parameters not mentioned in the list are forced to be 0. The ansätze are labeled by the PSG parameters $\xi\in \{-1,0,1\}$, $n_1, n_{C_3C_3'}, p_1 \in \{0,1\}$ and the parities $\epsI,\epsSig  \in \{-1,1\}$.}
\label{tab:WeaklySymmetricAnsaetze}
\begin{tabular}{lll}
$(\epsSig,\epsI)$-$n_1 $-$(n_{C_3C_3'},p_1)$-$(\xi)$ & NN & Constraints: \\ \hline
$( 1,1)$-$n_1$-$(1,p_1)$-$(0)_y $         & $a^h,c^h,c^p$        & $\phi_{a^h_1}=0, \phi_{c^h_1}=\frac{\pi}{2}$ \\
$( 1,1)$-$n_1$-$(1,p_1)$-$(0)_{xz}$       & $a^h,c^h,b^p,d^p$    & $\phi_{a^h}=0, \phi_{c^h_1}=\frac{\pi}{2}$ \\
$( 1,1)$-$n_1$-$(0,p_1)$-$(0)$            & $b^h,d^h,a^p$             &  $\phi_{b^h_1}= \phi_{d^h_1}=0$ \\[10pt]
$( 1,-1)$-$n_1$-$(1,0)$-$(0)_y $         & $a^h,c^h,c^p$        & $\phi_{a^h_1}=0, \phi_{c^h_1}=\frac{\pi}{2}$\\
$( 1,-1)$-$n_1$-$(1,0)$-$(0)_{xz}$       & $a^h,c^h,b^p,d^p$    & $\phi_{a^h_1}=0, \phi_{c^h_1}=\frac{\pi}{2}$ \\
$( 1,-1)$-$n_1$-$(0,0)$-$(0)$            & $b^h,d^h,a^p$             & $ \phi_{b^h_1}= \phi_{d^h_1}=0$  \\[10pt]
$(-1,1)$-$n_1$-$(0,p_1)$-$(\xi)$                      & $a^h,b^h,d^h,a^p$           & $\phi_{a^h_1}=\frac{\pi}{2}+\frac{2\xi\pi}{3}, \phi_{b^h_1}=\phi_{d^h_1} = \frac{2\xi\pi}{3}$ \\
$(-1,1)$-$n_1$-$(1,p_1)$-$(\xi)$                      & $a^h,b^h,d^h,b^p,c^p,d^p$        & $\phi_{a^h_1}=\frac{2\xi\pi}{3} , \phi_{b^h_1}=\phi_{d^h_1}=\frac{\pi}{2}+\frac{2\xi\pi}{3}, \phi_{b^p_1}=\phi_{d^p_1}=\phi_{c^p_1}-\frac{\pi}{2}$ \\[10pt]
$(-1,-1)$-$n_1$-$(0,p_1)$-$(0)$                      & $a^h,b^h,d^h,a^p$           & $\phi_{a^h_1}=\frac{\pi}{2}, \phi_{b^h_1}=\phi_{d^h_1} = 0$ \\
$(-1,-1)$-$n_1$-$(1,p_1)$-$(0)$                      & $a^h,b^h,d^h,b^p,c^p,d^p$        & $\phi_{a^h_1}=0 , \phi_{b^h_1}=\phi_{d^h_1}=\frac{\pi}{2}, \phi_{b^p_1}=\phi_{d^p_1}=\phi_{c^p_1}-\frac{\pi}{2}$ \\
General constraints: &  &  \\ \hline
$\phi_{a^h_2}=\epsI \phi_{a^h_1} + n_1\pi$ & $\phi_{a^p_2}=\epsI \phi_{a^p_1} +  p_1\pi$ &  \\
$\phi_{b^h_2}=\epsI \phi_{b^h_1} + n_1\pi$ & $\phi_{b^p_2}=\epsI \phi_{b^p_1} +  p_1\pi$ &  \\
$\phi_{c^h_2}=\epsI \phi_{c^h_1} + n_1\pi$ & $\phi_{c^p_2}=\epsI \phi_{c^p_1} +  p_1\pi$ &  \\
$\phi_{d^h_2}=\epsI \phi_{d^h_1} + n_1\pi$ & $\phi_{d^p_2}=\epsI \phi_{d^p_1} +  p_1\pi$ &  \\
\end{tabular}
\end{table*}
\noindent
The ansätze can be grouped into four groups corresponding to the parity of their underlying symmetries $(\epsSig,\epsI)$.\\
The 12 ansätze with $(\epsSig,\epsI)=(1,1)$ exactly reduce to the 12 fully symmetric ansätze that were characterized by Liu \emph{et al.} \cite{liu_competing_2019} once we impose time reversal symmetry which constrains $c^h=0$.  Comparison with the fully symmetric classification sheds light on why $(1,\epsI)$-$n_1$-$(1,p_1)$-$(0)$ support two families of ansätze. One with nonzero $c^p$ and one with nonzero $b^p$ and $d^p$. Imposing $\Sigma$ symmetry causes $(1,\epsI)$-$n_1$-$(1,p_1)$-$(0)$ to fractionalize into two new classes labeled by the $\Ztwo$ parameter $n_{S\rots}$. The fractionalized ansätze then fulfill the symmetry: $(\widetilde{\Sigma})^2 = (-\mathbb{1})^{n_{S\rots}}$. The same happens for $(1,\epsI)$-$n_1$-$(0,p_1)$-$(0)$. However, the corresponding ansatz families do not allow any nearest-neighbor pairing fields and therefore do not correspond to nearest-neighbor $\Ztwo$ spin liquids in the same way as for the fully symmetric equivalence classes $n_1$-(00$n_\rots$).\\
For $\epsI=1$ we can identify the $\Ztwo$ parameter $p_1+n_1$ with $n_\rots$ by comparing with the fully symmetric ansätze. The ansätze then fulfill $(\widetilde{I})^2 = (-\mathbb{1})^{p_1+n_1}$. The ansätze with $(\epsSig,\epsI)=(-1,-1)$ fractionalize similarly upon imposing screw symmetry $S= I \Sigma$. Only the ansätze with $(\epsSig,\epsI)=(1,-1)$ do not fractionalize in states labeled by $p_1$. Instead, they allow for continuously variable phases $\phi_{a^p_1},\phi_{b^p_1},\phi_{c^p_1},\phi_{d^p_1}$.

\section{Choosing ansätze}
\label{Sec:ChoosingAnsätze}
Now that we have characterized all symmetric and chiral mean-field ansätze, we have to choose meaningful ansätze as well as a suitable mean-field decoupling for the XXZ Hamiltonian \eqref{eq:ModelHamiltonian}. As mentioned previously the mean-field decouplings are ambiguous. We choose them in a way that preserves the $SU(2)$ spin rotation symmetry at the Heisenberg point:

\begin{align}
    H &= \sum_{<ij>}J_{\perp}(\hat{S}_i^x\hat{S}_j^x + \hat{S}_i^y \hat{S}_j^y)+ J_{zz} \hat{S}_i^z \hat{S}_j^z \\
    &= \sum_{<ij>}\frac{J_{\perp}+J_{zz}}{2}\Vec{\hat{S}_i}\Vec{\hat{S}_j} + \frac{\Delta J}{2} \hat{S}_i^z\hat{S}_j^z -   \frac{\Delta J}{2}(\hat{S}_i^x\hat{S}_j^x + \hat{S}_i^y\hat{S}_j^y)  \nonumber\\
    &=  \sum_{<ij>} \frac{J_{\perp}+J_{zz}}{2} :\hat{B}_{ij}^\dagger \hat{B}_{ij}: -  J_{\perp}\hat{A}_{ij}^\dagger \hat{A}_{ij} \nonumber\\
    & \phantom{=} -   \frac{\Delta J}{2}(:\hat{t}_{ij}^{h,x\dagger} \hat{t}^{h,x}_{ij}:  +  :\hat{t}_{ij}^{h,y\dagger} \hat{t}^{h,y}_{ij}: -:\hat{t}_{ij}^{h,z\dagger} \hat{t}^{h,z}_{ij}:  ) \nonumber\\
     & \phantom{=} + \sum_i\lambda_i(b_{i\alpha}^\dagger b_{i\alpha}-2S) .\nonumber
\end{align}
The mean-field decoupling is then defined by:
\begin{subequations}
\begin{align}
        J^h_{ij} &= \frac{1}{2} \begin{pmatrix}
        J_{zz}+J_\perp & & &\\
            & J_\perp-J_{zz} & &\\
            & & J_\perp-J_{zz} & \\
            & & & J_{zz}-J_\perp
        \end{pmatrix},\\
        J^p_{ij} &=  \begin{pmatrix}
        -J_\perp & & &\\
            &  0 & &\\
            & & 0  & \\
            & & &  0
        \end{pmatrix},\\
        C_{ij} &=  0 .
\end{align}
\label{eq:decouplingHeisenberg}
\end{subequations}
This particular decoupling allows the resulting spin liquid state to break the U(1) spin rotation symmetry by acquiring different expectation values for $t^{h,x}$ and $t^{h,y}$. This choice is motivated by findings of Benton \emph{et al.} \cite{benton_quantum_2018}, where a transition from a U(1) symmetric to a nematic state is observed at the Heisenberg point.\\
In the following sections we solve the mean-field equations for the ansätze $(1,1)$-$n_1$-$(0,p_1)$-$(0)$  and $(-1,\epsI)$-$n_1$-$(0,p_1)$-$(\xi)$. We choose them because they can include $SU(2)$ symmetric pairing fields $A_{ij}$ that capture the relevant physics at the antiferromagnetic Heisenberg point.\\

Ansätze $(1,1)$-$n_1$-$(0,p_1)$-$(0)$ are fully symmetric. Therefore, they fulfill all lattice symmetries and correspond to coplanar spin liquids.\\

Ansätze $(-1,-1)$-$n_1$-$(0,p_1)$-$(0)$ are chiral and they break $\mathcal{T}$, $I$ and $\Sigma$ but preserve $\mathcal{T}I$ and $\mathcal{T}\Sigma$. Therefore, they also respect screw symmetry. Every $SU(2)$ symmetric triangular flux operator has an expectation value of $\pm \frac{\pi}{2}$. Therefore, they have the same symmetry and flux structure as the monopole flux  and the $(\frac{\pi}{2},\pi)$ state considered by Burnell \emph{et al.} \cite{burnell_monopole_2009}. For $\mathcal{B} = 0$ the ansätze reduce to the fully symmetric $(1,1)$-$n_1$-$(0,p_1)$-$(0)$.\\

Ansätze $(-1,1)$-$n_1$-$(0,p_1)$-$(0)$ are chiral and they break $\mathcal{T}$ and $\Sigma$ while preserving $I$ and $\mathcal{T}\Sigma$. Therefore, they also break $S$ modulo time reversal. Every $SU(2)$ symmetric triangular flux has a value of $\frac{\pi}{2}$. Therefore, they have a similar flux structure as the monopole-antimonopole flux state $[\frac{\pi}{2},-\frac{\pi}{2},0]$ considered by Kim \emph{et al.} \cite{kim_chiral_2008}. For $\mathcal{B} = 0$ the ansätze reduce to the fully symmetric $(1,1)$-$n_1$-$(0,p_1)$-$(0)$.\\

Ansätze $(-1,\epsI)$-$n_1$-$(0,p_1)$-$(\xi \neq 0 )$ are chiral and they have the same symmetries as $(-1,1)$-$n_1$-$(0,p_1)$-$(0)$. Their characteristic property is that the expectation value of every rhombus flux operator has a value of  $\xi\frac{\pi}{3}$. This leads to the identity \cite{messio_time_2013} 
\begin{align}
    &\phantom{=}   \mathcal{S} \expval{(\mathbf{\hat{S}}_0-\mathbf{\hat{S}}_3)\cdot(\mathbf{\hat{S}}_1 \times \mathbf{\hat{S}}_2)+(\mathbf{\hat{S}}_2-\mathbf{\hat{S}}_1)\cdot(\mathbf{\hat{S}}_3 \times \mathbf{\hat{S}}_0)} \nonumber\\
    &= 8\mathcal{A}^4 \sin (\xi\frac{\pi}{3}),
\end{align} which implies non-zero scalar spin chirality also in the case $\mathcal{B}=0$. 
The ansätze for $\xi=1$ and $\xi=-1$ can be mapped onto each other by the action of $\mathcal{T}$.

\section{Calculation of the free energy}
\label{Sec:Diagonalization}
Now that a mean-field decoupling and ansätze are chosen we can apply the Fourier transform 
\begin{align}
    b_{\mathbf{r}_\mu}= \sqrt{\frac{N_{SL}}{ N}}\sum_{\kvec}b_{\kvec,\mu}e^{-i\kvec\mathbf{r}_{\mu}},
\end{align}
to the Hamiltonian in Eq.~\eqref{eq:MFH} and bring it to the form: 
\begin{equation}
\label{eq:HamiltonianAfterFourier}
    H = \frac{1}{2}\sum_\kvec \hat{\Psi}_\kvec^\dagger \mathcal{H}(\kvec) \hat{\Psi}_\kvec  + 3N  f(\mathbf{h},\mathbf{p}) -N\lambda(2\mathcal{S}+1),
\end{equation}
where $N$ is the number of atoms on the lattice and $\mathcal{H}(\kvec)$ has the form
\begin{equation}
\label{eq:HamiltonianStructure}
    \mathcal{H}(\kvec) = \begin{pmatrix}
    H^h(\kvec) & H^p(\kvec) \\
     (H^p(\kvec))^\dagger &  (H^h(-\kvec))^T  
    \end{pmatrix} +\lambda \mathbb{1}_{4N_{SL}}.
\end{equation}
\noindent
For ($n_1=1$) $n_1=0$ $\hat{\Psi}_\kvec$ is a (64) 16 component spinor and ($N_{SL} = 16$) $N_{SL} = 4$ is the number of sublattices in the (enlarged) unit cell: $\hat{\Psi}_\kvec = (\hat{\psi}_\kvec,\hat{\psi}_{-\kvec}^\dagger)$ with $\hat{\psi}_\kvec = (\hat{b}_{\vec{k},1},\hat{b}_{\vec{k},2},\ldots , \hat{b}_{\vec{k},N_{SL}})$. The explicit form of $H^h$ and $H^p$ are given in Appendix \ref{appendix:ExplicitHamiltonians}.
The Hamiltonian \eqref{eq:HamiltonianAfterFourier} can be diagonalized by a Bogoliubov transform \cite{xiao_theory_2009}. We therefore introduce Bogoliubov transformation matrices $V(\kvec)$ such that:
\begin{align}
    \hat{\psi}_\kvec &= V(\kvec) \hat{\Gamma}_\kvec,\\
    V(\kvec)^\dagger \tau^3 V(\kvec) &= \tau^3, \label{eq:BogoliubovCommutation}\\
    V(\kvec)^\dagger \mathcal{H}(\kvec) V(\kvec) &=  \Omega(\kvec).
\end{align}
Here $\hat{\Gamma}_\kvec = (\hat{\gamma}_{\kvec,1},\ldots,\hat{\gamma}_{\kvec,2N_{SL}},\hat{\gamma}^\dagger_{-\kvec,1},\ldots \hat{\gamma}^\dagger_{-\kvec,2N_{SL}})$ is the Bogoliubov spinor, $\tau^3 = \sigma^3 \otimes \mathbb{1}_{2N_{SL}}$ and $\Omega(\kvec)$ is a diagonal matrix where the first $2N_{SL}$ entries $\omega_i(\kvec)$ ($i\in \{1,2N_{SL}\}$) and the last $2N_{SL}$ entries are $\omega_i(-\kvec)$. They are the positive and negative eigenvalues of the matrix $\tau^3\mathcal{H}(k)$ respectively \cite{xiao_theory_2009}.  Eq.~\eqref{eq:BogoliubovCommutation} ensures that the new bosonic operators $\hat{\gamma}_{\kvec,i}$ preserve the bosonic commutation relations $\comm{\hat{\gamma}_{\kvec,i}}{\hat{\gamma}_{\kvec',j}} = 0, \;\comm{\hat{\gamma}_{\kvec,i}}{\hat{\gamma}^\dagger_{\kvec',j}} = \delta_{i,j}\delta_{\kvec,\kvec'}$. Thus, we can write \eqref{eq:HamiltonianAfterFourier} as 
\begin{subequations}
\begin{align}
  H(\kvec) &= \sum_\kvec\sum_i^{2N_{SL}} \omega_i(\kvec) \hat{\gamma}^\dagger_{k,i}\hat{\gamma}_{k,i} + \frac{1}{2}\sum_\kvec \sum_i^{2N_{SL}} \omega_i(\kvec)  \\
  & + 3Nf(\mathbf{h},\mathbf{p})-N\lambda(2\mathcal{S}+1).  \nonumber
\end{align}
\end{subequations}
The mean-field free energy per site is thus given by 
\begin{align}
    f_{MF} &=    \frac{1}{2 \text{Vol}_{\text{B.Z.}}  N_{SL}}\int_{\text{B.Z.}} \dd k^3  \sum_i^{2N_{SL}} \omega_i(\kvec) \label{eq:freeEnergy}\\
    &\phantom{=} + 3   f(\mathbf{h},\mathbf{p})- \lambda(2\mathcal{S}+1). \nonumber 
\end{align}
where we exchanged the sum by an integral in the thermodynamic limit $\sum_\kvec \xrightarrow[]{} \frac{N}{ \text{Vol}_{\text{B.Z.}}N_{SL}}\int_{\text{B.Z.}}\dd k^3$.\\
This reduces to 
\begin{equation}
     f_{MF} = 3\mathbf{h}^\dagger J_h\mathbf{h} + 3\mathbf{p}^\dagger J_p\mathbf{p} + 3C, \ 
 \end{equation}
after solving the self-consistency equations. \\

As a final step we find the correct values of the mean-field by solving the self-consistency equations \eqref{eq:selfconsistency}. 
The solutions are listed in Table \ref{tab:SelfConsistentParameters}. \\
 Within numerical accuracy the only bond operator that acquires a non-zero expectation value are the $SU(2)$ symmetric $\hat{A}$ and $\hat{B}$. Therefore, all considered ansätze reduce to $SU(2)$ symmetric states or are in a condensed phase at $\mathcal{S}=0.5$. The ansätze $(1,1)$-$n_1$-$(0,p_1)$-$(0)$  and $(-1,\epsI)$-$n_1$-$(0,p_1)$-$(0)$ describe the same four fully symmetric spin liquids as the fully symmetric ansätze $(1,1)$-$n_1$-$(0,p_1)$-$(0)$ since for them $\mathcal{B}=0$ . Only the saddle points of $(-1,1)$-$1$-$(0,1)$-$(\pm 1)$ and $(-1,1)$-$1$-$(0,0)$-$(\pm 1)$ are chiral spin liquids. 
 The resulting six distinct spin liquid states can solely be described by the fluxes that are enclosed by the  $\hat{A}$ operators on the hexagonal, bow tie and rhombus loops: $(\phi_{\mhexagon},\phi_{\Join},\phi_{\Diamond}) = (p_1,p_1+n_1,1+\frac{2\xi}{3}) \pi$. \\
     We have found the $SU(2)$ symmetry to be stable even beyond the Heisenberg point. We have explicitly checked this for coupling angles $\theta \in [0,\frac{\pi}{2}]$. The normalized free energies of the spin liquid states can be found in Table \ref{tab:SpinLiquidEnergies}. The main dependence of the free energy on $\theta$ comes from $J_{\perp}$.  While the non-chiral state $(\pi,0,\pi)$  has the lowest free energy in the present decoupling (Eq. \eqref{eq:decouplingHeisenberg}), choosing a decoupling based on $ \mathbf{\hat{S}}_i\mathbf{\hat{S}}_j    = -2\hat{A}_{ij}^\dagger \hat{A}_{ij} + \mathcal{S}^2$ for all considered ansätze results in $(\pi,0,\pm\frac{\pi}{3})$  being the lowest energy state.\footnote{In this case $(\pi,0,\pm\frac{\pi}{3})$ has a value of $\mathcal{A} = 0.4307$, $(0,\pi,\pm\frac{\pi}{3})$ condenses and the values for $\mathcal{A}$ stay the same for all other ansätze.} Therefore, inferences about the possible ground state have to be made with care. \\

\begin{table}[]
\caption{Self-consistent mean-field parameters. The values with asterisk are determined self-consistently in the gapped spin liquid phase for the physical spin value of $\mathcal{S} = 0.5$. All other values are set by symmetry of the particular ansatz. "c" labels states that are condensed at $\mathcal{S} = 0.5$, i.e.,~are magnetically ordered.}
\centering
\label{tab:SelfConsistentParameters}
\begin{tabular}{llllllll}
Ansatz & $\mathcal{A}$ & $\mathcal{B}$ & $t^{h,x}$ & $t^{h,y}$ & $t^{h,z}$ & $2\lambda/J_{\perp}$    \\ \hline
(\phantom{-}1,\phantom{-}1)-$0$-$(0,0)$-$(0) $ & $0.3901^*$   & 0 & $0^*$ & 0 & $0^*$ & $1.3593^*$   \\
(\phantom{-}1,\phantom{-}1)-$0$-$(0,1)$-$(0) $ & $0.3931^*$   & 0 & $0^*$ & 0 & $0^*$ & $1.3880^*$  \\
(\phantom{-}1,\phantom{-}1)-$1$-$(0,0)$-$(0) $ & $ 0.3895^*$   & 0 & $0^*$ & 0 & $0^*$ & $1.3466^*$  \\
(\phantom{-}1,\phantom{-}1)-$1$-$(0,1)$-$(0) $ & $0.3933^*$   & 0 & $0^*$ & 0 & $0^*$ & $1.3910^*$   \\
(-1,-1)-$0$-$(0,0)$-$(0) $ & $0.3901^*$   & $0^*$ & $0^*$ & 0 & $0^*$ & $1.3593^*$   \\
(-1,-1)-$0$-$(0,1)$-$(0) $ & $0.3931^*$   & $0^*$ & $0^*$ & 0 & $0^*$ & $1.3880^*$   \\
(-1,-1)-$1$-$(0,0)$-$(0) $ & $ 0.3895^*$  & $0^*$ & $0^*$ & 0 & $0^*$ & $1.3466^*$   \\
(-1,-1)-$1$-$(0,1)$-$(0) $ & $0.3933^*$   & $0^*$ & $0^*$ & 0 & $0^*$ & $1.3910^*$   \\
(-1,\phantom{-}1)-$0$-$(0,0)$-$(0) $ & $0.3901^*$  & $0^*$ & $0^*$ & 0 & $0^*$ & $1.3593^*$   \\
(-1,\phantom{-}1)-$0$-$(0,1)$-$(0) $ & $0.3931^*$   & $0^*$ & $0^*$ & 0 & $0^*$ & $1.3880^*$   \\
(-1,\phantom{-}1)-$1$-$(0,0)$-$(0) $ & $ 0.3895^*$ & $0^*$ & $0^*$ & 0 & $0^*$ & $1.3466^*$  \\
(-1,\phantom{-}1)-$1$-$(0,1)$-$(0) $ & $0.3933^*$  & $0^*$ & $0^*$ & 0 & $0^*$ & $1.3910^*$   \\
(-1,\phantom{-}1)-$0$-$(0,0)$-$(\pm1) $ & c  & c & c & c & c & c  \\
(-1,\phantom{-}1)-$0$-$(0,1)$-$(\pm1) $ & c  & c & c & c & c & c  \\
(-1,\phantom{-}1)-$1$-$(0,0)$-$(\pm1) $ & $0.4246^*$  & $-0.1676^*$ & $0^*$ & 0 & $0^*$ & $1.3710^*$ \\
(-1,\phantom{-}1)-$1$-$(0,1)$-$(\pm1) $ & $0.4265^*$  & $-0.1684^*$ & $0^*$ & 0 & $0^*$ & $1.3538^*$ \\ 
\end{tabular}
\end{table}

\begin{table}[]
\caption{Values of the $\mathcal{A}$ fields and the normalized free energy per site  $f_{MF}/J_{\perp}$ of the six spin liquid states.}
\label{tab:SpinLiquidEnergies}
\begin{tabular}{llll}
$(\phi_{\mhexagon},\phi_{\Join},\phi_{\Diamond})$ & $\mathcal{A}$& $\mathcal{B}$ & $f_{MF}/J_{\perp}$ \\ \hline
$(0,0,\pi)$                                     & 0.3901    & 0     & $-0.4565    $    \\
$(\pi,\pi,\pi)$                                 & 0.3931    & 0     & $-0.4635$            \\
$(0,\pi,\pi)$                                   & 0.3895    & 0     & $-0.4551   $      \\
$(\pi,0,\pi)$                                   & 0.3933    & 0     & $-0.4641   $     \\
$(0,\pi,\pm\frac{\pi}{3})$                      & 0.4246    & $-0.1676$     & $-0.4564$  \\      
$(\pi,0,\pm\frac{\pi}{3})$                      & 0.4265    & $-0.1684$     & $-0.4607$        
\end{tabular}
\end{table}

 
Also note that we have chosen a decoupling of the XXZ Hamiltonian in Eqs.~\eqref{eq:decouplingHeisenberg} which is expected to capture the relevant physics in the vicinity of the antiferromagnetic Heisenberg point and selected corresponding ansätze. However, other mean-field decouplings and ansätze with focus on the vicinity of the classical Ising limit $J_\perp=0$ or the easy-plane limit $J_{zz}=0$ can lead to stable non $SU(2)$ symmetric spin liquids. Indeed, in a preliminary study of the fully symmetric ansätze we have found the ansätze $( 1,1)$-$n_1$-$(1,p_1)$-$(0)_{xz}$ to have non zero $t^{p,z}$ in the vicinity of the Ising point. Here, the ansatz $(1,1)$-$0$-$(1,1)$-$(0)_{xz}$ has the lowest energy. Ansätze  $( 1,1)$-$n_1$-$(1,p_1)$-$(0)_{xz}$  and $(1,1)$-$n_1$-$(1,p_1)$-$(0)_y$ have non zero expectation values $t^{p,x}$ and $t^{p,y}$ in the vicinity of the easy plane limit, respectively. At this point ansätze $(1,1)$-$0$-$(1,0)$-$(0)_{xz}$ and $(1,1)$-$0$-$(1,0)$-$(0)_y$ have the lowest energy. These break the $U(1)$ spin rotation symmetry in accordance with the nematic spin liquids found by Benton \emph{et al.} \cite{benton_quantum_2018}. However, contrary to the nematic states of Benton \emph{et al.} they preserve the $C_3$ lattice symmetry by construction. The energies of these spin liquids at mean-field level are highly dependent on the choices of the mean-field decouplings. Therefore, the ambiguity of choosing non $SU(2)$ symmetric mean-field decouplings described in Sec.~\ref{subsec:SBMFT} prevents an appropriate comparison of mean-field energies from different mean-field decouplings. A generalization of the large-$N$ $Sp(N)$ approach \cite{Read_LargeN_1991} together with arguments from the symplectic $N$ approach \cite{Flint_Symplectic_2009} to non $SU(2)$ symmetric Hamiltonians might shed some light on this issue.

Finally, note that interactions beyond nearest neighbors could in principle stabilize the chiral saddle points of $(-1,\epsI)$-$n_1$-$(0,p_1)$-$(0)$. 

\section{Spin structure factors}
\label{sec:SpinStructureFactors}
To compare the mean-field states to experiment and other numerical studies of the XXZ model we calculate the spin-spin correlations in local and global basis. 
The spin-spin correlations in the local spin basis is given by the tensor

\begin{equation}
    \mathcal{S}^{\alpha,\beta}(\mathbf{q}) = \frac{1}{3N}\sum_{l,j}e^{i \mathbf{q}( \mathbf{r}_i-\mathbf{r}_j)} \expval{\hat{\mathbf{S}}^\alpha_l\cdot\hat{\mathbf{S}}^\beta_j },
\end{equation}
This can be expressed in terms of the components of the Bogoliubov transformation matrix $V(\kvec)$
\begin{widetext}
\begin{align}
    \mathcal{S}^{\alpha,\beta}(\mathbf{q}) = \frac{1}{3N}\sum_\kvec \text{Tr}\biggl(&V_{12}^\dagger(\kvec)\Sigma^\alpha V_{11}(\kvec-\mathbf{q})[V_{21}^\dagger(\kvec-\mathbf{q})(\Sigma^\beta)^\mathsf{T} V_{22}(\kvec) +V_{11}^\dagger(\kvec-\mathbf{q})\Sigma^\beta V_{12}(\kvec)]\biggr). \label{eq:StrucureFactorBogoliubov}
\end{align}
\end{widetext}
where 
\begin{align}
V(\kvec) =    \begin{pmatrix}
V_{11}(\kvec) & V_{12}(\kvec) \\
V_{21}(\kvec) & V_{22}(\kvec) 
    \end{pmatrix}, && 
        \Sigma^\gamma = \mathbb{1}_{N_{SL}} \otimes \sigma^\gamma.
\end{align}
Since all our spin liquid states turned out to be $SU(2)$ symmetric the only independent nonzero component of $\mathcal{S}^{\alpha,\beta}(\mathbf{q})$ is $\mathcal{S}^{z,z}(\mathbf{q})$.
We plot $\mathcal{S}^{z,z}$ along $[h,h,l]$ plane in Figure \ref{fig:localSpinSpinCorrelation}.\\
The spin-spin correlation in the local spin basis $\widetilde{\mathcal{S}}^{\alpha,\beta}(\mathbf{q})$ can be calculated by using Eq.~\eqref{eq:StrucureFactorBogoliubov} and replacing $ \Sigma^\gamma \xrightarrow{} U \Sigma^\gamma U^\dagger$ with 
\begin{equation}
    U = \mathbb{1}_{\frac{N_{SL}}{4}}\otimes\begin{pmatrix}
    U_0& & &\\
    &U_1& &\\
    & &U_2 &\\
    & & & U_3
    \end{pmatrix}.
\end{equation}
$U_\mu$ are the $SU(2)$ matrices that rotate from the global to the local basis (see Appendix \ref{appendix:SU2Matrices}). \\
Neutron scattering experiments do not directly measure components of  $\widetilde{\mathcal{S}}^{\alpha,\beta}(\mathbf{q})$ but instead measure the neutron scattering amplitude \cite{taillefumier_competing_2017}:
\begin{equation}
\label{eq:NeutronScatteringAmplitude}
    \mathcal{S}_\text{TOT}( \mathbf{q} ) = (\delta_{\alpha,\beta}-\frac{ \mathbf{q}^\alpha \mathbf{q}^\beta}{\abs{\mathbf{q}}^2}) \Tilde{\mathcal{S}}^{\alpha,\beta}(\mathbf{q}).
\end{equation}
Following Fennell \emph{et al.} \cite{fennell_magnetic_2009}, we calculate the neutron scattering amplitude along the $[h,h,l]$ plane and split the total scattering amplitude
$ \mathcal{S}_\text{TOT}( \mathbf{q} ) $ into a spin flip (SF) channel
\begin{equation}
    \mathcal{S}_\text{SF}( \mathbf{q} ) = \frac{( \mathbf{P}\times\mathbf{q})^\alpha ( \mathbf{P}\times\mathbf{q})^\beta}{\abs{\mathbf{q}}^2}  \Tilde{\mathcal{S}}^{\alpha,\beta}(\mathbf{q}), 
\end{equation}  and a no spin flip (NSF) channel 
\begin{equation}
    \mathcal{S}_\text{NSF}( \mathbf{q} ) = \mathbf{P}^\alpha  \mathbf{P}^\beta \Tilde{\mathcal{S}}^{\alpha,\beta}(\mathbf{q}), 
\end{equation}
where $ \mathbf{P} = \frac{1}{\sqrt{2}}(1,-1,0)$ is the polarization vector of the neutrons. In the $[h,h,l]$ plane they fulfill $\mathcal{S}_\text{TOT}( \mathbf{q} ) = \mathcal{S}_\text{SF}( \mathbf{q} ) +\mathcal{S}_\text{NSF}( \mathbf{q} ) $. Experimentally splitting up measurements into the polarization channels is advantageous since the pinch points that are characteristic for spin ice phase are only visible in the SF channel. When measuring $\mathcal{S}_\text{TOT}(  \mathbf{q} )$ the contributions from the NSF channel smear out the features \cite{fennell_magnetic_2009}.\\

 \subsection{Correlation results}
 
 \begin{figure}
\subfloat{%
\begin{tikzpicture}
\begin{axis}[
title style={font=\small, yshift= -1.5ex},
title = \text{a) $ (0,0,\pi)$ $\mathcal{S}^{z,z} $},
ylabel={ [0,0,l]},
xlabel={ [h,h,0]},
xmin=-6.05, xmax=6.05,
ymin=-6.05, ymax=6.05,
width=4.8cm,
height=4.8cm,
tick align=outside,
tick pos=left,
xtick={ -4,  0, 4},
ytick={ -4,  0, 4},
ytick style={draw=none},
xtick style={draw=none},
xticklabels={ $-4\pi$ , $0$ , $4\pi$ },
yticklabels={ $-4\pi$ , $0$ , $4\pi$ },
x tick label style={font=\small, yshift=1.5ex},
y tick label style={font=\small, xshift=1.5ex},
label style={font=\small},
y label style={yshift=-3ex},
x label style={yshift= +1ex},
point meta min=0,
point meta max=1,
]
\addplot graphics [includegraphics cmd=\pgfimage,xmin=-6, xmax=6, ymin=-6, ymax=6] {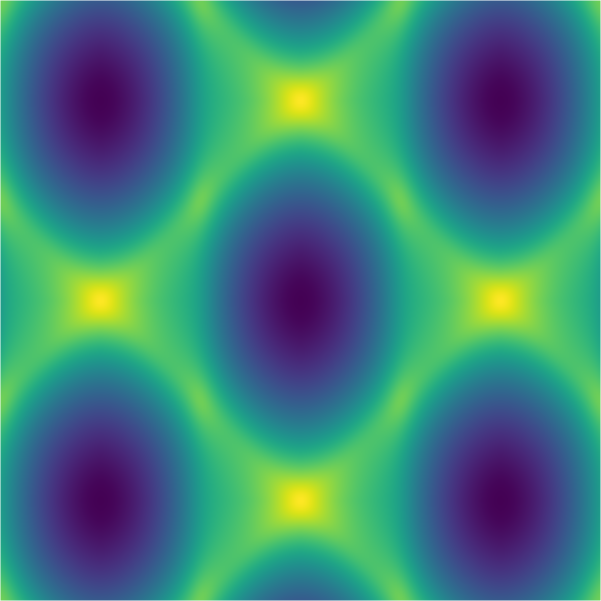};
\draw[white, thick] (axis cs:-2,4) -- node[above]{i)} (axis cs:2,4);
\draw[white, dashed, thick] (axis cs:-4,0) -- node[left]{ii)} (axis cs:0,4);
\end{axis}
\end{tikzpicture}%
}\hfill
\subfloat{%
  \begin{tikzpicture}
\begin{axis}[
title style={font=\small, yshift= -1.5ex},
title = \text{b) $ (\pi,\pi,\pi)$ $\mathcal{S}^{z,z} $},
ylabel={[0,0,l]},
xlabel={[h,h,0]},
xmin=-6.05, xmax=6.05,
ymin=-6.05, ymax=6.05,
width=4.8cm,
height=4.8cm,
tick align=outside,
tick pos=left,
xtick={ -4,  0, 4},
ytick={ -4,  0, 4},
ytick style={draw=none},
xtick style={draw=none},
xticklabels={ $-4\pi$ , $0$ , $4\pi$ },
yticklabels={ $-4\pi$ , $0$ , $4\pi$ },
x tick label style={font=\small, yshift=1.5ex},
y tick label style={font=\small, xshift=1.5ex},
label style={font=\small},
y label style={yshift= -3ex},
x label style={yshift= +1ex},
point meta min=0,
point meta max=1,
]
\addplot graphics [includegraphics cmd=\pgfimage,xmin=-6, xmax=6, ymin=-6, ymax=6] {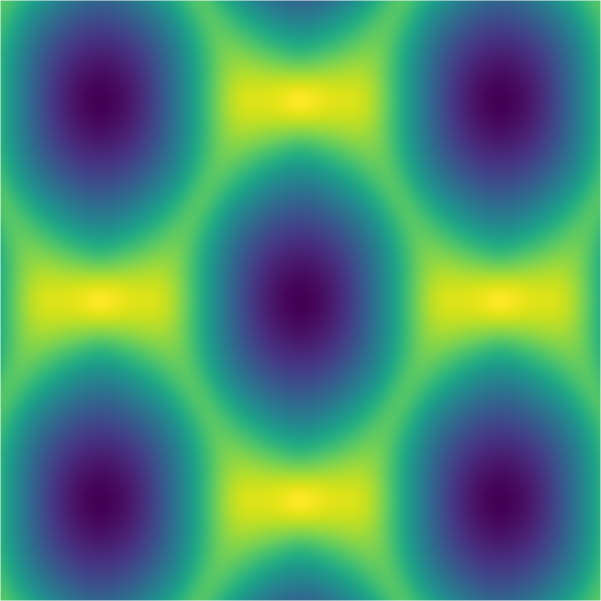};
\end{axis}
\end{tikzpicture}%
}
\hfill
\subfloat{%
  \begin{tikzpicture}
\begin{axis}[
title style={font=\small, yshift= -1.5ex},
title = \text{c)  $ (0,\pi,\pi)$ $\mathcal{S}^{z,z} $},
ylabel={[0,0,l]},
xlabel={[h,h,0]},
xmin=-6.05, xmax=6.05,
ymin=-6.05, ymax=6.05,
width=4.8cm,
height=4.8cm,
tick align=outside,
tick pos=left,
xtick={ -4,  0, 4},
ytick={ -4,  0, 4},
ytick style={draw=none},
xtick style={draw=none},
xticklabels={ $-4\pi$ , $0$ , $4\pi$ },
yticklabels={ $-4\pi$ , $0$ , $4\pi$ },
x tick label style={font=\small, yshift=1.5ex},
y tick label style={font=\small, xshift=1.5ex},
label style={font=\small},
y label style={yshift= -3ex},
x label style={yshift= +1ex},
point meta min=0,
point meta max=1,
]
\addplot graphics [includegraphics cmd=\pgfimage,xmin=-6, xmax=6, ymin=-6, ymax=6] {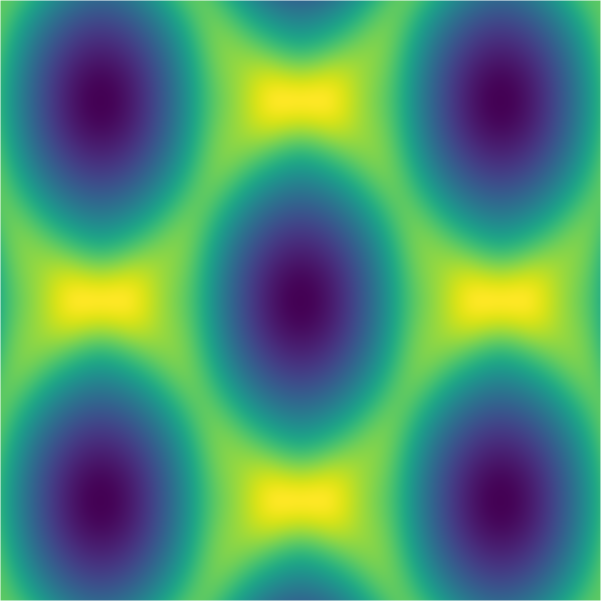};
\end{axis}
\end{tikzpicture}%
}\hfill
\subfloat{%
  \begin{tikzpicture}
\begin{axis}[
title style={font=\small, yshift= -1.5ex},
title = \text{d) $ (\pi,0,\pi)$ $\mathcal{S}^{z,z} $},
ylabel={[0,0,l]},
xlabel={[h,h,0]},
xmin=-6.05, xmax=6.05,
ymin=-6.05, ymax=6.05,
width=4.8cm,
height=4.8cm,
tick align=outside,
tick pos=left,
xtick={ -4,  0, 4},
ytick={ -4,  0, 4},
ytick style={draw=none},
xtick style={draw=none},
xticklabels={ $-4\pi$ , $0$ , $4\pi$ },
yticklabels={ $-4\pi$ , $0$ , $4\pi$ },
x tick label style={font=\small, yshift=1.5ex},
y tick label style={font=\small, xshift=1.5ex},
label style={font=\small},
y label style={yshift= -3ex},
x label style={yshift= +1ex},
point meta min=0,
point meta max=1,
]
\addplot graphics [includegraphics cmd=\pgfimage,xmin=-6, xmax=6, ymin=-6, ymax=6] {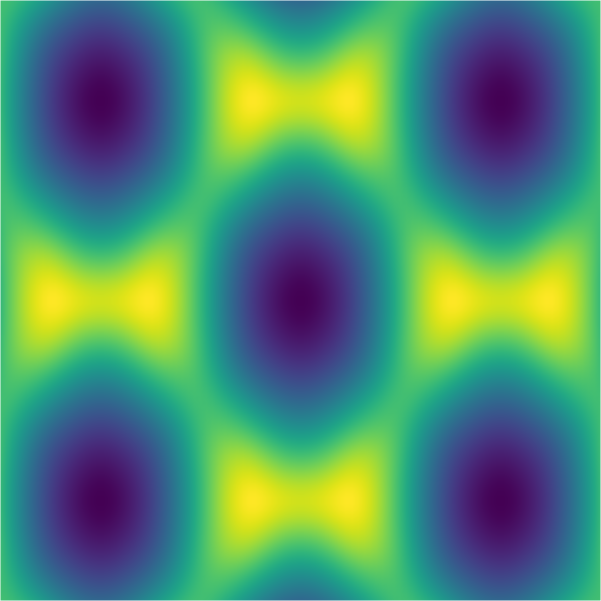};
\end{axis}
\end{tikzpicture}%
}\hfill
\subfloat{%
 \begin{tikzpicture}
\begin{axis}[
title style={font=\small, yshift= -1.5ex},
title = \text{e)  $(0,\pi, \pm\frac{\pi}{3})$ $\mathcal{S}^{z,z} $},
ylabel={[0,0,l]},
xlabel={[h,h,0]},
xmin=-6.05, xmax=6.05,
ymin=-6.05, ymax=6.05,
width=4.8cm,
height=4.8cm,
tick align=outside,
tick pos=left,
xtick={ -4,  0, 4},
ytick={ -4,  0, 4},
ytick style={draw=none},
xtick style={draw=none},
xticklabels={ $-4\pi$ , $0$ , $4\pi$ },
yticklabels={ $-4\pi$ , $0$ , $4\pi$ },
x tick label style={font=\small, yshift=1.5ex},
y tick label style={font=\small, xshift=1.5ex},
label style={font=\small},
y label style={yshift= -3ex},
x label style={yshift= +1ex},
point meta min=0,
point meta max=1
]
\addplot graphics [includegraphics cmd=\pgfimage,xmin=-6, xmax=6, ymin=-6, ymax=6] {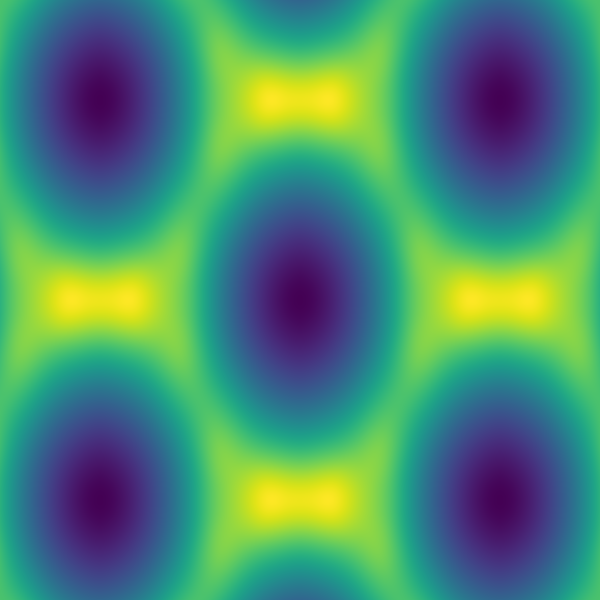};
\end{axis}
\end{tikzpicture}%
}
\subfloat{%
 \begin{tikzpicture}
\begin{axis}[
title style={font=\small, yshift= -1.5ex},
title = \text{f)  $(\pi,0, \pm\frac{\pi}{3})$ $\mathcal{S}^{z,z} $},
ylabel={[0,0,l]},
xlabel={[h,h,0]},
xmin=-6.05, xmax=6.05,
ymin=-6.05, ymax=6.05,
width=4.8cm,
height=4.8cm,
tick align=outside,
tick pos=left,
xtick={ -4,  0, 4},
ytick={ -4,  0, 4},
ytick style={draw=none},
xtick style={draw=none},
xticklabels={ $-4\pi$ , $0$ , $4\pi$ },
yticklabels={ $-4\pi$ , $0$ , $4\pi$ },
x tick label style={font=\small, yshift=1.5ex},
y tick label style={font=\small, xshift=1.5ex},
label style={font=\small},
y label style={yshift= -3ex},
x label style={yshift= +1ex},
point meta min=0,
point meta max=1
]
\addplot graphics [includegraphics cmd=\pgfimage,xmin=-6, xmax=6, ymin=-6, ymax=6] {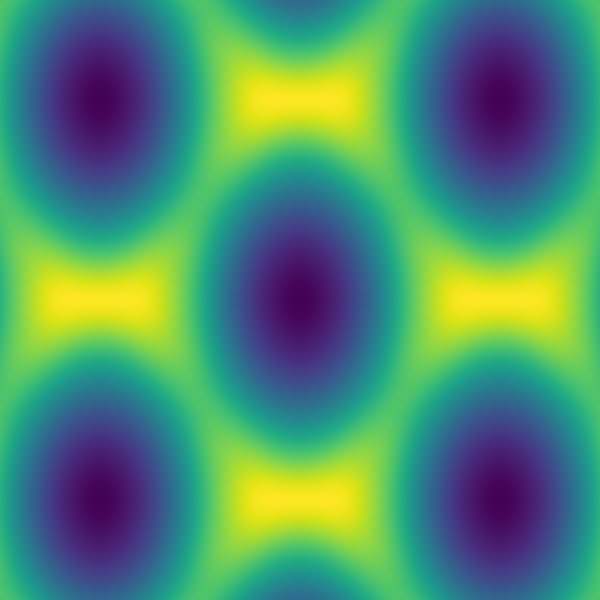};
\end{axis}
\end{tikzpicture}%
}\hfill
\subfloat{%
 \begin{tikzpicture}
    \begin{axis}[
    width=4.8cm,
    height=4.8cm,
    xmin=-6.05, xmax=6.05,
    ymin=-6.05, ymax=6.05,
    ,point meta min=0,
    point meta max=1,
    hide axis,
    colormap/viridis,
    colorbar horizontal,
    colorbar style={
        width= 8.2cm,
        height = 2ex,
        xtick={0,0.2,...,1},
        x tick label style={font=\small}}
    ]
    \end{axis}
\end{tikzpicture}%
}
 
\caption{(Color online). The normalized spin-spin correlation in local spin basis $\mathcal{S}^{z,z}(\mathbf{q})$ plotted along the $[h,h,l]$ plane. The spin liquid states are labeled by the fluxes of the $\hat{A}$ operators on the hexagonal, bow tie and rhombus loops: $(\phi_{\mhexagon},\phi_{\Join},\phi_{\Diamond}) = (p_1,p_1+n_1,1+\frac{2k}{3}) \pi$. The dashed and solid white lines in a) indicate the momentum cuts presented in Fig.~\ref{fig:PinchPoints}.   }
 \label{fig:localSpinSpinCorrelation}
\end{figure}

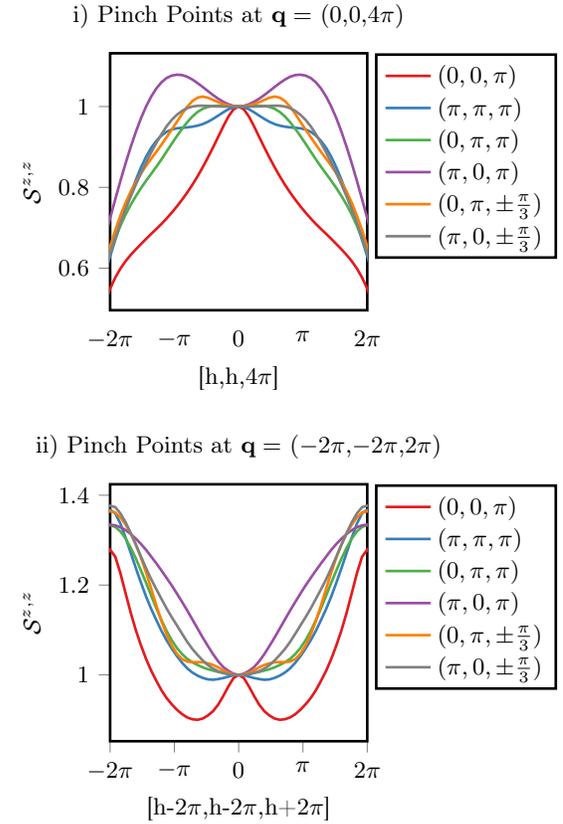
\begin{figure}
\subfloat{%
\begin{tikzpicture}
\begin{axis}[
legend pos=outer north east,
legend cell align={left},
line width=1pt,
cycle list/Set1,
no markers,
title = \text{i) Pinch Points at $\mathbf{q}=$ (0,0,$4\pi$)},
ylabel={ $\mathcal{S}^{z,z}$},
xlabel={[h,h,$4\pi$]},
xmin=-0.5, xmax=0.5,
width=5cm,
height=5cm,
tick align=outside,
tick pos=left,
xtick={ -0.5,-0.25, 0,0.25,0.5},
xtick style={draw=none},
xticklabels={ $-2\pi$ ,$-\pi$, $0$ ,$\pi$, $2\pi$ }]
\addplot table [x=a, y=c, col sep=comma] {horizontal4pi.csv};
\addlegendentry{ $ (0,0,\pi)$ }
\addplot table [x=a, y=b, col sep=comma] {horizontal4pi.csv};
\addlegendentry{ $ (\pi,\pi,\pi)$ }
\addplot table [x=a, y=e, col sep=comma] {horizontal4pi.csv};
\addlegendentry{ $ (0,\pi,\pi)$ }
\addplot table [x=a, y=d, col sep=comma] {horizontal4pi.csv};
\addlegendentry{ $ (\pi,0,\pi)$ }
\addplot table [x=a, y=f, col sep=comma] {horizontal4pi.csv};
\addlegendentry{ $ (0,\pi,\pm\frac{\pi}{3})$ }
\addplot[gray] table [x=a, y=g, col sep=comma] {horizontal4pi.csv};
\addlegendentry{ $ (\pi,0,\pm\frac{\pi}{3})$ }
\end{axis}
\end{tikzpicture}
}\hfill
\subfloat{%
\begin{tikzpicture}
\begin{axis}[
legend pos=outer north east,
legend cell align={left},
line width=1pt,
cycle list/Set1,
no markers,
title = \text{ii) Pinch Points at $\mathbf{q}=$ ($-2\pi$,$-2\pi$,$2\pi$)},
ylabel={ $\mathcal{S}^{z,z}$},
xlabel={[h-$2\pi$,h-$2\pi$,h+$2\pi$]},
xmin=-0.5, xmax=0.5,
width=5cm,
height=5cm,
tick align=outside,
tick pos=left,
xtick={ -0.5,-0.25, 0,0.25,0.5},
xticklabels={ $-2\pi$ ,$-\pi$, $0$ ,$\pi$, $2\pi$ }]
\addplot table [x=a, y=c, col sep=comma] {horizontal2pi.csv};
\addlegendentry{ $ (0,0,\pi)$ }
\addplot table [x=a, y=b, col sep=comma] {horizontal2pi.csv};
\addlegendentry{ $ (\pi,\pi,\pi)$ }
\addplot table [x=a, y=e, col sep=comma] {horizontal2pi.csv};
\addlegendentry{ $ (0,\pi,\pi)$ }
\addplot table [x=a, y=d, col sep=comma] {horizontal2pi.csv};
\addlegendentry{ $ (\pi,0,\pi)$ }
\addplot table [x=a, y=f, col sep=comma] {horizontal2pi.csv};
\addlegendentry{ $ (0,\pi,\pm\frac{\pi}{3})$ }
\addplot[gray] table [x=a, y=g, col sep=comma] {horizontal2pi.csv};
\addlegendentry{ $ (\pi,0,\pm\frac{\pi}{3})$ }
\end{axis}
\end{tikzpicture}
}
\caption{(Color online). Cut through the pinch points of the  spin-spin correlation $\mathcal{S}^{z,z}(\mathbf{q})$ in the local spin basis at  i)  $\mathbf{q}$=(0,0, $4\pi$) as indicated by the solid line in Figure \ref{fig:localSpinSpinCorrelation} a) and ii) $\mathbf{q}$=($-2\pi$,$-2\pi$,$+2\pi$) as indicated by the dashed line in Figure \ref{fig:localSpinSpinCorrelation} a) . For better comparability, the plots are normalized such that the pinch points have a magnitude of one.}
 \label{fig:PinchPoints}
\end{figure}

\begin{figure}
\hspace*{-.35in}
\subfloat{%
\begin{tikzpicture}
\begin{axis}[
title style={font=\small, yshift= -1.5ex},
title = \text{a)  $(0,0,\pi)$ $\mathcal{S}_{\text{TOT}}$},
ylabel={[0,0,l]},
xlabel={[h,h,0]},
xmin=-16.05, xmax=16.05,
ymin=-16.05, ymax=16.05,
width=4.8cm,
height=4.8cm,
tick align=outside,
tick pos=left,
xtick={ -16,-8,  0,8, 16},
ytick={ -16,-8,  0,8, 16},
ytick style={draw=none},
xtick style={draw=none},
xticklabels={ $-16\pi$ ,$-8\pi$, $0$ ,$8\pi$, $16\pi$},
yticklabels={ $-16\pi$ ,$-8\pi$, $0$ ,$8\pi$, $16\pi$},
x tick label style={font=\small, yshift=1.2ex},
y tick label style={font=\small, xshift=1.2ex},
label style={font=\small},
y label style={yshift= -3ex},
x label style={yshift= +1ex},
point meta min=0,
point meta max=1,
]
\addplot graphics [includegraphics cmd=\pgfimage,xmin=-16, xmax=16, ymin=-16, ymax=16] {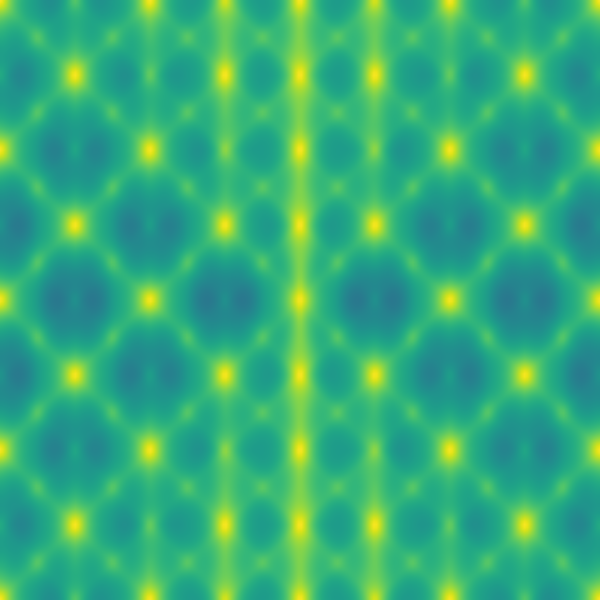};
\end{axis}
\end{tikzpicture}%
}
\subfloat{%
  \begin{tikzpicture}
\begin{axis}[
title style={font=\small, yshift= -1.5ex},
title = \text{b) $ (\pi,\pi,\pi)$ $\mathcal{S}_{\text{TOT}}$},
ylabel={[0,0,l]},
xlabel={[h,h,0]},
xmin=-16.05, xmax=16.05,
ymin=-16.05, ymax=16.05,
width=4.8cm,
height=4.8cm,
tick align=outside,
tick pos=left,
xtick={ -16,-8,  0,8, 16},
ytick={ -16,-8,  0,8, 16},
ytick style={draw=none},
xtick style={draw=none},
xticklabels={ $-16\pi$ ,$-8\pi$, $0$ ,$8\pi$, $16\pi$},
yticklabels={ $-16\pi$ ,$-8\pi$, $0$ ,$8\pi$, $16\pi$},
x tick label style={font=\small, yshift=1.2ex},
y tick label style={font=\small, xshift=1.2ex},
label style={font=\small},
y label style={yshift= -3ex},
x label style={yshift= +1ex},
point meta min=0,
point meta max=1,
]
\addplot graphics [includegraphics cmd=\pgfimage,xmin=-16, xmax=16, ymin=-16, ymax=16] {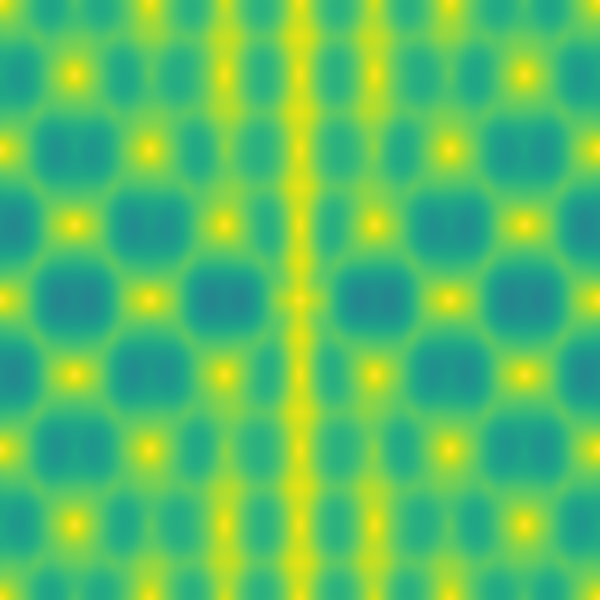};
\end{axis}
\end{tikzpicture}%
}
\\\hspace*{-.35in}
\subfloat{%
  \begin{tikzpicture}
\begin{axis}[
title style={font=\small, yshift= -1.5ex},
title = \text{c) $ (0,\pi,\pi)$ $\mathcal{S}_{\text{TOT}}$},
ylabel={[0,0,l]},
xlabel={[h,h,0]},
xmin=-16.05, xmax=16.05,
ymin=-16.05, ymax=16.05,
width=4.8cm,
height=4.8cm,
tick align=outside,
tick pos=left,
xtick={ -16,-8,  0,8, 16},
ytick={ -16,-8,  0,8, 16},
ytick style={draw=none},
xtick style={draw=none},
xticklabels={ $-16\pi$ ,$-8\pi$, $0$ ,$8\pi$, $16\pi$},
yticklabels={ $-16\pi$ ,$-8\pi$, $0$ ,$8\pi$, $16\pi$},
x tick label style={font=\small, yshift=1.2ex},
y tick label style={font=\small, xshift=1.2ex},
label style={font=\small},
y label style={yshift= -3ex},
x label style={yshift= +1ex},
point meta min=0,
point meta max=1,
]
\addplot graphics [includegraphics cmd=\pgfimage,xmin=-16, xmax=16, ymin=-16, ymax=16] {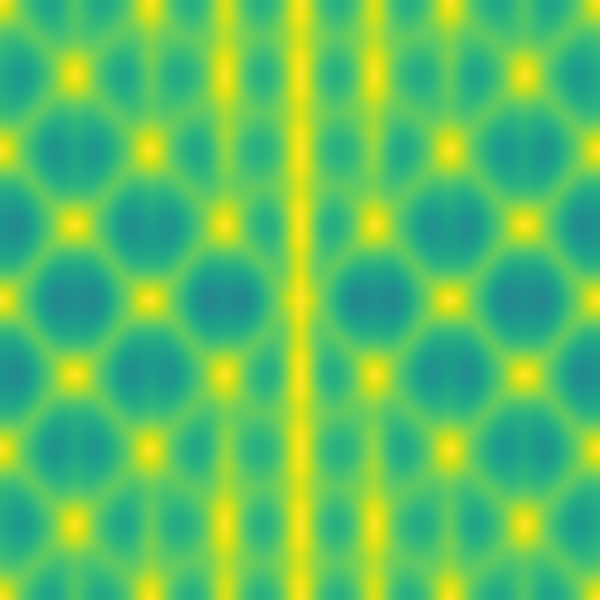};
\end{axis}
\end{tikzpicture}%
}
\subfloat{%
  \begin{tikzpicture}
\begin{axis}[
title style={font=\small, yshift= -1.5ex},
title = \text{d)  $ (\pi,0,\pi)$ $\mathcal{S}_{\text{TOT}}$},
ylabel={[0,0,l]},
xlabel={[h,h,0]},
xmin=-16.05, xmax=16.05,
ymin=-16.05, ymax=16.05,
width=4.8cm,
height=4.8cm,
tick align=outside,
tick pos=left,
xtick={ -16,-8,  0,8, 16},
ytick={ -16,-8,  0,8, 16},
ytick style={draw=none},
xtick style={draw=none},
xticklabels={ $-16\pi$ ,$-8\pi$, $0$ ,$8\pi$, $16\pi$},
yticklabels={ $-16\pi$ ,$-8\pi$, $0$ ,$8\pi$, $16\pi$},
x tick label style={font=\small, yshift=1.2ex},
y tick label style={font=\small, xshift=1.2ex},
label style={font=\small},
y label style={yshift= -3ex},
x label style={yshift= +1ex},
point meta min=0,
point meta max=1,
]
\addplot graphics [includegraphics cmd=\pgfimage,xmin=-16, xmax=16, ymin=-16, ymax=16] {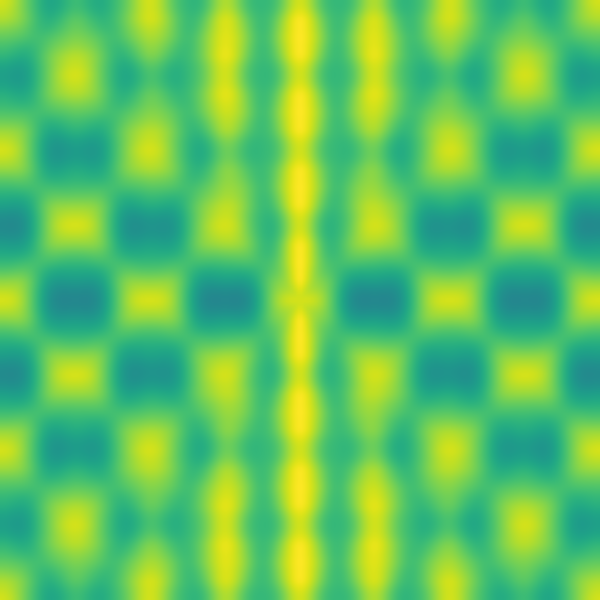};
\end{axis}
\end{tikzpicture}%
}\\\hspace*{-.35in}
\subfloat{%
 \begin{tikzpicture}
\begin{axis}[
title style={font=\small, yshift= -1.5ex},
title = \text{e)  $(0,\pi, \pm\frac{\pi}{3})$ $\mathcal{S}_{\text{TOT}}$},
ylabel={[0,0,l]},
xlabel={[h,h,0]},
xmin=-16.05, xmax=16.05,
ymin=-16.05, ymax=16.05,
width=4.8cm,
height=4.8cm,
tick align=outside,
tick pos=left,
xtick={ -16,-8,  0,8, 16},
ytick={ -16,-8,  0,8, 16},
ytick style={draw=none},
xtick style={draw=none},
xticklabels={ $-16\pi$ ,$-8\pi$, $0$ ,$8\pi$, $16\pi$},
yticklabels={ $-16\pi$ ,$-8\pi$, $0$ ,$8\pi$, $16\pi$},
x tick label style={font=\small, yshift=1.2ex},
y tick label style={font=\small, xshift=1.2ex},
label style={font=\small},
y label style={yshift= -3ex},
x label style={yshift= +1ex},
point meta min=0,
point meta max=1,
]
\addplot graphics [includegraphics cmd=\pgfimage,xmin=-16, xmax=16, ymin=-16, ymax=16] {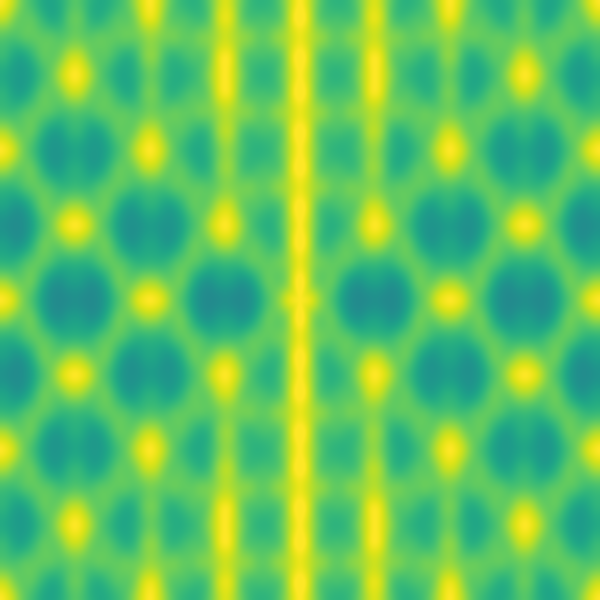};
\end{axis}
\end{tikzpicture}%
}
\subfloat{%
 \begin{tikzpicture}
\begin{axis}[
title style={font=\small, yshift= -1.5ex},
title = \text{f)  $(\pi,0, \pm\frac{\pi}{3})$ $\mathcal{S}_{\text{TOT}}$},
ylabel={[0,0,l]},
xlabel={[h,h,0]},
xmin=-16.05, xmax=16.05,
ymin=-16.05, ymax=16.05,
width=4.8cm,
height=4.8cm,
tick align=outside,
tick pos=left,
xtick={ -16,-8,  0,8, 16},
ytick={ -16,-8,  0,8, 16},
ytick style={draw=none},
xtick style={draw=none},
xticklabels={ $-16\pi$ ,$-8\pi$, $0$ ,$8\pi$, $16\pi$},
yticklabels={ $-16\pi$ ,$-8\pi$, $0$ ,$8\pi$, $16\pi$},
x tick label style={font=\small, yshift=1.2ex},
y tick label style={font=\small, xshift=1.2ex},
label style={font=\small},
y label style={yshift= -3ex},
x label style={yshift= +1ex},
point meta min=0,
point meta max=1,
]
\addplot graphics [includegraphics cmd=\pgfimage,xmin=-16, xmax=16, ymin=-16, ymax=16] {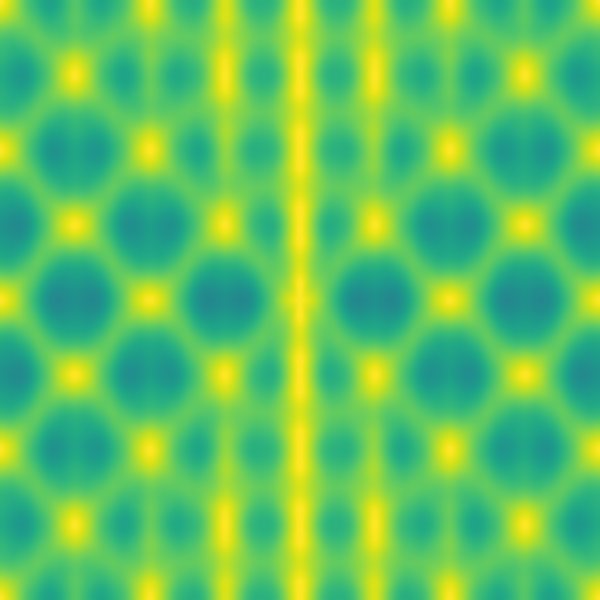};
\end{axis}
\end{tikzpicture}%
}\\
\subfloat{%
 \begin{tikzpicture}
    \begin{axis}[
    width=4.8cm,
    height=4.8cm,
    xmin=-6.05, xmax=6.05,
    ymin=-6.05, ymax=6.05,
    ,point meta min=0,
    point meta max=1,
    hide axis,
    colormap/viridis,
    colorbar horizontal,
    colorbar style={
        width= 8.2cm,
        height = 2ex,
        xtick={0,0.2,...,1},
        x tick label style={font=\small}}
    ]
    \end{axis}
\end{tikzpicture}%
}
 
\caption{(Color online). The normalized 
total neutron scattering amplitude $\mathcal{S}_{\text{TOT}}(\mathbf{q})$ plotted along the [h,h,l] plane. The spin liquid states are labeled by the fluxes of the $\hat{A}$ operators on the hexagonal, bow tie and rhombus loops: $(\phi_{\mhexagon},\phi_{\Join},\phi_{\Diamond}) = (p_1,p_1+n_1,1+\frac{2k}{3}) \pi$.   }
 \label{fig:SSFGlobalTOT}
\end{figure}
 
 \begin{figure}
 \hspace*{-.35in}
\subfloat{%
\begin{tikzpicture}
\begin{axis}[
title style={font=\small, yshift= -1.5ex},
title = \text{a) $(0,0,\pi)$ $\mathcal{S}_{\text{SF}}$},
ylabel={[0,0,l]},
xlabel={[h,h,0]},
xmin=-16.05, xmax=16.05,
ymin=-16.05, ymax=16.05,
width=4.8cm,
height=4.8cm,
tick align=outside,
tick pos=left,
xtick={ -16,-8,  0,8, 16},
ytick={ -16,-8,  0,8, 16},
ytick style={draw=none},
xtick style={draw=none},
xticklabels={ $-16\pi$ ,$-8\pi$, $0$ ,$8\pi$, $16\pi$},
yticklabels={ $-16\pi$ ,$-8\pi$, $0$ ,$8\pi$, $16\pi$},
x tick label style={font=\small, yshift=1.2ex},
y tick label style={font=\small, xshift=1.2ex},
label style={font=\small},
y label style={yshift= -3ex},
x label style={yshift= +1ex},
point meta min=0,
point meta max=1,
]
\addplot graphics [includegraphics cmd=\pgfimage,xmin=-16, xmax=16, ymin=-16, ymax=16] {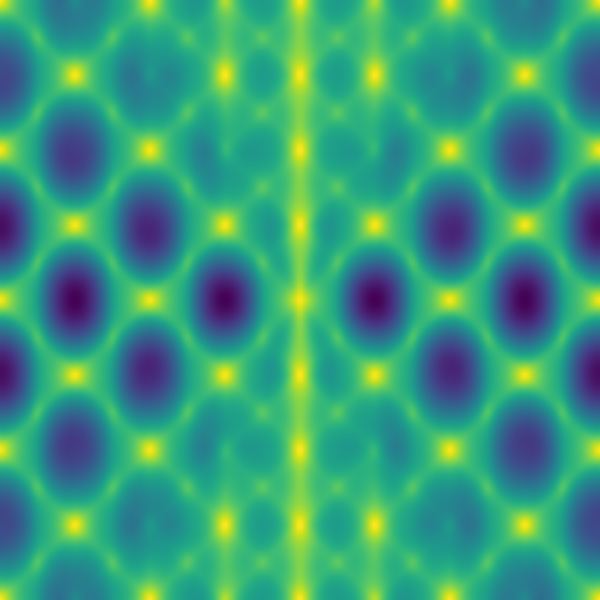};
\end{axis}
\end{tikzpicture}%
}
\subfloat{%
  \begin{tikzpicture}
\begin{axis}[
title style={font=\small, yshift= -1.5ex},
title = \text{b)  $(\pi,\pi,\pi)$ $\mathcal{S}_{\text{SF}}$},
ylabel={[0,0,l]},
xlabel={[h,h,0]},
xmin=-16.05, xmax=16.05,
ymin=-16.05, ymax=16.05,
width=4.8cm,
height=4.8cm,
tick align=outside,
tick pos=left,
xtick={ -16,-8,  0,8, 16},
ytick={ -16,-8,  0,8, 16},
ytick style={draw=none},
xtick style={draw=none},
xticklabels={ $-16\pi$ ,$-8\pi$, $0$ ,$8\pi$, $16\pi$},
yticklabels={ $-16\pi$ ,$-8\pi$, $0$ ,$8\pi$, $16\pi$},
x tick label style={font=\small, yshift=1.2ex},
y tick label style={font=\small, xshift=1.2ex},
label style={font=\small},
y label style={yshift= -3ex},
x label style={yshift= +1ex},
point meta min=0,
point meta max=1,
]
\addplot graphics [includegraphics cmd=\pgfimage,xmin=-16, xmax=16, ymin=-16, ymax=16] {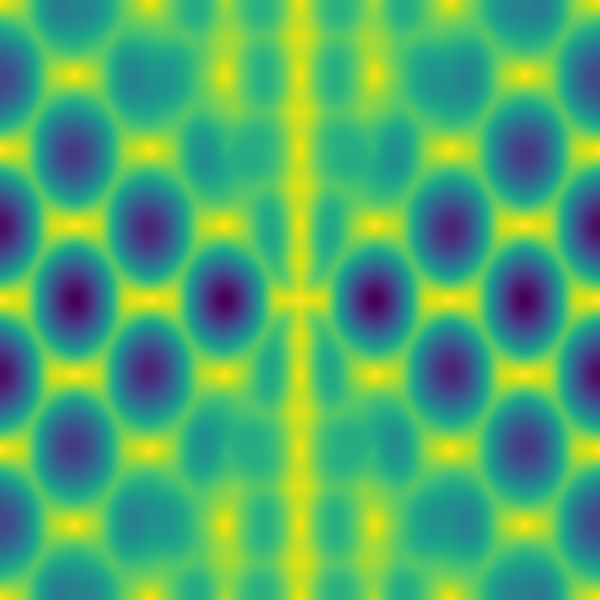};
\end{axis}
\end{tikzpicture}%
} \\\hspace*{-.35in}
\subfloat{%
  \begin{tikzpicture}
\begin{axis}[
title style={font=\small, yshift= -1.5ex},
title = \text{c) $(0,\pi,\pi)$ $\mathcal{S}_{\text{SF}}$},
ylabel={[0,0,l]},
xlabel={[h,h,0]},
xmin=-16.05, xmax=16.05,
ymin=-16.05, ymax=16.05,
width=4.8cm,
height=4.8cm,
tick align=outside,
tick pos=left,
xtick={ -16,-8,  0,8, 16},
ytick={ -16,-8,  0,8, 16},
ytick style={draw=none},
xtick style={draw=none},
xticklabels={ $-16\pi$ ,$-8\pi$, $0$ ,$8\pi$, $16\pi$},
yticklabels={ $-16\pi$ ,$-8\pi$, $0$ ,$8\pi$, $16\pi$},
x tick label style={font=\small, yshift=1.2ex},
y tick label style={font=\small, xshift=1.2ex},
label style={font=\small},
y label style={yshift= -3ex},
x label style={yshift= +1ex},
point meta min=0,
point meta max=1,
]
\addplot graphics [includegraphics cmd=\pgfimage,xmin=-16, xmax=16, ymin=-16, ymax=16] {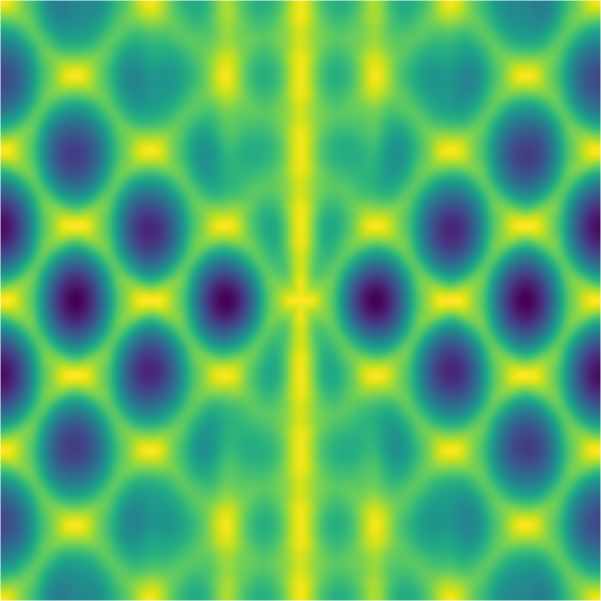};
\end{axis}
\end{tikzpicture}%
}
\subfloat{%
  \begin{tikzpicture}
\begin{axis}[
title style={font=\small, yshift= -1.5ex},
title = \text{d) $(\pi,0,\pi)$ $\mathcal{S}_{\text{SF}}$},
ylabel={[0,0,l]},
xlabel={[h,h,0]},
xmin=-16.05, xmax=16.05,
ymin=-16.05, ymax=16.05,
width=4.8cm,
height=4.8cm,
tick align=outside,
tick pos=left,
xtick={ -16,-8,  0,8, 16},
ytick={ -16,-8,  0,8, 16},
ytick style={draw=none},
xtick style={draw=none},
xticklabels={ $-16\pi$ ,$-8\pi$, $0$ ,$8\pi$, $16\pi$},
yticklabels={ $-16\pi$ ,$-8\pi$, $0$ ,$8\pi$, $16\pi$},
x tick label style={font=\small, yshift=1.2ex},
y tick label style={font=\small, xshift=1.2ex},
label style={font=\small},
y label style={yshift= -3ex},
x label style={yshift= +1ex},
point meta min=0,
point meta max=1,
]
\addplot graphics [includegraphics cmd=\pgfimage,xmin=-16, xmax=16, ymin=-16, ymax=16] {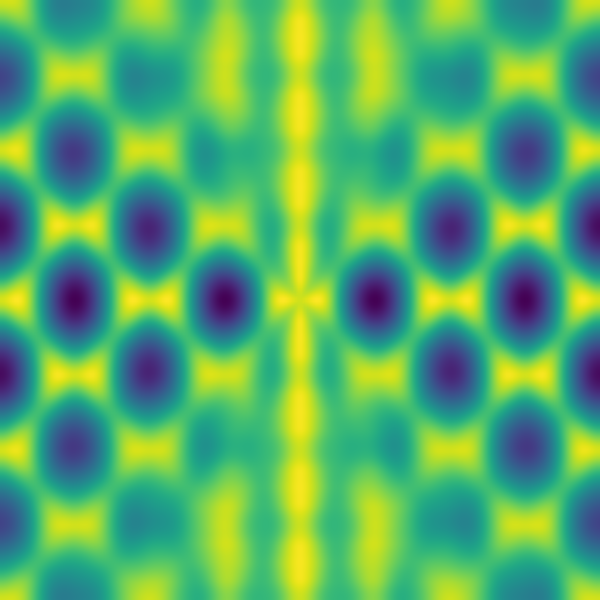};
\end{axis}
\end{tikzpicture}%
}\\\hspace*{-.35in}
\subfloat{%
 \begin{tikzpicture}
\begin{axis}[
title style={font=\small, yshift= -1.5ex},
title = \text{e)  $(0,\pi, \pm\frac{\pi}{3})$ $\mathcal{S}_{\text{SF}}$},
ylabel={[0,0,l]},
xlabel={[h,h,0]},
xmin=-16.05, xmax=16.05,
ymin=-16.05, ymax=16.05,
width=4.8cm,
height=4.8cm,
tick align=outside,
tick pos=left,
xtick={ -16,-8,  0,8, 16},
ytick={ -16,-8,  0,8, 16},
ytick style={draw=none},
xtick style={draw=none},
xticklabels={ $-16\pi$ ,$-8\pi$, $0$ ,$8\pi$, $16\pi$},
yticklabels={ $-16\pi$ ,$-8\pi$, $0$ ,$8\pi$, $16\pi$},
x tick label style={font=\small, yshift=1.2ex},
y tick label style={font=\small, xshift=1.2ex},
label style={font=\small},
y label style={yshift= -3ex},
x label style={yshift= +1ex},
point meta min=0,
point meta max=1,
]
\addplot graphics [includegraphics cmd=\pgfimage,xmin=-16, xmax=16, ymin=-16, ymax=16] {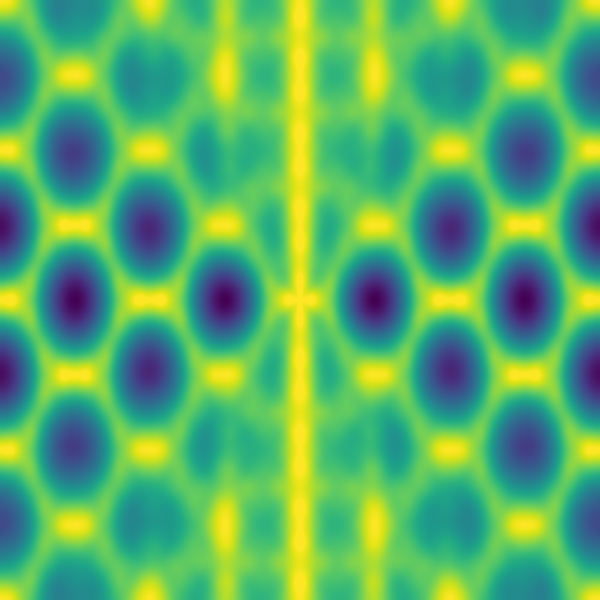};
\end{axis}
\end{tikzpicture}%
}
\subfloat{%
 \begin{tikzpicture}
\begin{axis}[
title style={font=\small, yshift= -1.5ex},
title = \text{f)  $(\pi,0, \pm\frac{\pi}{3})$ $\mathcal{S}_{\text{SF}}$},
ylabel={[0,0,l]},
xlabel={[h,h,0]},
xmin=-16.05, xmax=16.05,
ymin=-16.05, ymax=16.05,
width=4.8cm,
height=4.8cm,
tick align=outside,
tick pos=left,
xtick={ -16,-8,  0,8, 16},
ytick={ -16,-8,  0,8, 16},
ytick style={draw=none},
xtick style={draw=none},
xticklabels={ $-16\pi$ ,$-8\pi$, $0$ ,$8\pi$, $16\pi$},
yticklabels={ $-16\pi$ ,$-8\pi$, $0$ ,$8\pi$, $16\pi$},
x tick label style={font=\small, yshift=1.2ex},
y tick label style={font=\small, xshift=1.2ex},
label style={font=\small},
y label style={yshift= -3ex},
x label style={yshift= +1ex},
point meta min=0,
point meta max=1,
]
\addplot graphics [includegraphics cmd=\pgfimage,xmin=-16, xmax=16, ymin=-16, ymax=16] {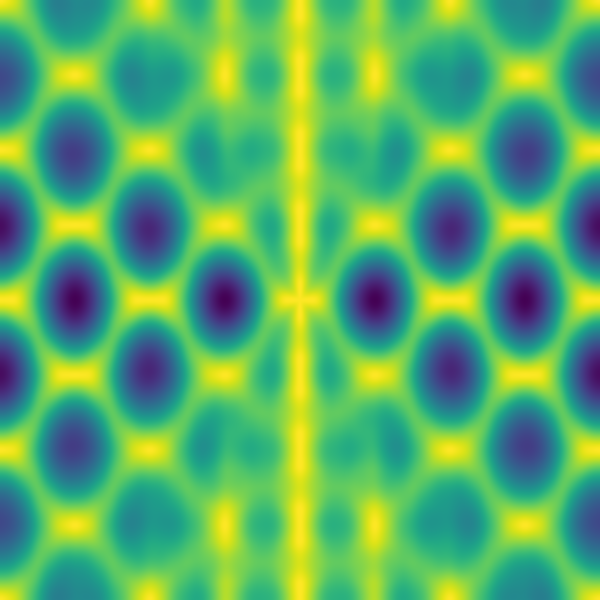};
\end{axis}
\end{tikzpicture}%
}\\
\subfloat{%
 \begin{tikzpicture}
    \begin{axis}[
    width=4.8cm,
    height=4.8cm,
    xmin=-6.05, xmax=6.05,
    ymin=-6.05, ymax=6.05,
    ,point meta min=0,
    point meta max=1,
    hide axis,
    colormap/viridis,
    colorbar horizontal,
    colorbar style={
        width= 8.2cm,
        height = 2ex,
        xtick={0,0.2,...,1},
        x tick label style={font=\small}}
    ]
    \end{axis}
\end{tikzpicture}%
}
 
\caption{(Color online). The normalized 
spin flip channel of the neutron scattering amplitude $\mathcal{S}_{\text{SF}}(\mathbf{q})$ plotted along the [h,h,l] plane. The spin liquid states are labeled by the fluxes of the $\hat{A}$ operators on the hexagonal, bow tie and rhombus loops: $(\phi_{\mhexagon},\phi_{\Join},\phi_{\Diamond}) = (p_1,p_1+n_1,1+\frac{2k}{3}) \pi$.     }
 \label{fig:SSFGlobalSF}
\end{figure}

  \begin{figure}
\hspace*{-.35in}
\subfloat{%
\begin{tikzpicture}
\begin{axis}[
title style={font=\small, yshift= -1.5ex},
title = \text{a) $ (0,0,\pi)$ $\mathcal{S}_\text{NSF} $},
ylabel={[0,0,l]},
xlabel={[h,h,0]},
xmin=-16.05, xmax=16.05,
ymin=-16.05, ymax=16.05,
width=4.7cm,
height=4.7cm,
tick align=outside,
tick pos=left,
xtick={ -16,-8,  0,8, 16},
ytick={ -16,-8,  0,8, 16},
ytick style={draw=none},
xtick style={draw=none},
xticklabels={ $-16\pi$ ,$-8\pi$, $0$ ,$8\pi$, $16\pi$},
yticklabels={ $-16\pi$ ,$-8\pi$, $0$ ,$8\pi$, $16\pi$},
x tick label style={font=\small, yshift=1.2ex},
y tick label style={font=\small, xshift=1.2ex},
label style={font=\small},
y label style={yshift= -3ex},
x label style={yshift= +1ex},
point meta min=0,
point meta max=1,
]
\addplot graphics [includegraphics cmd=\pgfimage,xmin=-16, xmax=16, ymin=-16, ymax=16] {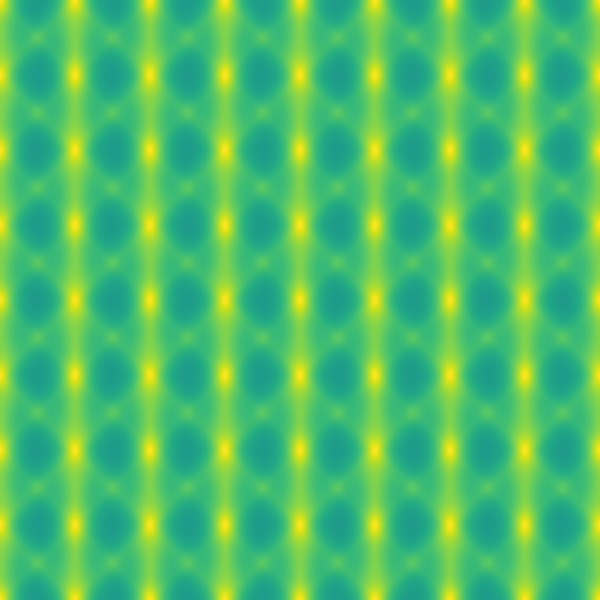};
\end{axis}
\end{tikzpicture}%
}
\subfloat{%
  \begin{tikzpicture}
\begin{axis}[
title style={font=\small, yshift= -1.5ex},
title = \text{b) $ (\pi,\pi,\pi)$ $\mathcal{S}_\text{NSF} $},
ylabel={[0,0,l]},
xlabel={[h,h,0]},
xmin=-16.05, xmax=16.05,
ymin=-16.05, ymax=16.05,
width=4.7cm,
height=4.7cm,
tick align=outside,
tick pos=left,
xtick={ -16,-8,  0,8, 16},
ytick={ -16,-8,  0,8, 16},
ytick style={draw=none},
xtick style={draw=none},
xticklabels={ $-16\pi$ ,$-8\pi$, $0$ ,$8\pi$, $16\pi$},
yticklabels={ $-16\pi$ ,$-8\pi$, $0$ ,$8\pi$, $16\pi$},
x tick label style={font=\small, yshift=1.2ex},
y tick label style={font=\small, xshift=1.2ex},
label style={font=\small},
y label style={yshift= -3ex},
x label style={yshift= +1ex},
point meta min=0,
point meta max=1,
]
\addplot graphics [includegraphics cmd=\pgfimage,xmin=-16, xmax=16, ymin=-16, ymax=16] {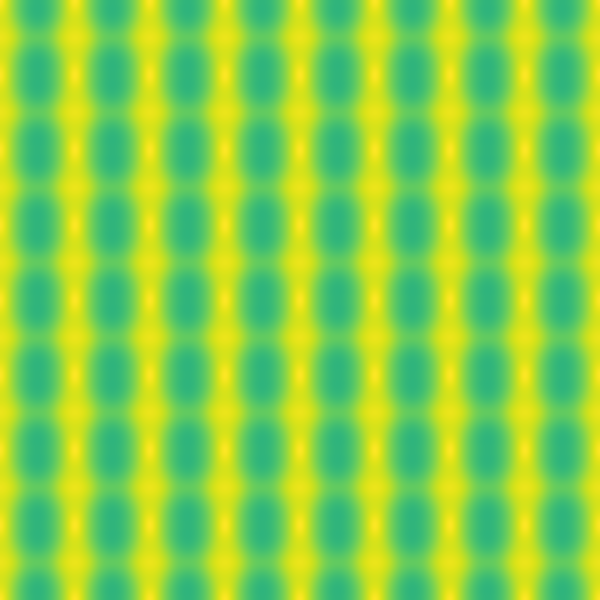};
\end{axis}
\end{tikzpicture}%
}
\\\hspace*{-.35in}
\subfloat{%
  \begin{tikzpicture}
\begin{axis}[
title style={font=\small, yshift= -1.5ex},
title = \text{c) $(0,\pi,\pi)$ $\mathcal{S}_\text{NSF} $},
ylabel={[0,0,l]},
xlabel={[h,h,0]},
xmin=-16.05, xmax=16.05,
ymin=-16.05, ymax=16.05,
width=4.7cm,
height=4.7cm,
tick align=outside,
tick pos=left,
xtick={ -16,-8,  0,8, 16},
ytick={ -16,-8,  0,8, 16},
ytick style={draw=none},
xtick style={draw=none},
xticklabels={ $-16\pi$ ,$-8\pi$, $0$ ,$8\pi$, $16\pi$},
yticklabels={ $-16\pi$ ,$-8\pi$, $0$ ,$8\pi$, $16\pi$},
x tick label style={font=\small, yshift=1.2ex},
y tick label style={font=\small, xshift=1.2ex},
label style={font=\small},
y label style={yshift= -3ex},
x label style={yshift= +1ex},
point meta min=0,
point meta max=1,
]
\addplot graphics [includegraphics cmd=\pgfimage,xmin=-16, xmax=16, ymin=-16, ymax=16] {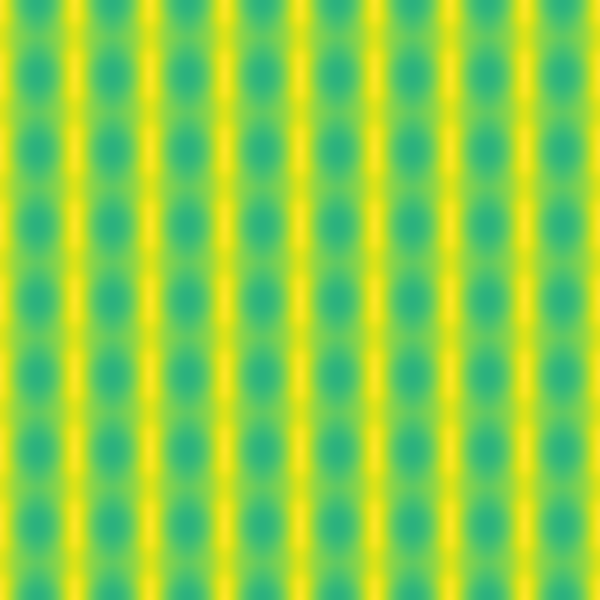};
\end{axis}
\end{tikzpicture}%
}
\subfloat{%
  \begin{tikzpicture}
\begin{axis}[
title style={font=\small, yshift= -1.5ex},
title = \text{d) $ (\pi,0,\pi)$ $\mathcal{S}_\text{NSF} $},
ylabel={[0,0,l]},
xlabel={[h,h,0]},
xmin=-16.05, xmax=16.05,
ymin=-16.05, ymax=16.05,
width=4.7cm,
height=4.7cm,
tick align=outside,
tick pos=left,
xtick={ -16,-8,  0,8, 16},
ytick={ -16,-8,  0,8, 16},
ytick style={draw=none},
xtick style={draw=none},
xticklabels={ $-16\pi$ ,$-8\pi$, $0$ ,$8\pi$, $16\pi$},
yticklabels={ $-16\pi$ ,$-8\pi$, $0$ ,$8\pi$, $16\pi$},
x tick label style={font=\small, yshift=1.2ex},
y tick label style={font=\small, xshift=1.2ex},
label style={font=\small},
y label style={yshift= -3ex},
x label style={yshift= +1ex},
point meta min=0,
point meta max=1,
]
\addplot graphics [includegraphics cmd=\pgfimage,xmin=-16, xmax=16, ymin=-16, ymax=16] {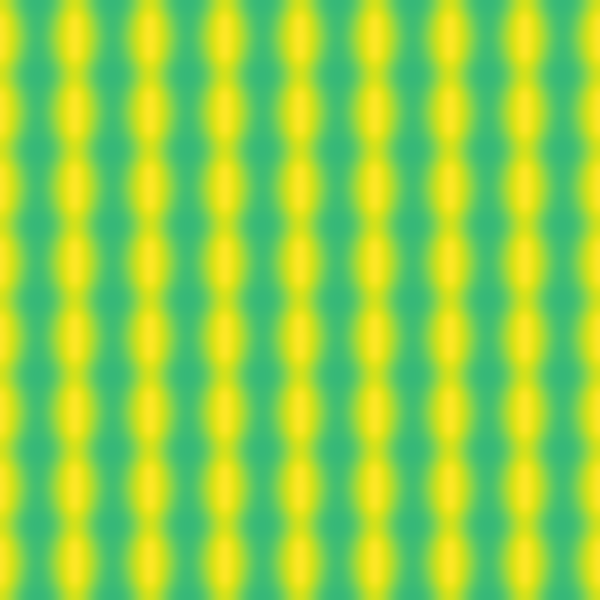};
\end{axis}
\end{tikzpicture}%
}\\\hspace*{-.35in}
\subfloat{%
 \begin{tikzpicture}
\begin{axis}[
title style={font=\small, yshift= -1.5ex},
title = \text{e)  $(0,\pi, \pm\frac{\pi}{3})$ $\mathcal{S}_{\text{NSF}}$},
ylabel={[0,0,l]},
xlabel={[h,h,0]},
xmin=-16.05, xmax=16.05,
ymin=-16.05, ymax=16.05,
width=4.8cm,
height=4.8cm,
tick align=outside,
tick pos=left,
xtick={ -16,-8,  0,8, 16},
ytick={ -16,-8,  0,8, 16},
ytick style={draw=none},
xtick style={draw=none},
xticklabels={ $-16\pi$ ,$-8\pi$, $0$ ,$8\pi$, $16\pi$},
yticklabels={ $-16\pi$ ,$-8\pi$, $0$ ,$8\pi$, $16\pi$},
x tick label style={font=\small, yshift=1.2ex},
y tick label style={font=\small, xshift=1.2ex},
label style={font=\small},
y label style={yshift= -3ex},
x label style={yshift= +1ex},
point meta min=0,
point meta max=1,
]
\addplot graphics [includegraphics cmd=\pgfimage,xmin=-16, xmax=16, ymin=-16, ymax=16] {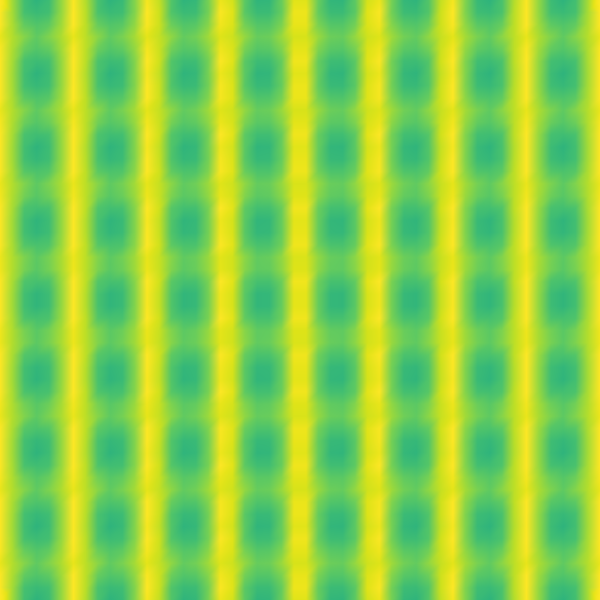};
\end{axis}
\end{tikzpicture}%
}
\subfloat{%
 \begin{tikzpicture}
\begin{axis}[
title style={font=\small, yshift= -1.5ex},
title = \text{f)  $(\pi,0, \pm\frac{\pi}{3})$ $\mathcal{S}_{\text{NSF}}$},
ylabel={[0,0,l]},
xlabel={[h,h,0]},
xmin=-16.05, xmax=16.05,
ymin=-16.05, ymax=16.05,
width=4.8cm,
height=4.8cm,
tick align=outside,
tick pos=left,
xtick={ -16,-8,  0,8, 16},
ytick={ -16,-8,  0,8, 16},
ytick style={draw=none},
xtick style={draw=none},
xticklabels={ $-16\pi$ ,$-8\pi$, $0$ ,$8\pi$, $16\pi$},
yticklabels={ $-16\pi$ ,$-8\pi$, $0$ ,$8\pi$, $16\pi$},
x tick label style={font=\small, yshift=1.2ex},
y tick label style={font=\small, xshift=1.2ex},
label style={font=\small},
y label style={yshift= -3ex},
x label style={yshift= +1ex},
point meta min=0,
point meta max=1,
]
\addplot graphics [includegraphics cmd=\pgfimage,xmin=-16, xmax=16, ymin=-16, ymax=16] {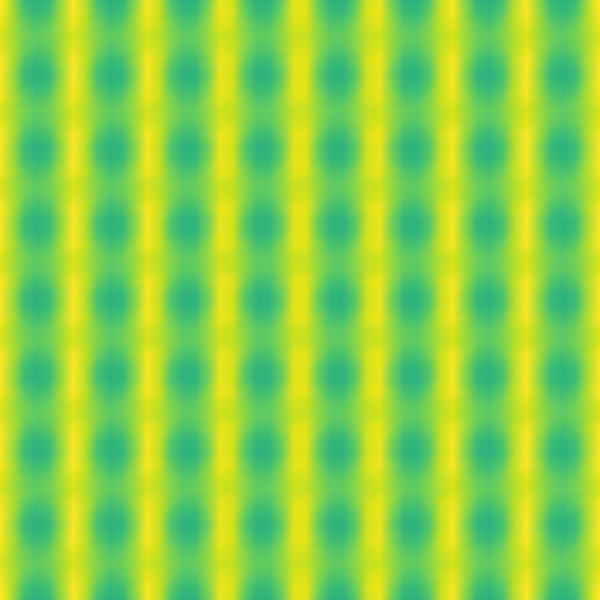};
\end{axis}
\end{tikzpicture}%
}\\
\subfloat{%
 \begin{tikzpicture}
    \begin{axis}[
    width=4.8cm,
    height=4.8cm,
    xmin=-6.05, xmax=6.05,
    ymin=-6.05, ymax=6.05,
    ,point meta min=0,
    point meta max=1,
    hide axis,
    colormap/viridis,
    colorbar horizontal,
    colorbar style={
        width= 8.2cm,
        height = 2ex,
        xtick={0,0.2,...,1},
        x tick label style={font=\small}}
    ]
    \end{axis}
\end{tikzpicture}%
}
 
\caption{(Color online). The normalized 
no spin flip channel of the neutron scattering amplitude $\mathcal{S}_{\text{NSF}}(\mathbf{q})$ plotted along the [h,h,l] plane. The spin liquid states are labeled by the fluxes of the $\hat{A}$ operators on the hexagonal, bow tie and rhombus loops: $(\phi_{\mhexagon},\phi_{\Join},\phi_{\Diamond}) = (p_1,p_1+n_1,1+\frac{2k}{3}) \pi$.     }
 \label{fig:SSFGlobalNSF}
\end{figure}

 The spin-spin correlations in the local spin basis can be seen in Figure \ref{fig:localSpinSpinCorrelation}. Their main features are the broadened pinch points at $\mathbf{q}=(0,0, \pm4 \pi)$, $\mathbf{q}=(\pm 2\pi,\pm 2\pi,\pm 2\pi)$ and symmetry related points. In case of the classical Heisenberg model these pinch points have been argued to be caused by the ice-rule: The sum of all spins on every tetrahedron has to vanish \cite{Isakov_Dipolar_2004,Moessner_Classical_1998,Moessner_Low-temperature_1998}.
The quantum fluctuations break this ice rule which result in smeared out pinch points \cite{Igbal_Quantum_2019, canals_pyrochlore_1998, Huang_SpinIce_2016, Schaefer_Pyrochlore_2020, Zhang_Dynamical_2019, plumb_continuum_2019, kiese_multiloop_2021}. For the Heisenberg model on the pyrochlore lattice  two types of pinch points have previously been reported. The spin-spin correlation either has a maximum \cite{Schaefer_Pyrochlore_2020, Hagymasi_Possible_2021} or a saddle point at the pinch points \cite{canals_pyrochlore_1998, kiese_multiloop_2021}. Plots of the spin correlations in the vicinity of the pinch points for the six spin-liquid states considered here can be found in Figure \ref{fig:PinchPoints}. For the $(0,\pi,\pi)$, $(\pi,0,\pi)$, $(\pi,0,\pm\frac{\pi}{3})$ and $(0,\pi,\pm\frac{\pi}{3})$ state, the  $\mathcal{S}^{z,z}$ correlator has a saddle point while for the $(0,0,\pi)$ state,  $(\pi,\pi,\pi)$  state, the  $\mathcal{S}^{z,z}$ correlator has a maximum at the pinch points.\\
 For $\mathcal{S}^{z,z}$ the state $(\pi,0,\pi)$ shows good qualitative agreement with previous theoretical work using pseudo fermion functional renormalization group (PFFRG) studies \cite{Igbal_Quantum_2019, kiese_multiloop_2021, ritter_multiloop_2021} and exact diagonalization of small clusters \cite{canals_pyrochlore_1998} of the Heisenberg and XXZ model.
 The state $(0,0,\pi)$ shows good qualitative agreement with a PFFRG result on the $J_1-J_2$ model for antiferromagnetic $J_2$ \cite{Igbal_Quantum_2019}.\\
 \noindent
 The results for $\mathcal{S}_{\text{TOT}}$ as well as the SF and NSF channel can be seen in Figure \ref{fig:SSFGlobalTOT}, \ref{fig:SSFGlobalSF}, \ref{fig:SSFGlobalNSF} , respectively.  
 Our results can be compared to the PFFRG results of Ritter \cite{ritter_multiloop_2021} and the classical Monte Carlo result of Taillefumier \emph{et al.} \cite{taillefumier_competing_2017}. We should emphasize that Ritters results are for a coupling angle of $\theta =20^\circ$. The resulting PFFRG ground state only slightly breaks $SU(2)$ symmetry and the neutron scattering amplitudes show the same features as the mean-field results presented here. This observation matches with the stability of the $SU(2)$ symmetric mean-field states beyond the Heisenberg point which we have found in our analysis.\\
 Since most numerical methods rule out time reversal symmetry breaking states by construction, we are not aware of numerical data to compare our results for the chiral $(\pi,0,\pm\frac{\pi}{3})$ , $(0,\pi,\pm\frac{\pi}{3})$ states.

\section{Conclusions}
\label{sec:Discussion}
Using a PSG approach, we classified all fully symmetric and chiral  $\Ztwo$ Schwinger boson mean-field ansätze on the pyrochlore lattice. To the best of our knowledge, this is the first time this has been done for a 3D lattice. We went beyond the formalism of Messio \emph{et al.} \cite{messio_time_2013} by including triplet fields into our chiral PSG analysis. Furthermore, we computed the ground-state energy for 16 chiral and four fully symmetric ansätze within a mean-field approximation for the XXZ model near the Heisenberg point. Remarkably, all of the ansätze where bosonic spinons do not condense reduce to six $SU(2)$ symmetric spin liquid states, regardless of the coupling angle $\theta$. The four states  $(0,0,\pi)$, $(\pi,\pi,\pi)$, $(0,\pi,\pi)$ and $(\pi,0,\pi)$ can be described by the fully symmetric ansätze previously characterized by Liu \emph{et al.} \cite{liu_competing_2019}.  Depending on the decoupling, we identified two lowest energy states near the Heisenberg point. The fully symmetric spin liquid state $(\pi,0,\pi)$ and the chiral spin liquid $(\pi,0,\pm\frac{\pi}{3})$. The former has previously been described by Liu \emph{et al.} \cite{liu_competing_2019} while the latter is new. Its characteristic feature is a $\pm\frac{\pi}{3}$ flux that is enclosed by the $\hat{A}_{ij}$ operators on rhombus loops of length four and differs from previously studied chiral states featuring a flux of $\pi/2$ through triangular loops \cite{kim_chiral_2008, burnell_monopole_2009}. It breaks time reversal symmetry $\mathcal{T}$, mirror symmetry $\Sigma$ and a screw symmetry $S$, while it is symmetric under $\mathcal{T}\Sigma$ and $\mathcal{T}S$.

It is important to note, however, that our analysis of the XXZ Heisenberg model is based on decoupling of the Hamiltonian which is expected to work well only in the vicinity of the $SU(2)$ symmetric Heisenberg point. Since the form of the decoupling is ambiguous, different decouplings will favor non $SU(2)$ symmetric spin liquids that might result in different ground states in the vicinity of the classical Ising limit $\theta=0$, as well as in the easy-plane limit $\theta = \pi/2$, for example. Indeed, within a preliminary analysis of the fully symmetric ansätze we find a $U(1)$ symmetric ground-state close to the classical limit and a $U(1)$ symmetry breaking ansatz as ground state close to the easy plane limit, in accordance with the results of Benton \emph{et al.} \cite{benton_quantum_2018}.

Moreover, we computed spin-spin correlations as well as neutron scattering amplitudes and compared them to previously published work.
Our new chiral states  $(\pi,0,\pm\frac{\pi}{3})$, $(0,\pi,\pm\frac{\pi}{3})$ may be further explored by calculating the dynamical structure factor, where time reversal symmetry breaking can be explicitly seen. 
It might also be worthwhile to further study the ansätze  (-1, 1)-$0$-$(0,0)$-$(1)$, (-1, 1)-$0$-$(0,1)$-$(1)$, where the spinons condense and give rise to magnetic order. \\
While ansätze $(-1,\epsI)$-$n_1$-$(0,p_1)$-$(0)$ have similar flux structures as the chiral states considered by Burnell \emph{et al.} \cite{burnell_monopole_2009} and Kim \emph{et al.} \cite{kim_chiral_2008}, the saddle points for these states that we found in our mean-field analysis give rise to fully symmetric spin liquids. Here, further neighbor interactions might stabilize the chirality.
Finally, in the present study we have not solved the mean-field equation for the ansätze  
$(1,-1)$-$n_1$-$(0,0)$-$(0)$ which allow a continuous flux $2\phi_{A_1}$  that is enclosed by the $\hat{A}_{ij}$ operators on bow-tie loops. It remains to be seen if their mean-field saddle points describe non-chiral, fully symmetric states, or if $\phi_{A_1}$ acquires a non-trivial value, giving rise to a different chiral spin liquid. 

\acknowledgements
We thank Y. Iqbal for helpful discussions.
J.C.H.~acknowledges support by the Provincia Autonoma di Trento, the ERC Starting Grant StrEnQTh (project ID 804305), the Google Research Scholar Award ProGauge, and Q@TN — Quantum Science and Technology in Trento.


\appendix

\section{Local spin basis vectors}
\label{appendix:SpinBasis}
The local spin basis vectors first introduced in Eq.~\eqref{eq:GeneralSpinBasis} are \cite{yan_theory_2017}:
\begin{align}
 &\mathbf{s}^z_\mu = \frac{1}{\sqrt{3}}(1,1,1)-\frac{4}{\sqrt{3}}\mathbf{a_\mu},  &&\mathbf{s}^y_\mu= \mathbf{s}^z_\mu\times\mathbf{s}^x_\mu, \label{equation:localbasis}\\
    &\mathbf{s}^x_0 =\frac{1}{\sqrt{6}}(-2,1,1), &&\mathbf{s}^x_1 =\frac{1}{\sqrt{6}}(-2,-1,-1),\nonumber\\
    &\mathbf{s}^x_2 =\frac{1}{\sqrt{6}}(2,1,-1), &&\mathbf{s}^x_3 =\frac{1}{\sqrt{6}}(2,-1,1). \nonumber
\end{align}

\section{Transformation of sublattice coordinates and local spin basis}
\vspace*{-3mm}The space group generators transform the different coordinates as follows:\\
\begin{subequations}

\begin{align}
   T_i\mathbf{r}_\mu  &= (r_1+\delta_{i,1},r_2+\delta_{i,2},r_3+\delta_{i,3})_\mu,\\
    \rots\mathbf{r}_\mu &= -\left(r_3+\delta_{\mu,3},r_1+\delta_{\mu,1},r_2+\delta_{\mu,2}\right)_{\pi_{123}(\mu)},\\
    S \mathbf{r}_\mu &= (-r_1-\delta_{\mu,1},-r_2-\delta_{\mu,2},\nonumber\\
    & \qquad \qquad r_1+r_2+r_3+1-\delta_{\mu,0})_{\pi_{03}(\mu)},\\
    I \mathbf{r}_\mu &= (-r_1-\delta_{\mu,1},-r_2-\delta_{\mu,2},-r_3-\delta_{\mu,3})_\mu\\
    \Sigma \mathbf{r}_\mu &= (r_1,r_2,-r_1-r_2-r_3)_{\pi_{03}(\mu)},\\
     C_3 \mathbf{r}_\mu &= (r_3, r_1,r_2)_{\pi_{123}(\mu)}\\
     C_3' \mathbf{r}_\mu &= (1-r_1-r_2-r_3, r_1, r_3)_{\pi_{012}(\mu)},\\
     \mathcal{T} \mathbf{r}_\mu &= \mathbf{r}_\mu,
\end{align}
\end{subequations}
where $\pi_{123}(\mu)$ and $\pi_{03}(\mu)$ cyclically permute sites 1,2,3 and 0,3 respectively.
The local spins  transforms like:
\begin{subequations}
\begin{align}
    T_i\mathbf{S}_\mu &= (S^x,S^y,S^z)_\mu,\\
    \rots\mathbf{S}_\mu  &= (-\frac{S^x}{2}-\frac{\sqrt{3}S^y}{2},\frac{\sqrt{3}S^x}{2}-\frac{S^y}{2},S^z)_{\pi_{123}(\mu)}, \\
     S \mathbf{S}_\mu &= -(-\frac{S^x}{2}+\frac{\sqrt{3}S^y}{2},\frac{\sqrt{3}S^x}{2}+\frac{S^y}{2},S^z)_{\pi_{03}(\mu)},\hspace*{-5mm}\\
     I \mathbf{S}_\mu &= \mathbf{S}_\mu, \\
      \mathcal{T} \mathbf{S}_\mu &= (-S^x,-S^y,-S^z)_\mu,\label{equation:TranformLocalSpinVec}
\end{align}
\end{subequations}
\noindent

The symmetry group generators fulfill the following algebraic group relations
\begin{subequations}
 \label{eq:algebraicRelations}
\begin{align}
    T_iT_{i+1}T_i\inv T_{i+1}\inv   & = 1,\\
                        \rots^6     & = 1,\\
                        S^2T_3\inv  & = 1,\\
        \rots T_i \rots\inv T_{i+1} & = 1,\\
        S T_i S\inv T_3\inv T_i     & = 1, \qquad i\in\{1,2\},\\
        S T_3 S\inv T_3\inv         & = 1,\\
                    (\rots S)^4     & = 1,\\ 
                    (\rots^3 S)^2   & = 1,\\
                    \mathcal{T}^2   & = -1,\\
\mathcal{T}\mathcal{O}\mathcal{T}\inv \mathcal{O}\inv &= 1,
\end{align}
\end{subequations}
where $i=1,2,3$ and $i+3 = i$. $\mathcal{O}$ is a placeholder for an arbitrary space group generator: $\mathcal{O}\in\{T_1,T_2,T_3,\rots,S\}$.

\section{$SU(2)$ matrices}
\label{appendix:SU2Matrices}
\vspace*{-3mm} The $SU(2)$ matrices $U_\mathcal{O}$ associated with the symmetry operations that appear in Eq.~\eqref{eq:TransformationMeanFieldAnsatz} are:
\begin{align}
         U_{T_i} &= \sigma_0, \\
    U_{\rots} &= U_{C_3} = e^{-\frac{i}{2}\frac{2 \pi}{3}(0,0,1)\vec{\sigma}}, \\ 
    U_{S,\mu} &= (-1)^{1-\delta_{\mu,1}}e^{-\frac{i}{2}\frac{2 \pi}{2}(\frac{-\sqrt{3}}{2},\frac{1}{2},0)\vec{\sigma}}.
\end{align}
The matrix for the screw operation depends on which sublattice it acts on. Spins on sublattice $1$ are rotated the other way around than spins on sublattice $2$. Spins on sublattices $0$ and $3$ are rotated and then projected onto the local spin basis of the other sublattice. This results in an effective $\pi$ rotation about the $(\frac{-\sqrt{3}}{2},\frac{1}{2},0)$ axis. The sign of the effective rotation can be chosen freely and different signs correspond to different gauges. Here we chose the signs of rotation to be equal on sublattice $0,2,3$.\\
The $SU(2)$ matrices  $U_\mu$ that transform from global to local basis on sublattice $\mu$ (See Eq.~\eqref{eq:FromGlobalToLocal}) are given by 
\begin{subequations}
\begin{align}
    &U_0 = 
    \begin{pmatrix} 
    \frac{1}{\sqrt{3-\sqrt{3}}}e^{i\frac{2\pi 11}{48}} & \frac{1}{\sqrt{3+\sqrt{3}}}e^{i\frac{2\pi 5}{48}}\\
    \frac{1}{\sqrt{3+\sqrt{3}}}e^{i\frac{2\pi 19}{48}} & \frac{1}{\sqrt{3-\sqrt{3}}}e^{-i\frac{2\pi 11}{48}}
    \end{pmatrix},  
     \\&  U_1 = 
    \begin{pmatrix} 
    \frac{1}{\sqrt{3+\sqrt{3}}}e^{-i\frac{2\pi 7}{48}} & \frac{1}{\sqrt{3-\sqrt{3}}}e^{-i\frac{2\pi  }{48}}\\
     \frac{1}{\sqrt{3-\sqrt{3}}}e^{-i\frac{2\pi 23}{48}} & \frac{1}{\sqrt{3+\sqrt{3}}}e^{i\frac{2\pi 7}{48}}
    \end{pmatrix},   \\
      &U_2 = 
    \begin{pmatrix} 
    \frac{1}{\sqrt{3+\sqrt{3}}}e^{i\frac{2\pi 5}{48}} & \frac{1}{\sqrt{3-\sqrt{3}}}e^{-i\frac{2\pi 13}{48}}\\
    \frac{1}{\sqrt{3-\sqrt{3}}}e^{-i\frac{2\pi 11}{48}} & \frac{1}{\sqrt{3+\sqrt{3}}}e^{-i\frac{2\pi 5}{48}}
    \end{pmatrix},     
    \\&  U_3 = 
    \begin{pmatrix} 
    \frac{1}{\sqrt{3-\sqrt{3}}}e^{-i\frac{2\pi }{48}} & \frac{1}{\sqrt{3+\sqrt{3}}}e^{i\frac{2\pi 17}{48}}\\
    \frac{1}{\sqrt{3+\sqrt{3}}}e^{i\frac{2\pi 7}{48}} & \frac{1}{\sqrt{3-\sqrt{3}}}e^{i\frac{2\pi }{48}}
    \end{pmatrix}.
\end{align}
\end{subequations}

 \section{Solution of the chiral algebraic PSG}
\label{Appendix:SolChiralPSG}
\vspace*{-3mm}
The symmetry enriched algebraic relations of $\chi_e$ are 
\begin{subequations}
\label{eq:GaugeEnrichedAlgebraicRelations}
\begin{align}
   (G_{T_i}T_i)(G_{T_{i+1}}T_{i+1}) (G_{T_i}T_i)\inv (G_{T_{i+1}}T_{i+1})\inv \in  \Ztwo&,\\
   (G_{C_3}C_3)^3 \in  \Ztwo&, \label{equation:SpaceGroupPSGC3}\\
    (G_{C_3'}C_3')^3 \in  \Ztwo&,\label{equation:SpaceGroupPSGC3p}\\
    (G_{C_3}C_3)(G_{C_3'}C_3')(G_{C_3}C_3)(G_{C_3'}C_3') \in  \Ztwo&\\
   (G_{C_3}C_3)(G_{T_i}T_i) (G_{C_3}C_3)\inv (G_{T_{i+1}}T_{i+1})\inv \in  \Ztwo&,\label{equation:SpaceGroupPSGC3Ti}\\
    (G_{C_3'}C_3')(G_{T_1}T_1) (G_{C_3'}C_3')\inv (G_{T_{2}}T_{2})\inv (G_{T_{1}}T_{1}) \in  \Ztwo&, \label{equation:SpaceGroupPSGC3pT1}\\
    (G_{C_3'}C_3')(G_{T_2}T_2) (G_{C_3'}C_3')\inv (G_{T_{1}}T_{1}) \in  \Ztwo&,   \\
    (G_{C_3'}C_3')(G_{T_3}T_3) (G_{C_3'}C_3')\inv (G_{T_{3}}T_{3})\inv (G_{T_{1}}T_{1}) \in  \Ztwo&. 
    \label{equation:SpaceGroupPSGC3pT3}
\end{align}
\end{subequations}

\begin{widetext}

These can be rewritten into the following phase equations
\begin{subequations}
\begin{align}
    \phase{T_i}{}+\phase{T_{i+1}}{T_i\inv}-\phase{T_{i}}{T_{i+1}\inv}- \phase{T_{i+1}}{} &=  \pi n_i,  \label{equation:PhaseEquationTranslations}\\
    \phase{C_3}{} + \phase{C_3'}{(C_3)\inv} +\phase{C_3 }{(C_3C_3')^{-1}} 
    +\phase{C_3' }{C_3'} &= \pi n_{C_3,C_3'}, \label{equation:PhaseEquationC3C3prime}\\
    \phase{C_3}{} + \phase{C_3}{C_3\inv} +\phase{C_3}{C_3^{-2}} &= \pi n_{C_3}, \label{equation:PhaseEquationC3}\\
        \phase{C_3'}{} + \phase{C_3'}{(C_3')\inv} +\phase{C_3'}{(C_3')^{-2}} &= \pi n_{C_3'},  \label{equation:PhaseEquationC3prime}\\
    \phase{C_3}{} + \phase{T_i}{C_3\inv}- \phase{C_3}{T^{-1}_{i+1}} -\phase{T_{i+1}}{} &= \pi n_{C_3 T_i}, \label{equation:PhaseEquationC3Ti}\\
        \phase{C_3'}{} + \phase{T_{1}}{(C_3')\inv} -\phase{C_3'}{T_1\inv T_{2}}
        -\phase{T_{2}}{T_1\inv}+\phase{T_{1}}{T_1} &= \pi n_{C_3'T_1 },\label{equation:PhaseEquationC3primeT1}\\
        \phase{C_3'}{} + \phase{T_{2}}{(C_3')\inv} -\phase{C_3'}{T_1\inv}
        +\phase{T_{1}}{T_1} &= \pi n_{C_3'T_2 }, \\
       \phase{C_3'}{} + \phase{T_{3}}{(C_3')\inv} -\phase{C_3'}{T_1\inv T_{3} }
        -\phase{T_{3}}{T_1\inv}+\phase{T_{1}}{T_1} &= \pi n_{C_3'T_3 }.\label{equation:PhaseEquationC3primeT3}
\end{align}
\end{subequations}

\end{widetext}
where $n_{X} \in \{0,1\}$.  \\
\\
Our goal is to find all phases $\phase{\mathcal{O}}{}$ as functions of $\mathbf{r}_\mu$ and $n_{X}$. 
However, a general gauge transformation $G$ (See Eq.~\eqref{eq:GaugeTransformation}) changes the phases given by the PSG phase equations like\cite{liu_competing_2019}:
\begin{equation}
    \phase{\mathcal{O}}{} \xrightarrow{} \phase{G}{} + \phase{\mathcal{O}}{}-\phase{G}{\mathcal{O}\inv}
\end{equation} 
To ultimately get an unambiguous result we have to fix the gauge in the process of solving the phase equations. Since we have four FCC-sublattices we have freedom to choose 16 independent local gauges. Four for every direction $r_1,r_2,r_3$ and a constant one for every sublattice $\mu=0,1,2,3$:
\begin{subequations}
\begin{align}
    &G_1:\phase{G_1}{} = n_{G1,\mu}\pi r_1, \label{gaugeT1}\\
    &G_2:\phase{G_1}{} = n_{G2,\mu}\pi r_2, \label{gaugeT2}\\
    &G_3:\phase{G_1}{} = n_{G3,\mu}\pi r_3, \label{gaugeT3}\\
    &G_4:\phase{G_1}{} = \phi_\mu \label{gaugeconst}.
\end{align}
\end{subequations}

Due to IGG $=\Ztwo$, we are also free to add a site independent $\Ztwo$ phase to any of our 5 phases  $\phase{\mathcal{O}}{}$ \cite{wang_spin_2006}. That makes 16 local gauge  and 5 IGG choices in total. 
With the first 12 gauge choices (equation \eqref{gaugeT1}-\eqref{gaugeT3}) we can fix the phases associated with the translation operators to $\phase[(r_1,r_2,r_3)_\mu]{T_1}{}=  \phase[(0,r_2,r_3)_\mu]{T_2}{}= \phase[(0,0,r_3)_\mu]{T_3}{}= 0$.  Note, that this can only be satisfied for open boundary conditions \cite[Appendix A]{wang_spin_2006}.\\
\\
Using this choice, equation \eqref{equation:PhaseEquationTranslations} is solved by: 
\begin{align}
        \phase{T_1}{} &= 0,\\
        \phase{T_2}{} &=  n_1\pi r_1,\\
    \phase{T_3}{} &= n_3\pi r_1 + n_2\pi r_2.
\end{align}
\\
Writing out equation \eqref{equation:PhaseEquationC3Ti} we get:
\begin{subequations}
\begin{align}
    &\phase{C_3}{} - \phase[(r_1,r_2+1,r_3)_\mu]{C_3}{}\\
    =& (n_{C_3 T_1} +n_1r_1)\pi, \nonumber \\[10pt]
     & \phase{C_3}{} - \phase[(r_1,r_2,r_3+1)_\mu]{C_3}{} \nonumber\\
    =& (n_{C_3 T_2} + n_3r_1 + n_2r_2 + n_1 r_2)\pi, \\[10pt]
     & \phase{C_3}{} - \phase[(r_1+1,r_2,r_3)_\mu]{C_3}{}\nonumber\\
    =& (n_{C_3 T_3}  + n_3  r_2 + n_2  r_3 )\pi.
\end{align}
\end{subequations}
This is solved by 
\begin{subequations}
\begin{align}
    \phase{C_3}{} &= f_1(r_1,r_3) -r_2(n_{C_3 T_1}\pi + n_1\pi r_1),\label{equation:RotnTrans1}\\
    \phase{C_3}{} &= f_2(r_1,r_2) \\\nonumber &-r_3(n_{C_3 T_2}\pi
    + n_3\pi r_1 -n_2\pi r_2 - n_1\pi r_2), \label{equation:RotnTrans3}\\
    \phase{C_3}{} &= f_3(r_2,r_3) \\& -r_1(n_{C_3 T_3}\pi + n_3\pi r_2 -n_2\pi r_3),\nonumber 
\end{align}
\end{subequations}
where $f_{C_3}(r_1,r_3)$ is some function of $r_1$ and $r_3$. 
Since the function $f_{1}(r_1,r_3)$ in \eqref{equation:RotnTrans1} can not include any terms that feature $r_2$ it can not include terms like $r_1 r_2 n_3\pi $ that have to appear in $\phase{C_3}{}$ due to \eqref{equation:RotnTrans3}. To fulfill equations \eqref{equation:RotnTrans1} - \eqref{equation:RotnTrans3} we have to infer a relationship between $n_1,n_2,n_3$. With $n_1=n_2=n_3$ we have the following solution:

\begin{align}
    \phase{C_3}{} &=  \phase[\mathbf{0}_\mu]{C_3}{}    
    -  n_1\pi(r_1r_2+r_1r_3)\nonumber    \\        
    &-  (r_1 n_{C_3 T_3} + r_2 n_{C_3 T_1} + r_3 n_{C_3 T_2})\pi \label{equation:PhaseC31}
\end{align}

\noindent
Plugging equation \eqref{equation:PhaseC31} into \eqref{equation:PhaseEquationC3} gives:
\begin{align}
 &\phase{C_3}{} + \phase[\left(r_2,r_3,r_1\right)_{\pi_{321}(\mu)}]{T_1}{} + \phase[\left(r_3,r_1,r_2 \right)_{\pi_{123}(\mu)}]{T_1}{} \nonumber\\
   = & \phase[\mathbf{0}_\mu]{C_3}{}+\phase[\mathbf{0}_{\pi_{123}(\mu)}]{C_3}{}+\phase[\mathbf{0}_{\pi_{321}(\mu)}]{C_3}{} + \sum_{i,j}r_in_{C_3T_j}\pi \nonumber\\
   =& n_{C_3}\pi \label{equation:C30constraint}
\end{align}
which constrains $ \sum_{j}n_{C_3T_j}=0$. $\pi_{123}(\mu)$ permutes $\mu$ in the cycle (123).
\\
Writing out Eqs.~\eqref{equation:PhaseEquationC3primeT1}-\eqref{equation:PhaseEquationC3primeT3} we get:
\begin{subequations}
\begin{align}
    &\phase{C_3'}{} - \phase[(r_1-1,r_2+1,r_3)_\mu]{C_3'}{}\label{eq:c3'T1} \\
   =\;& n_{C_3'T_1}\pi + n_1\pi(r_1+1),\nonumber\\[10pt]
    &\phase{C_3'}{} - \phase[(r_1-1,r_2,r_3)_\mu]{C_3'}{}  \label{eq:c3'T2}\\
   =\;&n_{C_3'T_2}\pi+ n_1\pi r_2, \nonumber\\[10pt]
    & \phase{C_3'}{} - \phase[(r_1-1,r_2,r_3+1)_\mu]{C_3'}{}\label{eq:c3'T3} \\
   =\;&n_{C_3'T_3}\pi + n_1\pi(r_2+r_3).\nonumber 
\end{align}
\end{subequations}
 From Eq.~\eqref{eq:c3'T2} we can infer that
 \begin{equation}
     \phase{C_3'}{} = f_{C_3'}(r_2,r_3) + n_1\pi r_1r_2 + n_{C_3'T_2}\pi r_1, 
 \end{equation}
where $f_{C_3'}(r_2,r_3)$ is some function of $r_2$ and $r_3$. Using this and Eq.~\eqref{eq:c3'T3} we get 
 \begin{align}
     \phase{C_3'}{} =\; & f_{C_3'}(r_2) + n_1\pi r_1r_2 + n_{C_3'T_2}\pi r_1 \\
     & + r_3 \pi(\frac{r_3-1}{2}n_1+n_{C_3'T_2}+n_{C_3'T_3}).\nonumber
 \end{align}
Plugging this into Eq.~\eqref{eq:c3'T1} finally gives
 \begin{align}
    \phase{C_3'}{} &=  \phase[\mathbf{0}_\mu]{C_3'}{}  + r_1\pi n_{C_3'T_2}  + n_1\pi r_1r_2 \label{eq:phaseC3one}\\
    &+  r_3 \pi(\frac{r_3-1}{2}n_1+n_{C_3'T_2}+n_{C_3'T_3}) \nonumber\\
    &  +  r_2 \pi(\frac{r_2-1}{2}n_1+n_{C_3'T_2}+n_{C_3'T_1}).\nonumber
 \end{align}

\noindent
Inserting Eq.~\eqref{eq:phaseC3one} into Eq.~\eqref{equation:PhaseEquationC3prime} gives
\begin{align}
   &\phase[\mathbf{0}_\mu]{C_3'}{}+\phase[\mathbf{0}_{\pi_{021}(\mu)}]{C_3'}{}+\phase[\mathbf{0}_{\pi_{120}(\mu)}]{C_3'}{} \nonumber\\ & + r_3\pi(n_{C_3'T_1}+n_{C_3'T_2}+n_{C_3'T_3}) \nonumber\\ \nonumber
    =\;& (n_{C_3'} + n_{C_3'T_1}+n_1)\pi, \label{eq:nC3p'}
\end{align}
which gives two constraints 
\begin{align}
    & n_{C_3'T_1}+n_{C_3'T_2}+n_{C_3'T_3} =0,\\[10pt]
    & \phase[\mathbf{0}_\mu]{C_3'}{}+\phase[\mathbf{0}_{\pi_{021}(\mu)}]{C_3'}{}+ \phase[\mathbf{0}_{\pi_{120}(\mu)}]{C_3'}{}  \label{eq:Co3primeconstr}\\
    =\;& (n_{C_3'T_1}+n_1+n_{C_3'})\pi.\nonumber
\end{align}
\noindent
The last phase equation \eqref{equation:PhaseEquationC3C3prime} is then
\begin{align}
   & \phase[\mathbf{0}_\mu]{C_3 }{}+\phase[\mathbf{0}_{\pi_{132}(\mu)}]{C_3' }{}+\phase[\mathbf{0}_{\pi_{(20)(13)}(\mu)}]{C_3}{}+\phase[\mathbf{0}_{\pi_{120}(\mu)}]{C_3' }{}  \nonumber\\
    =\;& (n_{C_3 C_3'}+ n_{C_3'T_2}+ n_{C_3T_1})\pi. \label{eq:phaseeqC3C3prime}
\end{align}
$\pi_{(20)(13)}(\mu)$ permutes $\mu$ in the cycles (20) and (13).
Since equations \eqref{equation:SpaceGroupPSGC3}, \eqref{equation:SpaceGroupPSGC3p}, \eqref{equation:SpaceGroupPSGC3Ti} and \eqref{equation:SpaceGroupPSGC3pT1} have operators that appear an odd number of times, we can use our 5 IGG choices of  $T_1,T_2,T_3,C_3,C_3'$ to set $n_{C_3'T_1}=n_{C_3T_2}=n_{C_3T_3}=n_{C_3} = 0$ and $n_{C_3'}=n_1$. Using $ \sum_{j}n_{C_3T_j}=0$ this also implies $n_{C_3T_1}=0$. \\
As a last step, we find $\phase[\mathbf{0}_\mu]{C_3}{}$ and $\phase[\mathbf{0}_\mu]{C_3'}{}$.
We have the four constant sublattice gauge choices left (Eq.~\eqref{gaugeconst}). By fixing the IGG choices Eqs.~\eqref{equation:C30constraint} and \eqref{eq:Co3primeconstr} are reduced to
\begin{subequations}
\begin{align}
  3\phase[\mathbf{0}_{0}]{C_3}{}  &= 0, \\ 
   \phase[\mathbf{0}_{1}]{C_3}{} + \phase[\mathbf{0}_{2}]{C_3}{}+ \phase[\mathbf{0}_{3}]{C_3}{} & = 0, \\
     3\phase[\mathbf{0}_{3}]{C_3'}{}  &= 0, \\ 
   \phase[\mathbf{0}_{1}]{C_3'}{} + \phase[\mathbf{0}_{2}]{C_3'}{}+ \phase[\mathbf{0}_{0}]{C_3'}{} & = 0. 
\end{align}
\label{equation:C30123constraint}
\end{subequations}\\
The form of equations \eqref{equation:C30123constraint} is invariant under gauge transformations. That is why we can fix the constant gauge on sublattices 0,1,2,3 to set $\phase[\mathbf{0}_{2}]{C_3}{}$ = $ \phase[\mathbf{0}_{3}]{C_3}{} = 0$ as well as $\phase[\mathbf{0}_{1}]{C_3'}{} = 0$. Eqs.~\eqref{equation:C30123constraint} and \eqref{eq:Co3primeconstr} then also imply $\phase[\mathbf{0}_{1}]{C_3}{} =0$ and $ \phase[\mathbf{0}_{2}]{C_3'}{}= - \phase[\mathbf{0}_{0}]{C_3'}{}$.
Eq.~\eqref{eq:phaseeqC3C3prime} then reduces to 
\begin{align}
    \phase[\mathbf{0}_{0}]{C_3}{}-\phase[\mathbf{0}_{2}]{C_3'}{}&= (n_{C_3'T_2} + n_{C_3 C_3'})\pi, \\
    \phase[\mathbf{0}_{2}]{C_3' }{} + \phase[\mathbf{0}_{3}]{C_3'}{} &= (n_{C_3'T_2} + n_{C_3 C_3'})\pi.
\end{align}
Therefore, $\phase[\mathbf{0}_{0}]{C_3}{} =-\phase[\mathbf{0}_{3}]{C_3'}{} = \frac{2\pi \xi}{3}$
where $\xi\in\{-1,0,1\}$ and $\phase[\mathbf{0}_{2}]{C_3' }{}=\phase[\mathbf{0}_{0}]{C_3}{}+ (n_{C_3'T_2} + n_{C_3 C_3'})\pi$.
\\
The final solution is then:
\begin{subequations}
 \begin{align}
        \phase{T_1}{} &= 0,\\
        \phase{T_2}{} &= n_1\pi r_1,\\
        \phase{T_3}{} &= n_1\pi(r_1 +r_2),\\
    \phase{C_3}{} &= \frac{2\pi \xi}{3} \delta_{\mu, 0} +  n_1\pi(r_1r_2+r_1r_3),\\
    \phase{C_3'}{} &= -\frac{2\pi \xi}{3} \delta_{\mu, 3} +  r_2 \pi(\frac{r_2-1}{2}n_1+n_{C_3'T_2})  \\
                    &  + r_1\pi n_{C_3'T_2} +  r_3\pi\frac{r_3-1}{2}n_1 + n_1\pi r_1r_2 \nonumber\\
    & +(\frac{2\pi \xi}{3}+n_{C_3C_3'}+n_{C_3'T_2}) (-\delta_{\mu, 0}+\delta_{\mu 2})\pi,\nonumber
\end{align}
\end{subequations}
where $\xi\in\{-1,0,1\}$ and $n_1,n_{C_3C_3'},n_{C_3'T_2} \in \{0,1\}$.


\section{Classification of chiral ansätze}
\label{Appendix:FluxTransform}
We use the short notation $\mathcal{B}_{\mathbf{0}_\mu\mathbf{0}_\mu} = \mathcal{B}_{\mu\nu}$ for bonds on the main tetrahedron and  $\mathcal{B}_{I(\mathbf{0}_\mu)I(\mathbf{0}_\mu)} = \mathcal{B}_{I\mu\nu}$ for bonds on the inverse tetrahedron. \\
\noindent
 As described by Messio  \emph{et al.} \cite{messio_time_2013} we can classify all possible ansätze by looking at the transformation of the minimal set of linearly independent fluxes under elements in  $\chi_o$. All elements of $\chi_o$ can be written as compositions of $I$, $\Sigma$ and elements of $\chi_e$. The elements of $\chi_e$ leave the fluxes invariant, so we only have to consider the action of $I$, $\Sigma$ on the fluxes. It is equally valid to consider the actions of $\rots$ and $S$ on the fluxes, but the calculations are a bit more involved. Since $\rots =I C_3$ and $\Sigma= S \rots^3$ their parities are related like $\epsI = \epsilon_\rots$ and $\epsilon_\Sigma = \epsilon_\rots \epsS$. \\
 Fluxes are independent if they can not be mapped onto each other by symmetry operations in $\chi_e$ and can not be created by adding other independent fluxes. The number of independent fluxes depends on the number of present mean-field parameters as well.
\\
To find out how many independent fluxes there are we start with how many independent loops of even and odd length there are in the pyrochlore lattice independent of possible bond variables.

\begin{figure}
\subfloat[]{%
\includegraphics[width=.49\linewidth]{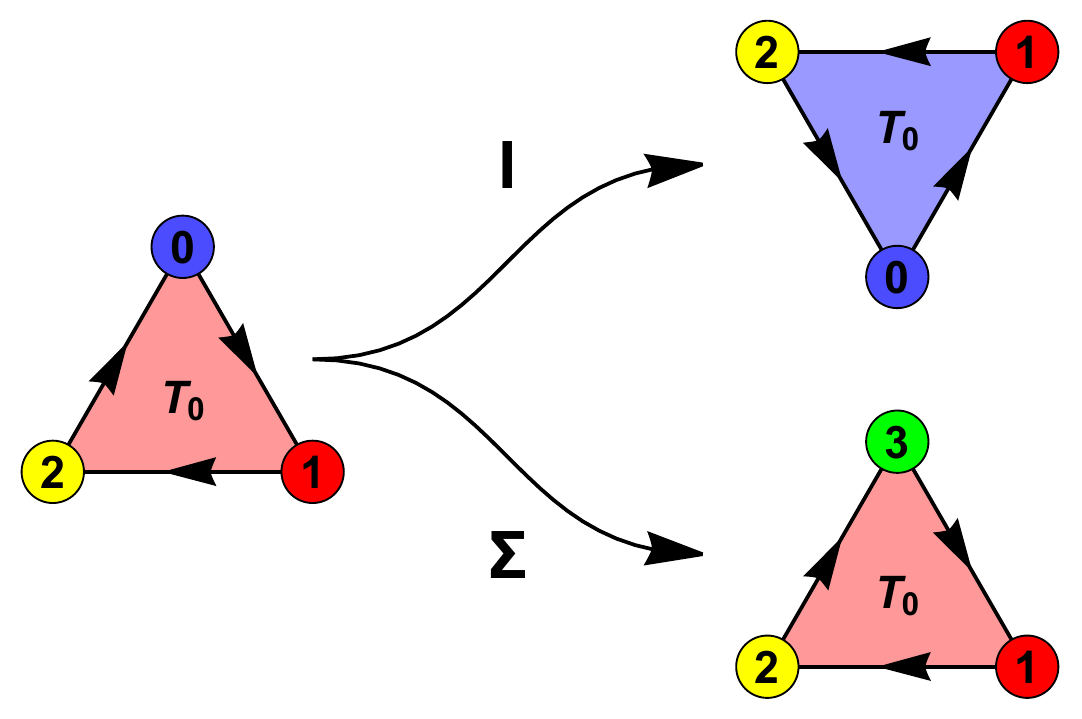}%
}\hfill
 \subfloat[]{%
\includegraphics[width=.49\linewidth]{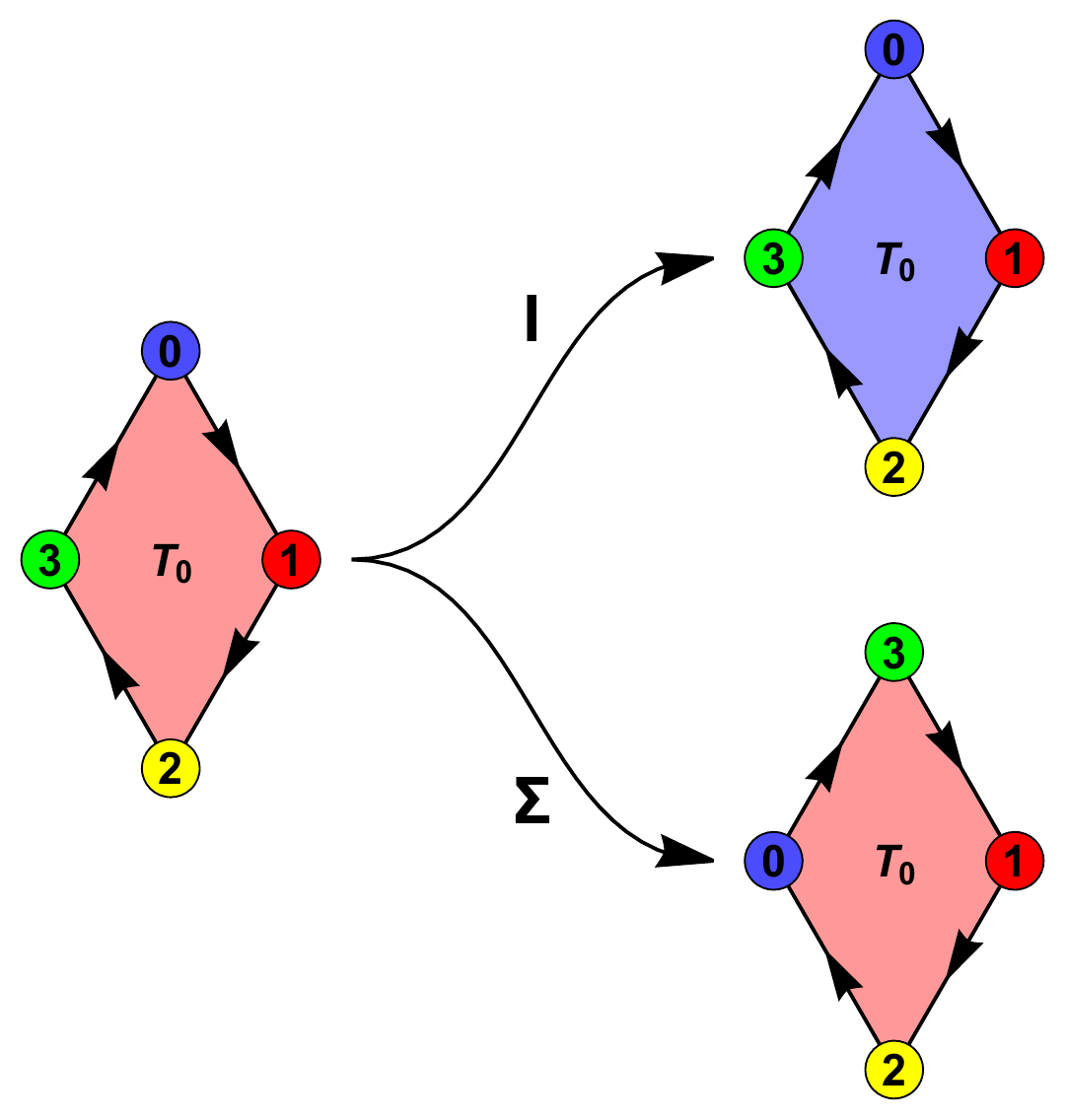}%
}\hfill
\subfloat[]{%
\includegraphics[width=.49\linewidth]{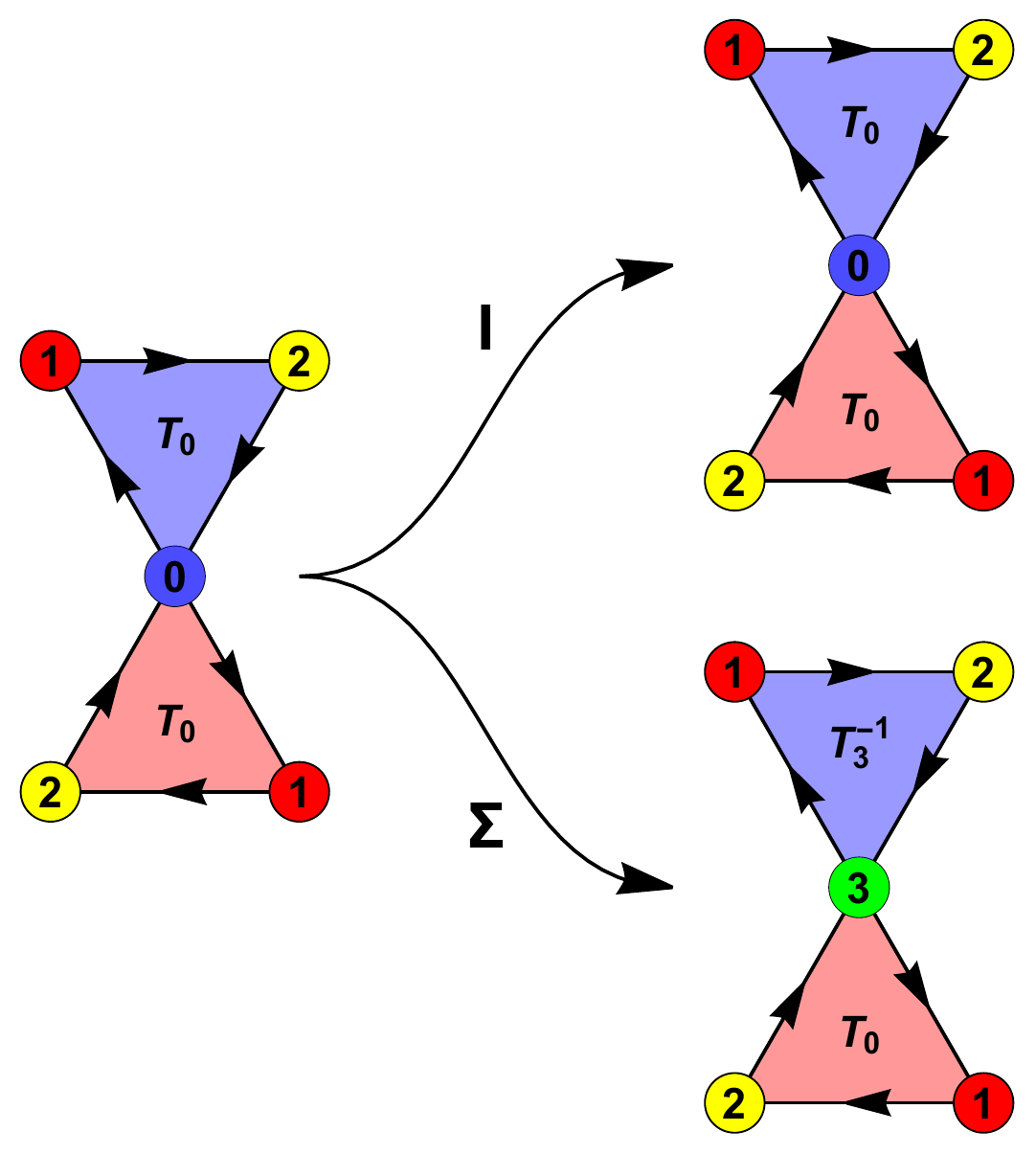}%
}
 
\caption{(Color online). Transformations of the triangle, rhombus and bow tie loops under $I$ and $\Sigma$ symmetry. $T_0$ labels the tetrahedra of the (0,0,0) unit cell, $T_3^{-1}$ labels the tetrahedra of the (0,0,-1) unit cell. }
    \label{fig:Loops}
\end{figure}

\begin{enumerate}
    \item Trivial Loop (Loop size = 2): There are two independent bonds. One on the main and one on the inverse tetrahedron. The trivial loop is going back and forth along a bond. 
    
    \item Triangle  (Loop size = 3): There are 8 triangles in the pyrochlore unit cell. There are two sets of three triangles that can be mapped onto each other by $C_3$ which leaves us with 2 triangle loops on the inverse and main tetrahedron. These can be mapped onto each other by $C_3'$ rotation and translation. In total, we therefore have two independent triangle loops.
    
    \item Rhombus (Loop size = 4): There are 6 rhombi in the pyrochlore unit cell. Three on each tetrahedron. All rhombi on a tetrahedron can be mapped onto each other by $C_3$ which leaves us with two independent rhombi.
    
    \item Bow tie  (Loop size = 6): There are 12 bow ties and 24 "bent" bow ties in the unit cells (0,0,0) and the three main tetrahedra of the cells (0,0,-1), (0,-1,0), (-1,0,0): 9 per two adjacent tetrahedra. By $C_3$ mapping we can reduce the number to 12. 3 in the tetrahedra of (0,0,0) and 9 in the tetrahedra of e.g. (0,0,0) and (-1,0,0). \\
    We can further reduce the number by realizing that if we add a rhombus to a bow tie we get a bent bow tie. This reduces the number of loops to $\frac{12}{6}=2$. These can finally be mapped onto each other by $C_3'$ rotation which leaves one independent bow tie flux.
    
    \item Hexagon (Loop size = 6): Four unit cells always enclose a hexagon. There are 4 different hexagons that can not be mapped onto each other by translations. By $C_3$ symmetry we can reduce this to 2 and by $C_3'$ symmetry to one independent hexagon. 
 
    \item Bigger loops (Loop size $>$ 6): All loops with size larger than 6 can be created by adding loops of smaller size and therefore do not add to the linearly independent loops.
\end{enumerate}
\noindent
The trivial, triangle and rhombus loops on the main tetrahedron can be mapped to the same loops on the inverse tetrahedron by $I$. From the algebraic relations we can see that $\Sigma I = S^2I\Sigma$. So $\Sigma$ and $I$ commute modulo $S^{-2}\in \chi_e$ which leaves fluxes invariant. Therefore, transformation of trivial, triangle and rhombus fluxes on the main and inverse tetrahedra give the same constraints. We therefore only consider them on the inverse tetrahedron. \\
Transformation of hexagonal loops give the same constraints as the bow ties. We  therefore not consider the hexagonal fluxes here explicitly. 
Figure~\ref{fig:Loops} show how all independent loops transform under $\Sigma$ and $I$. When we now specify fluxes by adding bond operators we can transform these loop diagrams into equations. Fluxes can in principle consist of one, two or many different types of bond operators. E.g. $\Arg(\mathcal{B}_{12}\mathcal{B}_{23}\mathcal{B}_{31})$, $\Arg(\mathcal{A}_{12}\mathcal{B}^{h,z}_{23}\mathcal{A}^*_{31})$ , $\Arg(\mathcal{B}_{12}t^{h,x}_{23}t^{h,z}_{31})$.
We only have to consider fluxes with one or two fields. Fluxes with three or more fields can be constructed from these. \\ 
Before we can start turning the diagrams into equations we have to define our bond fields.
Since there are two independent bonds we have two independent phases for each field.\\
We fix these as $\mathcal{B}_{01}= \mathcal{B} e^{-i \phi_{B_1}}$, $ \mathcal{A}_{01}= \mathcal{A} e^{-i \phi_{A_1}}$ (on the main tetrahedron) and
 $\mathcal{B}_{I01}= \mathcal{B} e^{-i \phi_{B_2}}$, $ \mathcal{A}_{I01}= \mathcal{A} e^{-i \phi_{A_2}}$  (on the inverse tetrahedron) and equivalently for $t^{h,z}_{ij}$ and $t^{p,z}_{ij}$.\\
Transformation of x and y triplet operators is not as trivial because symmetry operations change the direction.
For example: 
\begin{subequations}
\begin{align}
    C_3(C_3(\hat{t}^{x}_{01}))&=C_3(-\frac{1}{2}\hat{t}^{x}_{02}+\frac{\sqrt{3}}{2}\hat{t}^{y}_{02})= -\frac{1}{2}\hat{t}^{x}_{0,3}-\frac{\sqrt{3}}{2}\hat{t}^{y}_{03}, \\
    C_3(C_3(\hat{t}^{y}_{01}))&=C_3(-\frac{\sqrt{3}}{2}\hat{t}^{x}_{02}-\frac{1}{2}\hat{t}^{y}_{02})=  \frac{\sqrt{3}}{2}\hat{t}^{x}_{03} -\frac{1}{2}\hat{t}^{y}_{03}. 
\end{align} 
\end{subequations}
We define new operators with easier transformation properties:
\begin{subequations}
\begin{align}
   \hat{t}^{h,x'}_{03} &:= C_3(\hat{t}^{h,x'}_{0,2}):=  C_3(C_3(\hat{t}^{h,x'}_{0,1})) := C_3(C_3(\hat{t}^{h,x}_{0,1})),\\
   t^{h,x'}_{1,2} &:= C_3(\hat{t}^{h,x'}_{3,1}):=  C_3(C_3(\hat{t}^{h,x'}_{2,3})) := C_3(C_3(\hat{t}^{h,x}_{2,3})),
\end{align}
\end{subequations}
\\
\noindent
and equivalently for the bonds on the inverse tetrahedron, the pairing triplet operators and the y-triplet fields.  The operators $\hat{t}^{h,x'}_{ij}$ and $\hat{t}^{h,y'}_{ij}$ are linearly independent on every bond. We fix their expectation values as $t^{p,x'}_{01}= t^{p,x} e^{-i \phi_{t^{p,x}_1}}$,  $t^{p,x'}_{I01}= t^{p,x} e^{-i \phi_{t^{p,x}_2}}$, $t^{p,y'}_{01}= t^{p,y} e^{-i \phi_{t^{p,y}_1}}$, $t^{p,y'}_{I01}= t^{p,y} e^{-i \phi_{t^{p,y}_2}}$ 
and equivalently for the hopping triplet fields. \\
The operators transform as 
\begin{align}
    \hat{B}_{ji}  = \hat{B}_{ij}^\dagger, &&
    \hat{t}^{h,\gamma}_{ji}  = - (\hat{t}^{h,\gamma}_{ij})^\dagger,  &&
   \hat{A}_{ji}  = -\hat{A}_{ij}, && 
    \hat{t}^{p,\gamma}_{ji}  =   \hat{t}^{p,\gamma }_{ij}. \label{eq:TransformationProperties}
\end{align}
To keep calculations short we use the superscript $\gamma \in \{x',y',z\}$
to label the triplet operators. Also note that the triplet x' and z operators pick up an extra $\pi$ phase when acted upon by $\Sigma$ compared to the singlet and triplet y operators. This is not due to a gauge transformation added to the screw operation but solely due to the spin rotation  (see Table \ref{tab:ScrewTransformation}).

\begin{table}[]
\centering
\caption{The signs that fields pick up when being transformed under $\Sigma$. Note, that singlet (s) and y' - triplet fields pick up the same signs.}
\label{tab:ScrewTransformation}
\begin{tabular}{ccccc}
Bonds &  \;s &  \;x' &  \;y' &  \;z \\ \hline
01 $\xrightarrow{}$ 31              & \;$-$       & \;+ & \;$-$  & \;+ \\
02 $\xrightarrow{}$ 32              & \;+       & \;$-$  & \;+ & \;$-$  \\
03 $\xrightarrow{}$ 30              & \;+       & \;$-$  & \;+ & \;$-$  \\
12 $\xrightarrow{}$ 12              & \;$-$        & \;+ & \;$-$  & \;+ \\
23 $\xrightarrow{}$ 20             & \;+       & \;$-$  & \;+ & \;$-$  \\
31 $\xrightarrow{}$ 01             & \;$-$        & \;+ & \;$-$  & \;+ \\ 
\end{tabular}
\end{table}
\noindent
In the following subsections we consider all one and two operator fluxes and translate their transformation behavior into constraints for the phases $\phi_{\mathcal{O}_i}$. The solutions to the equations can be found in Tab.~\ref{tab:WeaklySymmetricAnsaetze}.

\subsection{Fluxes with $\mathcal{B}_{ij}$ fields}
\noindent
Hopping operators can be written in gauge invariant loops of odd size. For $\mathcal{B}_{ij}$ fields, the only independent loops we have to consider are the triangles (Fig.~\ref{fig:Loops}(a)).
\\
\begin{align}
\Arg(\mathcal{B}_{I01} \mathcal{B}_{I12}   \mathcal{B}_{I20})&.\label{equation:BFluxGeneral}
\end{align}
If we use the transformation property $\mathcal{B}_{ji} = \mathcal{B}_{ij}^*$ (see Eq.~\eqref{eq:TransformationProperties}) we can turn  Fig.~\ref{fig:Loops}(a) into equations for the phases:
\begin{subequations}
\begin{align}
\phi_{B_2}+\frac{4\xi\pi}{3}+ n_1\pi&=\epsI(\phi_{B_1}+\frac{4\xi\pi}{3}),\\
\phi_{B_2}+\frac{4\xi\pi}{3} &=3\epsSig\phi_{B_2}.
\end{align}
\label{eq:Bconstraints}
\end{subequations}


\subsection{Fluxes with $\mathcal{A}_{ij}$ fields}\noindent
Pairing operators can only be written in gauge invariant loops of even length, where   $\mathcal{A}$ and $\mathcal{A}^*$ alternate. Therefore, we have to consider the rhombus and bow tie loops of Fig.~\ref{fig:Loops}(b,c). 
\begin{subequations}
\begin{align}
     \Arg(\mathcal{A}_{I01} \mathcal{A}^*_{I12} \mathcal{A}_{I23} \mathcal{A}^*_{I30})&,\\
      \Arg(\mathcal{A}_{I01} \mathcal{A}^*_{I12} \mathcal{A}_{I20} \mathcal{A}^*_{01} \mathcal{A}_{12} \mathcal{A}^*_{20})&.
\end{align}
\end{subequations}
The rhombus fluxes give constraints for the PSG parameter $\xi$:
\begin{subequations}
\label{eq:kconstraints}
\begin{align}
    \frac{4\xi\pi}{3} &= \epsI \frac{4\xi\pi}{3},\\
    \frac{4\xi\pi}{3} &= \epsSig\frac{2\xi\pi}{3}. 
\end{align}
\end{subequations}
Therefore, $\xi \neq 0$ only if $\epsI = -\epsSig =1$.\\
The bow tie loops give constraints for the phases

\begin{subequations}
\label{eq:Aconstraints}
\begin{align}
    ( \phi_{A_1}- \phi_{A_2})&=\epsI(\phi_{A_2}- \phi_{A_1}),\\
     ( \phi_{A_1}- \phi_{A_2})&=\epsSig(\phi_{A_1}- \phi_{A_2}).
\end{align}
\end{subequations}


\subsection{Fluxes with $t^{h,\gamma}_{ij}$ fields}\noindent
As for the $\mathcal{B}_{ij}$ fields we only have to consider triangle flux: 
\begin{align}
\Arg(t^{h,\gamma}_{I01} t^{h,\gamma}_{I12}   t^{h,\gamma}_{I20})&.\label{equation:thFluxGeneral}
\end{align}
Using $t^{h,\gamma}_{ji}= - t^{h,\gamma*}_{ij}$ we get the phase equations
\begin{subequations}
\label{eq:thconstraints}
\begin{align}
\phi_{t^{h,\gamma}_2}+\frac{4\xi\pi}{3}+ n_1\pi&=\epsI(\phi_{t^{h,\gamma}_1}+\frac{4\xi\pi}{3}),\\
\phi_{t^{h,\gamma}_2}+\frac{4\xi\pi}{3} &= 3\epsSig\phi_{t^{h,\gamma}_2} +  \delta_{\gamma,y'}  \pi.
\end{align}
\end{subequations}
The term $ \delta_{\gamma,y'} \pi$ comes from the spin rotation part of $\Sigma$.

\subsection{Fluxes with $t^{p,\gamma}_{i,j}$ fields}\noindent
Since $t^{p}$ are pairing fields, we have to consider the even rhombi and bow tie loops.
\begin{subequations}
\begin{align}
  \Arg(t^{p,\gamma}_{I01} t^{p,\gamma*}_{I12} t^{p,\gamma}_{I23} t^{p,\gamma*}_{I30})&, \\
    \Arg(t^{p,\gamma}_{I01} t^{p,\gamma*}_{I12} t^{p,\gamma}_{I20} t^{p,\gamma*}_{01} t^{p,\gamma}_{12} t^{p,\gamma*}_{20})&.
\end{align}
\end{subequations}
The rhombus fluxes give the same constraints ($ \xi \neq 0$ only if $(\epsI,\epsSig) = (1,-1)$) as Eq.~\eqref{eq:kconstraints}.
The bow tie fluxes give:
\begin{subequations}
\label{eq:tpconstraints}
\begin{align}
    ( \phi_{t^{p,\gamma}_1}- \phi_{t^{p,\gamma}_2})&=\epsI(\phi_{t^{p,\gamma}_2}- \phi_{t^{p,\gamma}_1}),\\
     ( \phi_{t^{p,\gamma}_1}- \phi_{t^{p,\gamma}_2})&=\epsSig(\phi_{t^{p,\gamma}_1}- \phi_{t^{p,\gamma}_2}).
\end{align}
\end{subequations}


\subsection{Fluxes with $\mathcal{A}_{ij}$ and $\mathcal{B}_{ij}$ fields }\noindent
We only have to consider triangle loops with one $\mathcal{B}_{ij}$ field. 

\begin{align}
 \Arg(\mathcal{B}_{I01} \mathcal{A}^*_{I12}   \mathcal{A}_{I20}). \label{equation:ABFluxGeneral} 
\end{align}
These give the constraints:
\begin{subequations}
\label{eq:ABconstraints1}
\begin{align}
    \phi_{B_2} + (1 + n_1)\pi &= \epsI\phi_{B_1} +  \pi,\\
    \phi_{B_2} + \pi &= \epsSig(\phi_{B_2} + \frac{2\xi\pi}{3}).
\end{align}
\end{subequations}
\noindent
Using Eq.~\eqref{eq:Bconstraints} and \eqref{eq:kconstraints} we can reduce this to:
\begin{subequations}
\label{eq:ABconstraints}
\begin{align}
      \epsSig &= -1, \\
    2\phi_{B_1} &= \pi + \frac{4\xi\pi}{3}, \\
    \phi_{B_2} &= \epsI \phi_{B_1}+\pi n_1.
\end{align}
\end{subequations}


\subsection{Fluxes with $\mathcal{A}_{ij}$ and $t^{h,\gamma}_{i,j}$ fields}\noindent
We only have to consider the triangle loops with one $t^{h,\gamma}_{i,j}$ field:
\begin{align}
 \Arg(t^{h,\gamma}_{I01} \mathcal{A}^*_{I12}   \mathcal{A}_{I20}).\label{equation:AthFluxGeneral}
\end{align}
They give constraints
\begin{subequations}
\label{eq:Athconstraints1}
\begin{align}
    \phi_{t^{h,\gamma}_2} + (1+ n_1) \pi &= \epsI\phi_{t^{h,\gamma}_1} +\pi,\\
    \phi_{t^{h,\gamma}_2}  &= \epsSig(\phi_{t^{h,\gamma}_2} + \frac{2\xi\pi}{3}) +  \delta_{\gamma,y'}\pi.
\end{align}
\end{subequations}
\noindent
Using Eq.~\eqref{eq:thconstraints} we can rewrite this to 
\begin{subequations}
\label{eq:Athconstraints}
\begin{align}
    \delta_{\gamma,y'} &= 0, \label{eq:thNoSolFory'}\\
    2\phi_{t^{h,\gamma}_2} &= \frac{4\xi\pi}{3},\\
    \phi_{t^{h,\gamma}_2} &= \epsI  \phi_{t^{h,\gamma}_1}+n_1 \pi.
\end{align}
\end{subequations}
Eq.~\eqref{eq:thNoSolFory'} says that there are no valid chiral  ansätze with both $\mathcal{A}_{ij}$ and $t^{h,y'}_{ij}$ fields.

\subsection{Fluxes with $\mathcal{A}_{ij}$ and $t^{p,\gamma}_{i,j}$ fields}\noindent
As established in the main text $\mathcal{A}_{ij}$ and $t^{p,\gamma}_{i,j}$ fields can not appear simultaneously in an ansatz. Therefore, we do not have to consider loops with both of these fields.

\subsection{Fluxes with $\mathcal{B}_{ij}$ and $t^{p,\gamma}_{i,j}$ fields}\noindent
Here we have to consider similar triangle loops as for $\mathcal{A}_{ij}$ and  $\mathcal{B}_{ij}$ fields
\begin{align}
 \Arg(\mathcal{B}_{I01} t^{h,p*}_{I12}   t^{h,p}_{I20}).\label{equation:BtpFluxGeneral}
\end{align}
These lead to the constraints: 
\begin{subequations}
\label{eq:tpBconstraints1}
\begin{align}
     \phi_{B_2}+ n_1\pi &= \epsI\phi_{B_1},  \\
    \phi_{B_2}  &= \epsSig(\phi_{B_2} + \frac{2\xi\pi}{3}).
\end{align}
\end{subequations}
With Eq.~\eqref{eq:Bconstraints} these can be reduced to: 
\begin{subequations}
\begin{align}
    2\phi_{B_1} &=  \frac{4\xi\pi}{3}, \label{eq:PhiB1constraintBtp}\\
    \phi_{B_1} &= \epsI \phi_{B_2}+ n_1\pi. 
\end{align}
\end{subequations}


\subsection{Fluxes with $\mathcal{B}_{ij}$ and $t^{h,\gamma}_{i,j}$ fields}
\noindent
For $\mathcal{B}_{ij}$ and $t^{h,\gamma}_{ij}$ fields we only have to consider the trivial flux:
\begin{equation}
     \Arg(\mathcal{B}_{01}t^{h,\gamma}_{10}).
\end{equation}
which gives the constraints:
\begin{align}
    (\phi_{t^{h,\gamma}_2}-\phi_{B_2}) &=   \epsI(\phi_{t^{h,\gamma}_1}-\phi_{B_1}) \\
     (1-\epsSig)(\phi_{t^{h,\gamma}_2}-\phi_{B_2}) &=   \pi(1- \delta_{\gamma,y'})  \label{eq:bth}.
\end{align}


\subsection{Fluxes with $t^{p,\gamma}$ and $t^{h,\gamma'}$ fields}\noindent
We have to consider the triangle loop
\begin{align}
    \Arg ( t^{h,\gamma'}_{I01} t^{p,\gamma*}_{I12} t^{p,\gamma}_{I20}).
\end{align}
which give constraints: 
\begin{subequations}
\begin{align}
     \phi_{t^{h,\gamma'}_2}+ n_1\pi &= \epsI\phi_{t^{h,\gamma'}_1},  \\
    \phi_{t^{h,\gamma'}_2}   &= \epsSig(\phi_{t^{h,\gamma'}_2} + \frac{2\xi\pi}{3})+\pi(1- \delta_{\gamma',y'} ).
\end{align}
\end{subequations}
which can be reduced with Eq.~\eqref{eq:thconstraints} to
\begin{subequations}
\begin{align}
     2\phi_{t^{h,\gamma'}_1} &= \pi+  \frac{4\xi\pi}{3}, \label{eq:Phith1constraintthtp}\\
    \phi_{t^{h,\gamma'}_1} &= \epsI\phi_{t^{h,\gamma'}_4}+ n_1\pi.
\end{align} 
\end{subequations}


\subsection{Fluxes with $t^{p,\gamma}$ and $t^{p,\gamma'}$ fields}\noindent
We have to consider the trivial flux:
\begin{align}
    \Arg( t^{p,\gamma }_{I01}t^{p,\gamma'*}_{I10}).
\end{align}
which gives the constraints:

\begin{subequations}
\begin{align}
    \phi_{t^{p,\gamma }_2} -\phi_{t^{p,\gamma'}_2} &= \epsI( \phi_{t^{p,\gamma }_1} -\phi_{t^{p,\gamma'}_1}),\\
    (1-\epsSig)(\phi_{t^{p,\gamma }_1} -\phi_{t^{p,\gamma' }_1}) &= \pi(\delta_{\gamma ,y'} + \delta_{\gamma',y'}).
\end{align}
\end{subequations}
When $\epsSig  = 1$ there are no solutions for  $\gamma \neq \gamma' = y'$.


\subsection{Fluxes with $t^{h,\gamma}$ and $t^{h,\gamma'}$ fields}
\noindent
We have to consider the trivial flux:
\begin{align}
   \Arg(t^{h,\gamma'}_{I01}t^{h,\gamma}_{I10}).
\end{align}
which gives the constraints:

\begin{subequations}
\begin{align}
 (\phi_{t^{h,\gamma}_2}-\phi_{t^{h,\gamma'}_2})  &=  \epsI(\phi_{t^{h,\gamma}_1}-\phi_{t^{h,\gamma'}_1}),\\
    (1-\epsSig )(\phi_{t^{h,\gamma}_2}-\phi_{t^{h,\gamma'}_2})  &=   \pi(\delta_{\gamma,y'} + \delta_{\gamma',y'}),   \label{eq:ththconst}
\end{align}
\end{subequations}


\subsection{Solutions of phase equations} \noindent
We organize the solutions to the phase equations in Table \ref{tab:WeaklySymmetricAnsaetze}. We only list $\Ztwo$ spin liquid ansätze with at least one pairing field. Ansätze with only hopping fields can also be derived by the phase equations. They are, however, behaving like  U(1) spin liquids and are thus subjected to the Higgs mechanism. 
The phases of the mean-field parameters $\phi_{a^p_1}, \phi_{b^p_1}, \phi_{c^p_1},\phi_{d^p_1}$  have to be related to the phases of the mean-fields $\phi_{A_1},\phi_{t^{p,x}_1},\phi_{t^{p,y}_1},\phi_{t^{p,z}_1}$  like 
\begin{align}
(\phi_{a^p_1}, \phi_{b^p_1}, \phi_{c^p_1},\phi_{d^p_1})  &=  (\phi_{A_1},\phi_{t^{p,x}_1},\phi_{t^{p,y}_1},\phi_{t^{p,z}_1}), \\ 
(\phi_{a^p_2}, \phi_{b^p_2}, \phi_{c^p_2},\phi_{d^p_2})  &=  (\phi_{A_2},\phi_{t^{p,x}_2},\phi_{t^{p,y}_2},\phi_{t^{p,z}_2}), \\ 
(\phi_{a^h_1}, \phi_{b^h_1}, \phi_{c^h_1},\phi_{d^h_1})  &=  (\phi_{B_1},\phi_{t^{h,x}_1},\phi_{t^{h,y}_1},\phi_{t^{h,z}_1}), \\ 
(\phi_{a^h_2}, \phi_{b^h_2}, \phi_{c^h_2},\phi_{d^h_2})  &=  (\phi_{B_2},\phi_{t^{h,x}_2},\phi_{t^{h,y}_2},\phi_{t^{h,z}_2}).
\end{align}
for the mean-field Hamiltonian to have the same symmetry as the state that we want to construct.


\begin{figure}[h!]
    \includegraphics[scale = 0.4]{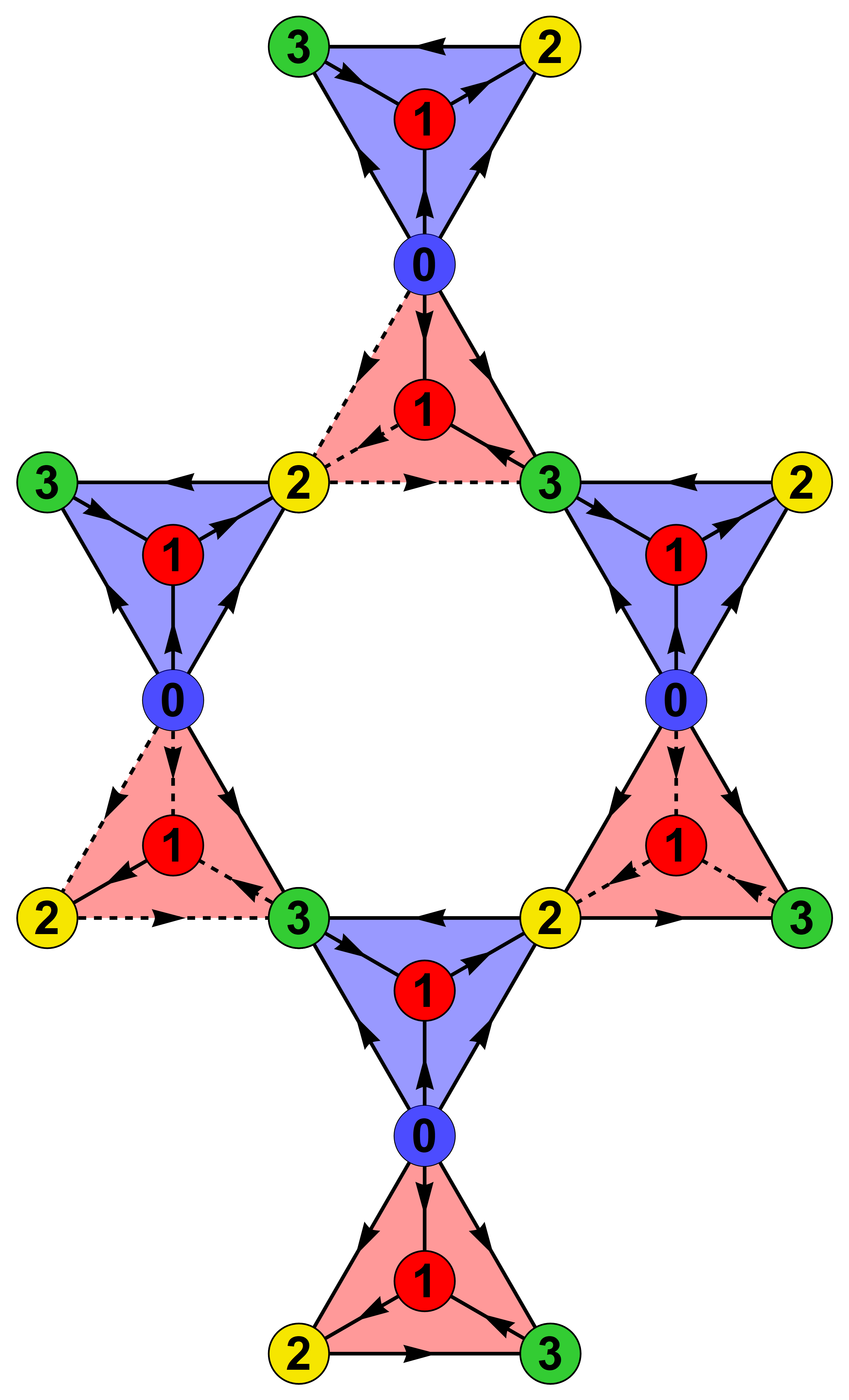}
    \caption{(Color online). The 16-site unit cell of the $n_1=1$ ansätze. The expectation values of the bond operators are defined by the  $u^t_{\mu\nu}$ matrices given in Eq. \eqref{eq:umatrixinstruction}. The direction $\mu \xrightarrow[]{} \nu$ is indicated by the arrowheads. On each dashed bond the $u^t_{\mu\nu}$ matrices have to be multiplied by an extra phase factor of $\exp (i\pi n_1)$. For $n_1=0$ the mean field ansatz is fully described by the four-site unit cell without any dashed lines. The dependence of the $u^t_{\mu\nu}$ matrices on the mean-fields is described in Eq. \eqref{eq:uMatrixShortNotation}.}
    \label{fig:my_label}
\end{figure}

\section{Hamiltonians for $n_1 = 0$ and for $n_1 = 1$}
\label{appendix:ExplicitHamiltonians}

We give the explicit form of the submatrices of Eq.~\eqref{eq:HamiltonianStructure} for both $n_1 = 0$ and for $n_1 = 1$. To keep things compact we introduce the notation $t\in \{h,p\}$ :
\begin{align}
    &u_{\mu\nu}= u^t_{ \mu  \nu}e^{\frac{i}{2}(\mathbf{a}_\mu-\mathbf{a}_\nu)\kvec}= u^t_{\mathbf{0}_\mu \mathbf{0}_\nu}e^{\frac{i}{2}(\mathbf{a}_\mu-\mathbf{a}_\nu)\kvec}, \\
    & u_{I\mu\nu}=u^t_{I \mu \nu}e^{-\frac{i}{2}(\mathbf{a}_\mu-\mathbf{a}_\nu)\kvec}= u^t_{I(\mathbf{0}_\mu) I(\mathbf{0}_\nu)}e^{-\frac{i}{2}(\mathbf{a}_\mu-\mathbf{a}_\nu)\kvec}.  
\end{align}
The submatrices fulfill $H^h(\kvec)=(H^h(\kvec))^\dagger$ and $H^t(\kvec)=(H^t(-\kvec))^T$  so we only need to give the upper triangular part to fully determines the whole matrices. For $n_1 = 0$, $H^t(k)$ are $8\times8$ matrices given by
\begin{equation}
  H^t(\kvec)  = \begin{pmatrix} 
                0 &  u_{01} + u_{I01}  & u_{02} + u_{I02}& u_{03} + u_{I03} \\
                   & 0  & u_{12} + u_{I23}& u_{13} + u_{I23}  \\
                  &     & 0 & u_{23} + u_{I23} \\
                  &     &   & 0  \\
                    \end{pmatrix}.
\end{equation}
For $n_1 = 1$, $H^t(k)$ are $32\times 32$ matrices:
\begin{equation}
    H^t(\kvec) = 
    \begin{pmatrix}
    H^t_{11}(\kvec) & H^t_{12}(\kvec) & H^t_{13}(\kvec) & H^t_{14}(\kvec) \\
                  & H^t_{22}(\kvec) & H^t_{23}(\kvec) & H^t_{24}(\kvec) \\
             &    & H^t_{33}(\kvec) & H^t_{34}(\kvec) \\
             &        &         & H^t_{44}(\kvec) \\
    \end{pmatrix}.
\end{equation}
The unit cell consists out of four main tetrahedra $q~\in~\{1,2,3,4\}$ and the submatrices $H^t_{q_1,q_2}$ include all bonds between main tetrahedron $q_1$ and $q_2$. They are given by:
\begin{equation}
  H^t_{qq}(\kvec)  = \begin{pmatrix} 
                 0&  u_{01} + u_{I01}e^{in_1\pi(\delta_{q,2}+\delta_{q,3})}  & u_{02} & u_{03}  \\
                                                & 0  & u_{12} & u_{13}  \\
                                                  &        & 0  & u_{23}  \\
                                                            &     &    &  0  \\
                    \end{pmatrix},
                    \end{equation}
\begin{widetext}

\begin{align*}
  H^t_{12}(\kvec)  = &\begin{pmatrix} 
                0       & 0      & u_{I02} & 0  \\
                0       &   0  & u_{I12} & 0  \\
                 u_{I20}   &  u_{I21}e^{i n_1\pi} &    0  &    0\\
                0       &   0   &   0   &   0   \\
                    \end{pmatrix}, &
  H^t_{13}(\kvec)   =  &\begin{pmatrix} 
                0       & 0         &        0  & u_{I03}  \\
                0       &   0       &       0    & u_{I13}  \\
                0       &       0   &    0      &    0\\
                 u_{I30}   &    u_{I31}e^{i n_1\pi}    &   0   &   0   \\
                    \end{pmatrix},
 \nonumber \\
  H^t_{14}(\kvec)  = &\begin{pmatrix} 
                0       & 0         &        0  & 0  \\
                0       &   0       &    0       & 0  \\
                0       &       0   &    0      &     u_{I23} \\
                0  &   0   &    u_{I32}e^{i n_1\pi}     & 0    \\
                    \end{pmatrix}, &
  H^t_{23}(\kvec)  = &\begin{pmatrix} 
                0       & 0         &        0  & 0  \\
                0       &   0       &    0       & 0  \\
                0       &       0   &    0      &    u_{I23}\\
                0  &   0   &    u_{I32}e^{i n_1\pi}     & 0    \\
                    \end{pmatrix},
 \\
  H^t_{24}(\kvec)  = &\begin{pmatrix} 
                0       & 0         &        0  &  u_{I03}   \\
                0       &   0       &       0    &  u_{I13}e^{i n_1\pi}  \\
                0       &       0   &    0      &    0\\
                 u_{I30}   &    u_{I31}    &   0   &   0   \\
                    \end{pmatrix}, &
  H^t_{34}(\kvec)  = &\begin{pmatrix} 
                0       & 0      &  u_{I02}e^{i n_1\pi}  & 0  \\
                0       &   0   &  u_{I12}  & 0  \\
                 u_{I20}e^{i n_1\pi}   &   u_{I21}e^{i n_1\pi}  &    0  &    0\\
                0       &   0   &   0   &   0   \\
                    \end{pmatrix}.
\end{align*}

For the chiral PSG the $u^t_{ij}$ matrices are given by
\begin{subequations}
\begin{align}
    u^t_{01} &= (a^t_1,b^t_1,c^t_1,d^t_1),\\
    u^t_{02} &= (a^t_1,-\frac{1}{2}(b^t_1 +\sqrt{3}c^t_1),\frac{1}{2}(\sqrt{3}b^t_1-c^t_1),d^t_1 )e^{-i\frac{2\xi}{3}\pi},\\
    u^t_{03} &= (a^t_1,-\frac{1}{2}(b^t_1 -\sqrt{3}c^t_1),-\frac{1}{2}(\sqrt{3}b^t_1+c^t_1),d^t_1 ) e^{-i\frac{4\xi}{3}\pi},\\
    u^t_{12} &= ( a_1^t,-\frac{1}{2}(b_1^t -\sqrt{3}c_1^t),-\frac{1}{2}(\sqrt{3}b_1^t + c_1^t),d_1^t )e^{\pm_t i\frac{2\xi}{3}\pi} e^{i n_{C_3C_3'}\pi}, \\
    u^t_{31} &= (a^t_1,-\frac{1}{2}(b^t_1 +\sqrt{3}c^t_1),\frac{1}{2}(\sqrt{3}b^t_1-c^t_1),d^t_1 )e^{\pm_t i\frac{2\xi}{3}\pi}e^{i n_{C_3C_3'}\pi},  \\
    u^t_{23} &=(a^t_1,b^t_1,c^t_1,d^t_1)^te^{\pm_t i\frac{2\xi}{3}\pi} e^{i n_{C_3C_3'}\pi}, \\
    u^t_{I01} &= (a^t_2,b^t_2,c^t_2,d^t_2),  \\
    u^t_{I02} &= (a^t_2,-\frac{1}{2}(b^t_2 +\sqrt{3}c^t_2),\frac{1}{2}(\sqrt{3}b^t_2-c^t_2),d^t_2 ) e^{-i\frac{2\xi}{3}\pi},  \\
    u^t_{I03} &= (a^t_2,-\frac{1}{2}(b^t_2 -\sqrt{3}c^t_2),-\frac{1}{2}(\sqrt{3}b^t_2 + c^t_2),d^t_2 )e^{-i\frac{4\xi}{3}\pi}, \\
    u^t_{I12} &=( a_2^t,-\frac{1}{2}(b_2^t -\sqrt{3}c_2^t),-\frac{1}{2}(\sqrt{3}b_2^t + c_2^t),d_2^t )e^{\pm_t i\frac{2\xi}{3}\pi}e^{i(n_1+n_{C_3C_3'})\pi},  \\
    u^t_{I31} &= (a^t_2,-\frac{1}{2}(b^t_2 +\sqrt{3}c^t_2),\frac{1}{2}(\sqrt{3} b^t_2-c^t_2) ,d^t_2)e^{\pm_t i\frac{2\xi}{3}\pi}e^{i(n_1+n_{C_3C_3'})\pi},  \\
    u^t_{I23} &=(a^t_2,b^t_2,c^t_2,d^t_2) e^{\pm_t i\frac{2\xi}{3}\pi} e^{i(n_1+n_{C_3C_3'})\pi}.
\end{align}
\label{eq:umatrixinstruction}
\end{subequations}
where for $t=h$, $\pm_h=+$ and for $t=p$, $\pm_p=-$.

\end{widetext}

\bibliographystyle{apsrev4-1}
\bibliography{references}

\end{document}